\documentclass[useAMS,usenatbib]{mn2e}
\usepackage{graphicx}
\usepackage[hang]{subfigure}
\usepackage{ amssymb }
\usepackage{color}
\usepackage{ amsmath }
\usepackage{fmtcount}
\usepackage{array,multirow}
\usepackage{booktabs}
\usepackage{appendix}

\usepackage[dvipsnames]{xcolor}

\newcommand{\txn}[1]{\textnormal{#1}}

\newcommand{\HST}{\emph{HST}}
\newcommand{\JWST}{\emph{JWST}}
\newcommand{\ALMA}{\emph{ALMA}}
\newcommand{\Spitzer}{\emph{Spitzer}}
\newcommand{\beagle}{\textsc{beagle}}
\newcommand{\tlusty}{\textsc{tlusty}}
\newcommand{\powr}{\textsc{powr}}
\newcommand{\starburst}{\textsc{starburst99}}

\newcommand{\pandeia}{\textsc{pandeia}}

\newcommand{\bagpipes}{\textsc{bagpipes}}
\newcommand{\prospector}{\textsc{prospector}}
\newcommand{\leopy}{\textsc{LEO-Py}}

\newcommand{\ideal}{ideal-G}
\newcommand{\constGauss}{const-G}
\newcommand{\delGauss}{delexp-G}
\newcommand{\constMF}{const-MF}
\newcommand{\sfrFree}{const+SF10}

\newcommand{\PM}{PM}
\newcommand{\PMSalmon}{PM-R12-\tauV}
\newcommand{\PMSantini}{PM-K12-UVS}
\newcommand{\Kelly}{BH}

\newcommand{\M}{\hbox{$\txn{M}$}}

\newcommand{\Mstar}{\hbox{$\M_{\star}$}}
\newcommand{\barMstar}{\hbox{$\bar{\M}_{\star}$}}
\newcommand{\MstarInLog}{\hbox{$\txn{M}$}}

\newcommand{\MtotInLog}{\hbox{$\txn{M}_\txn{tot}$}}

\newcommand{\yr}{\hbox{$\txn{yr}$}}
\newcommand{\Msun}{\hbox{$\txn{M}_{\odot}$}}

\renewcommand{\d}{\hbox{$d$}}
\renewcommand{\t}{\hbox{$t$}}

\newcommand{\tausfr}{\hbox{$\tau_\textsc{sfr}$}}

\newcommand{\sfr}{\hbox{${\Psi}$}}
\newcommand{\barsfr}{\hbox{$\bar{\Psi}$}}
\newcommand{\sfrInLog}{\hbox{${\psi}$}}

\newcommand{\SFR}{\hbox{${\Psi}$}}

\newcommand{\Z}{\hbox{$\txn{Z}$}}
\newcommand{\Zism}{\hbox{$\Z_\textsc{ism}$}}

\newcommand{\Zsun}{\hbox{$\Z_{\odot}$}}

\newcommand{\MstarSFR}{\hbox{$\sfr-\Mstar$}}

\newcommand{\Db}{\hbox{$\mathbf{D}$}}

\newcommand{\HII}{\mbox{H\,{\sc ii}}}
\newcommand{\OIII}{\mbox{[O\,{\sc iii}]$\lambda4959,5007\,$\AA}}
\newcommand{\HI}{\mbox{H\,{\sc i}}}
\newcommand{\nH}{\hbox{$n_\textsc{h}$}}

\newcommand{\logUs}{\hbox{$\log U_\textsc{s}$}}
\newcommand{\xid}{\hbox{$\xi_\txn{d}$}}

\newcommand{\tauV}{\hbox{$\hat{\tau}_\textsc{v}$}}
\newcommand{\tauBC}{\hbox{$\hat{\tau}^\textsc{bc}_\textsc{v}$}}
\newcommand{\tauHII}{\hbox{$\hat{\tau}^\textsc{hii}_\textsc{v}$}}
\newcommand{\tauHI}{\hbox{$\hat{\tau}^\textsc{hi}_\textsc{v}$}}

\newcommand{\mud}{\hbox{$\mu_\mathrm{d}$}}
\newcommand{\Auv}{\hbox{$\mathrm{A}_{\mathrm{UV}}$}}

\newcommand{\slope}{\hbox{$\beta$}}
\newcommand{\intercept}{\hbox{$\alpha$}}
\newcommand{\scatter}{\hbox{$\sigma$}}

\newcommand{\modelThetaB}{\hbox{$\bmath{\Theta}$}}

\newcommand{\prob}{\hbox{$\txn{P}$}}
\newcommand{\conditional}[2]{\hbox{$\prob(#1 \mid #2)$}}

\newcolumntype{L}[1]{>{\raggedright\let\newline\\\arraybackslash\hspace{0pt}}m{#1}}
\newcolumntype{C}[1]{>{\centering\let\newline\\\arraybackslash\hspace{0pt}}m{#1}}
\newcolumntype{R}[1]{>{\raggedleft\let\newline\\\arraybackslash\hspace{0pt}}m{#1}}

\title[M$_{*}$-SFR]{Modelling the M*-SFR relation at high redshift: untangling factors driving biases in the intrinsic scatter measurement}
\author[E. Curtis-Lake et al.]{E. Curtis-Lake$^{1,2,3}$\thanks{Email: ec296@cam.ac.uk},
 J. Chevallard$^{1}$, S. Charlot$^{1}$, L. Sandles$^{2,3}$\\
$^{1}$ Sorbonne Universit\'e, UPMC-CNRS, UMR7095, Institut d’Astrophysique de Paris, F-75014, Paris, France\\
$^{2}$ Cavendish Astrophysics, University of Cambridge,  Cambridge, CB3 0HE, UK \\
$^{3}$ Kavli Institute for Cosmology, University of Cambridge, Madingley Road, Cambridge CB3 0HA, UK}

\begin{document}

\date{}

\pagerange{\pageref{firstpage}--\pageref{lastpage}} \pubyear{2002}

\maketitle

\label{firstpage}

\begin{abstract}

We present a method to self-consistently propagate stellar-mass [$\Mstar=\log(\MstarInLog/\Msun)$]
 and star-formation-rate [$\sfr=\log(\sfrInLog/\Msun\,\txn{yr}^{-1}$)]
 uncertainties onto intercept ($\alpha$), slope ($\beta$) and intrinsic-scatter ($\sigma$) estimates for a simple model of the main sequence of star-forming galaxies, where $\sfr = \alpha + \beta\Mstar + \mathcal{N}(0,\sigma)$.  To test this method and compare it with other published methods, we construct mock photometric samples of galaxies at $z\sim5$ based on idealised models combined with broad- and medium-band filters at wavelengths 0.8--5\,$\mu$m. Adopting simple \sfr\ estimates based on dust-corrected ultraviolet luminosity can under-estimate $\sigma$. We find that broad-band fluxes alone cannot constrain the contribution from emission lines, implying that strong priors on the emission-line contribution are required if no medium-band constraints are available.  Therefore at high redshifts, where emission lines contribute a higher fraction of the broad-band flux, photometric fitting is sensitive to \sfr\ variations on short ($\sim10$ Myr) timescales.  Priors on age imposed with a constant (or rising) star formation history (SFH) do not allow one to investigate a possible dependence of $\sigma$ on \Mstar\ at high redshifts. Delayed exponential SFHs have less constrained priors, but do not account for \sfr\ variations on short timescales, a problem if $\sigma$ increases due to stochasticity of star formation.  A simple SFH with current star formation decoupled from the previous history is appropriate. We show that, for simple exposure-time calculations assuming point sources, with low levels of dust, we should be able to obtain unbiased estimates of the main sequence down to $log(\MstarInLog/\Msun)\sim8$ at $z\sim5$ with the James Webb Space Telescope while allowing for stochasticity of star formation.
\end{abstract}


\newpage

\section{Introduction}

Trends and correlations between different properties of observed galaxies provide vital clues for decoding the underlying physical processes that govern galaxy evolution.  One such trend is the tight correlation between star formation rate (SFR) and  stellar mass (\Mstar) for `normal' star forming galaxies (e.g. \citealt{Brinchmann2004} at $z\sim0$, \citealt{Elbaz2007} at $z\sim1$ and \citealt{Daddi2007} at $z\sim2$).  This is the so-called `main sequence of star-forming galaxies', labelled as such by \cite{Noeske2007}, which we will refer to simply as the `main sequence', or \MstarSFR\ relation throughout this paper.   The physical origin of this trend is still an active area of debate \citep[e.g.,][]{Kelson2014,Lin2019,Matthee2019}.  Perhaps it is only a population average, with objects above and below the relation having very different evolutionary histories.  Or perhaps physical processes, such as gas infall and feedback from stars and active galactic nuclei, draw galaxies back to the main sequence if they venture too far from it in either direction before the eventual cessation of star formation that causes them to drop off the main sequence altogether.  Either of these scenarios would produce a trend between \Mstar\ and SFR, but the origin of the scatter about the relation would be very different.  In the first scenario the distribution of SFR values at given stellar mass and redshift would be determined by the range of possible evolutionary paths, while in the second scenario it would be shaped by feedback duty-cycles that produce short-timescale variations in SFR.  Whatever the physical reason for the relation, it must simultaneously explain the existence of the correlation as well as its tightness, with an observed scatter of $\sigma$(log \SFR/\Msun yr$^{-1}$)$\sim0.20$--0.35\,dex \citep{Daddi2007,Speagle2014,Shivaei2015b,Salmon2015,Kurczynski2016a,Santini2017}.  

A major challenge in understanding the physical origin of the main-sequence is the difficulty of linking progenitor and descendent galaxy populations at different epochs. This has motivated the use of galaxy formation models to guide the interpretation of the main sequence. For example, the results of \citet[][using the semi-analytic models of \citealt{Dutton2009}]{Dutton2010a} support the first scenario where the position of a galaxy within the main sequence is set by the gas accretion history of its parent halo. Halos which started to accrete gas early lie above the relation, and vice-versa.  A limitation of the approach of \cite{Dutton2010a} is their simple treatment of the mass accretion histories, which are modelled to be smoothly evolving and neglect the impact of mergers. By introducing halo merger driven fluctuations in gas accretion in the analytical equilibrium model of \cite{Mitra2015}, \cite{Mitra2017} show that the scatter produced is larger than that found by \cite{Dutton2010a} ($\sim0.25$\,dex compared to $\sim0.12$\,dex), more in line with observations \citep[see also][]{Forbes2014}.  \cite{Mitra2017} predict only a modest redshift dependence of the scatter, increasing to higher redshifts ($\sim0.2$\,dex at $z\sim0.5$ to $\sim0.25$\,dex at $z\sim4$). 

The second scenario is supported by \cite{Tacchella2016a}, who used cosmological zoom-in simulations to follow the star formation and gas accretion histories of 26 simulated galaxies.  By tracking galaxy positions on the main sequence across cosmic time, \cite{Tacchella2016a} find that their simulated galaxies oscillate about the main sequence due to intricate gas dynamical effects. Although zoom-in simulations allow smaller scale processes to be resolved, they lack the statistics to compare the impact of these processes to those driven by halo mass accretion histories (the main driver of scatter in the \citealt{Dutton2010} SAM and \citealt{Mitra2017} semi-analytic model).

Most likely the main sequence is shaped by a combination of the two scenarios, but to disentangle their relative contributions requires cosmological hydrodynamic simulations of galaxy formation that include prescriptions of sub-grid physics that cannot be resolved in the simulation itself.  For instance,  \cite{Matthee2019} find that SFR variations over both long and short timescales contribute to the scatter about the main sequence within the EAGLE simulation of galaxy formation and evolution \citep{Crain2015,Schaye2015}.  They find that galaxies may cross the main sequence many times over their lifetime (as in the simulations studied in \citealt{Tacchella2016a}), but galaxies fluctuate about tracks that depend on the halo properties, most noticeably the halo formation time (reminiscent of the \citealt{Dutton2010} results).

While at $z\sim0$ the scatter in the EAGLE simulation is shown to modestly increase with decreasing stellar mass, at $0\lesssim z\lesssim4$, \cite{Matthee2019} find a greater decrease in scatter with increasing redshift at $\sim10^9\,\Msun$ than at higher masses.  The higher scatter at $\sim10^{10}\,\Msun$ compared to $\sim10^9\,\Msun$ at $\sim2-4$ is attributed to the enhanced influence of AGN feedback at the higher stellar masses.  At lower stellar masses, it is likely that \textit{stellar} feedback prescriptions have a greater impact on the scatter.  \cite{Sparre2015} investigate the form of the main-sequence in the Illustris simulations \citep{Vogelsberger2014}, finding constant scatter of $\scatter\sim0.2$--0.3\,dex at $\Mstar\lesssim10^{10.5}\,\Msun$ at $0\lesssim z\lesssim4$. The FIRE simulations \citep[Feedback in Realistic Environments,][]{Hopkins2014} use zoom-in simulations to resolve the ISM in galaxies down to scales required to properly model stellar feedback.  \cite{Sparre2017} show that these simulations predict an increase in scatter to low stellar masses at $z\sim2$ (see their figs~3 and 4), and predict much higher scatter in SFR than seen in the EAGLE simulation at these stellar masses ($\sim0.4\,$dex compared to $\sim0.2\,$dex in the EAGLE simulations at $10^9\Msun$). High-redshift simulations reveal bursty star formation histories (SFHs) of low-mass galaxies \citep[e.g.][]{Dayal2013,Yajima2017,Rosdahl2018}, which would likely lead to higher scatter in the main-sequence to low stellar masses, as seen in the FIRE simulations \citep{Sparre2017}. 

Measuring the evolution of the main sequence and its intrinsic scatter across a wide range of masses and redshifts can therefore constrain different galaxy evolution scenarios, and help to disentangle what physical mechanisms shape the main sequence (e.g. the gas dynamical properties, stellar and AGN feedback prescriptions or halo mass accretion rate distributions). In practice, the measurements are limited by the significant uncertainties affecting SFR and stellar-mass estimates,  especially at high redshifts.  Robust measurements of galaxy SFR can be obtained from measurements of H$\alpha$ line flux (using H$\beta$ to help correct for dust attenuation), which traces the emission of massive stars re-processed by ionized gas; or through the sum of ultra-violet (UV) and infra-red (IR) emission, which probes both the direct (un-obscured) and attenuated (reprocessed by dust)  UV emission from these young stars.  At high redshifts, both types of measurements become difficult as H$\alpha$ is no-longer visible from the ground at $z\gtrsim2.5$, and bolometric IR measurements are challenging except in the most luminous objects \citep[e.g. with \textit{Herschel};][]{Gruppioni2013}.  The Atacama Large Millimeter Array (\ALMA) provides the sensitivity to search for dust emission from `normal' star-forming galaxies at $z\gtrsim3$, but direct detections were only acquired for a very small fraction of the $z\gtrsim3$ objects in the Hubble Ultra-Deep field \citep{Bouwens2016_alma,Dunlop2017}. Given these limitations, one must rely on fitting broad-band UV-to-near-IR photometry with galaxy spectral evolution models to derive redshifts, SFRs and stellar masses for large samples of galaxies at high redshifts. This approach has pushed constraints of the main sequence to redshifts as high as $z\sim7$, with studies showing that such a relation may have been in place since $\sim900$\,Myr after the Big Bang (e.g. \citealt{Gonzalez2011,Salmon2015,Santini2017}). 

Various studies have attempted to constrain the intrinsic scatter of the main sequence. For example, \citet[hereafter K16]{Kurczynski2016a} derived the slope, intercept and intrinsic scatter of the main sequence at $0.5<z<3$ by fitting the broad-band photometry from the \textit{Hubble} Ultra Deep Field \citep[HUDF12;][]{Ellis2013,Koekemoer2013}, Ultraviolet Ultra Deep Field \citep[UVUDF;][]{Teplitz2013} and CANDELS/GOODS-S \citep{Grogin2011,Koekemoer2011} campaigns. They find a moderate increase of the intrinsic scatter with increasing cosmic time (which is in direct contradiction to the simulations of \citealt{Mitra2017}), but no significant dependence on stellar mass. At higher redshifts, \citet[hereafter Sal15]{Salmon2015} used CANDELS data to provide the first estimates of the intrinsic scatter in 3 redshift bins at $3.5<z<6.5$, finding a scatter of $\sim$0.2--0.3\,dex (once deconvolved from measurement uncertainties). \citet[hereafter San17]{Santini2017} exploited the effect of gravitational lensing in the HFF to reach lower mass-completeness limits than those reached in standard blank-field surveys. Adopting a sophisticated modelling approach to account for the complicated selection function of their sample, they find tantalising, but uncertain, evidence that the intrinsic scatter of the main sequence increases at low stellar masses.

In practice, comparing the constraints on the main sequence obtained in the various studies mentioned above is complicated since they rely on different SED modelling prescriptions: K16 and San17 adopt delayed exponential star-formation histories (SFHs), while Sal15 use a constant SFH; K16 derive SFRs directly from SED fits, while Sal15 and San17 estimate SFRs using rest-frame UV magnitudes corrected with different dust-attenuation prescriptions. Finally, all three studies adopt different approaches to model the galaxy main sequence in the presence of uncertainties on stellar mass and SFR.

Regardless of the models used, the stellar masses and SFRs estimated in the above studies suffer from significant uncertainties arising from the faintness of high-redshift sources, the sparse sampling of their SEDs (especially at $z\gtrsim3.5$, where often the only detections redward of the Balmer break are supplied by the 3.6 and 4.5\,$\mu$m \textit{Spitzer} IRAC filters) and emission-line contamination to photometry (\citealt{Schaerer2009,Curtis-Lake2013,DeBarros2013,Smit2014}).  Emission-line contamination is particularly harmful, since it affects the photometric bands sampling the SED redward of the Balmer break, which more closely traces emission from stars accounting for the bulk of a galaxy's stellar mass. Intrinsic-scatter measurements are highly sensitive to this myriad of uncertainties, as well as to the methods employed to account for them.

In recent years, there has been a significant drive toward developing Bayesian spectral-modelling tools able to provide robust uncertainties on estimates of galaxy physical parameters  (e.g. \beagle\ \citealt{Chevallard2016}, \bagpipes\ \citealt{Carnall2018} and \prospector\ \citealt{Leja2017,Johnson2019}). In addition, efforts to formalise the inclusion of nebular (both continuum and line) emission in spectral models (e.g. \citealt{Gutkin2016,Byler2017,Plat2019}, following the method of \citealt{Charlot2001}) have led to the possibility of explicitly accounting for variations in \textit{nebular} parameters (e.g, interstellar metallicity, ionization parameter, metal depletion onto dust grains) and thus to self-consistently include the effect of emission lines on broad-band fluxes in the fitting process. These new spectral modelling techniques allow a better quantification of the uncertainties affecting physical-parameter estimates in individual galaxies, while the propagation of these uncertainties on population-wide parameters (e.g., intercept, slope and intrinsic scatter of the main sequence) still requires the development of innovative modelling approaches.

In this paper we present an original Bayesian Hierarchical approach to model the main sequence based on SED-fitting results obtained with \beagle.  This modelling allows us to self-consistently propagate the uncertainties of stellar mass and SFR estimates of individual galaxies to the measurement of the slope, intercept and intrinsic scatter of the main sequence.  We also use idealised mock galaxy catalogues to compare our approach to other methods employed in the literature. These tests allow us to understand how common SED-fitting approaches can affect estimates of the intercept, slope and intrinsic scatter of the main sequence. In Section~\ref{section:scenarios} we introduce the idealized scenarios that we set up to test different approaches to model the main sequence.  These approaches are described in Section~\ref{section:modelling}.  The results of our analysis are presented in Section~\ref{section:results} {and prospects for \JWST\ are discussed in Section~\ref{section:JWST}. In Section~\ref{section:Discussion} we discuss our findings in the context of previous and future studies, as well as the limitations in our approach.  Throughout the paper we adopt a \cite{Chabrier2003} initial mass function with upper-mass cutoff of 100\Msun\, and employ a flat $\Lambda$CDM cosmology with $\Omega_\Lambda=0.7$, $\Omega_M = 0.3$, and H$_0 = 70\,h_{70}\,$kms$^{-1}\,$Mpc$^{-1}$.

\section{Modelling the main sequence with beagle}
\label{section:scenarios}

In this paper we present a Hierarchical Bayesian approach to measuring slope, intercept and scatter of the main sequence and compare it to other methods employed in the literature.  To provide a basis for these comparisons, we construct a number of simple scenarios consisting of populations of mock galaxies with a known input main sequence.  In this section, we describe  the set of scenarios we consider, and in the next section, we will provide details about the different methods used to measure the slope, intercept and scatter.  This approach allows us to compare directly the strengths and weaknesses of these methods.

We set up four different scenarios with known input \MstarSFR\  relation,
which is modelled as a linear relation with Gaussian scatter in the dependent variable, \sfr:

\begin{equation}
\label{eq:main_sequence}
\conditional{\sfr}{\Mstar,\intercept,\slope,\scatter}\sim\intercept+\slope\Mstar+\mathcal{N}(0,\scatter^2)\\
\end{equation}
\noindent
where:
\begin{equation}
\begin{aligned}
\label{eq:mtot_and_sfr}
\Mstar = & \log(\MstarInLog/\Msun)\\
\sfr = & \log(\sfrInLog/\Msun\,\textrm{yr}^{-1})\\
\end{aligned}
\end{equation}

\noindent 
In equation~\eqref{eq:main_sequence}, \intercept, \slope\ and \scatter\ are the intercept, slope and intrinsic scatter, respectively. Throughout this paper we use the notation $\mathcal{N}(0,\sigma^2)$ to denote a Gaussian distribution with zero mean and standard deviation, $\sigma$ (where $\sigma^2$ denotes the variance).   In each of the four scenarios we use input values of $[\intercept,\slope,\scatter]=[-8,1,0.3]$. 

The four scenarios increase in complexity by progressively including more of the effects that we see in real galaxy samples. The first scenario (the \ideal\ scenario) is designed as an ideal test-case, where the measurement errors on \Mstar\ and \sfr\ are modelled to be correlated, bi-variate Gaussians: $\mathcal{N}_2\bigl([\Mstar,\sfr],\bigl[ \begin{smallmatrix}0.015 & -0.01\\ -0.01 & 0.015\end{smallmatrix}\bigr]\bigr)$, where $\mathcal{N}_2$ denotes a bi-variate Gaussian. The values of this covariance matrix were chosen to provide uncertainties that co-vary with maximum variance perpendicular to the input relation (which is chosen to have a slope, $\slope=1$) in order to resemble the uncertainties found by Sal15 (see their fig. 7).   The constant covariance matrix means that the measurement errors in this scenario are `homoskedastic'.  We produce 10 realizations, each containing 100 objects with \Mstar\ and \sfr\ values drawn from the input \MstarSFR\ relation.  These \Mstar\ and \sfr\ values are then perturbed by the Gaussian uncertainties described by the covariance matrix.  The other three scenarios (we will call these scenarios \constGauss, \delGauss\ and \constMF\ for reasons that will become apparent) use mock photometry of 1000 objects with known input physical properties that follow the chosen input \MstarSFR\ relation, before performing SED fitting in order to provide posterior probabilities of \Mstar\ and \sfr. 

To produce and fit the mock photometry, we use the new-generation Bayesian galaxy spectral modelling tool \beagle\ \citep{Chevallard2016}. This employs the most recent version of the \cite{Bruzual2003} stellar population synthesis models, incorporated self-consistently in the photoionization models of \cite{Gutkin2016}. These stellar plus photoionization models allow us to explore a wide range of galaxy SEDs, which are described by the physical parameters reported in Table \ref{tab:physical_parameter_list}.  We do not exploit the full range of nebular parameters allowed by the models, choosing to keep the hydrogen density fixed to $\nH=100$\,cm$^{-3}$, which is close to the typical value measured for galaxies at $z\sim2-3$ \citep[e.g.][]{Sanders2016,Strom2017}.  We also fix C/O ratio to the solar value of (C/O)$_\odot=0.44$, since lines that are strong enough to significantly affect broad-band photometry are unaffected by C/O.  Inter-galactic medium absorption is applied following the model of \cite{Inoue2014}.

We include dust attenuation following the physically-motivated  prescription of  \citet[hereafter CF00]{Charlot2000}, which accounts for the enhanced attenuation suffered by young stars still embedded in their stellar birth clouds.  In the case where the stellar birth clouds are ionization-bounded, they can be described by an inner ionised \HII\ region surrounded by a neutral \HI\ envelope.  The $V$-band dust attenuation in stellar birth clouds can then be written as $\tauBC=\tauHII+\tauHI$, where $\tauHII$ and $\tauHI$ are the contributions by dust in the ionized and neutral gas components. It is important to note that \HII-region dust is already included in the nebular-emission models, and that \tauHII\ depends on the nebular parameters of the \HII\ region. In \beagle, it is possible that values of $\tauBC=(1-\mud)\tauV$ computed from the total $V$-band attenuation optical depth of a galaxy, \tauV, and the fraction of this arising from dust in the diffuse ISM, \mud\ (the CF00 model parameters; see Table~\ref{tab:physical_parameter_list}), be inconsistent with a given nebular model, if $(1-\mud)\tauV<\tauHII$. In order to avoid this unphysical region of parameter space when producing our scenarios, we impose relations between the dust and nebular parameters of Table~\ref{tab:physical_parameter_list} following the approach of \cite{Williams2018}.  In particular, we impose the relation between \Mstar, \sfr\ and \Z\ measured by \cite{Hunt2016} from a compilation of $\sim1000$ galaxies at redshifts up to $z\sim3.7$, adopting a Gaussian scatter of $\sigma=0.1$ about this relation:
\begin{equation}
\label{eq:FMR}
12+\log\mathrm{(O/H)} = -0.14\sfr+0.37\Mstar+4.82 + \mathcal{N}(0,0.1)\,,
\end{equation}
where $12+\log\mathrm{(O/H)}\approx\log(Z/\Zsun)+8.79$ at $\xid=0.1$. In addition we impose a relation between \Z\ and \logUs, which is a polynomial fit to the data presented in \citet[priv. communication]{Carton2017}, again including some scatter about the relation: 
\begin{equation}
\logUs = -3.638+0.055Z+0.68Z^2 + \mathcal{N}(0,0.1)
\label{eq:logU_Z}
\end{equation}
Finally, values of \tauV\ are assigned using the prescription of \citet[]{Williams2018}, which ensures that objects with lower metallicity harbour less dust (we refer the reader to their section 3.4.3 for details). The distributions of parameters within the \constMF\ scenario are displayed in Fig.~\ref{fig:scenario_iv_params}.

When fitting to the mock SEDs with \beagle, we self-consistently account for the dust already present in the \HII\ regions.\footnote{As of BEAGLE v0.27.0.}   We do not fit using models that have more dust in the \HII\ region than allowed by the CF00 dust attenuation parameters \tauV\ and \mud. For a given single stellar population (SSP), the size of the \HII\ region will vary with age as the ionising photon flux varies, and so the dust optical depth within the \HII\ region, \tauHII(\t), also varies.  We assume that the dust in the birth cloud remains constant with age, and so for ages with lower ionising flux, and hence smaller \HII\ regions, the remaining dust is present in the \HI\ envelope.  When assessing whether a given nebular model is consistent with the CF00 dust parameters, we check that the maximum dust optical depth, $\tauHII_\mathrm{max}=\mathrm{max}[\tauHII(t)]$, is smaller than $\tauBC=(1-\mud)\tauV$ and apply the residual \HI\ attenuation individually to each single \HII-region age step which contributes to the final SED.

\begin{table*}
\centering
  \caption{List of physical parameters available to vary when producing \beagle\ simulated SEDs and when SED-fitting.}
  \begin{tabular}{l|l}
  \hline
  Parameter & Description\\
  \hline
  $z$ & Redshift fixed to $z=5$\\
  $\MtotInLog/\Msun$ & Integrated SFH\\
  \MstarInLog/\Msun & Stellar mass, including stellar remnants, \MstarInLog = \MtotInLog - mass returned to the ISM\\
  $t/\mathrm{year}$ & Age of oldest stars in galaxy\\
  $\tausfr/\mathrm{year}$ & Timescale of star formation for a delayed exponential SFH where $\sfrInLog(t)\propto t\,\mathrm{exp}(-t/\tausfr)$\\
  $Z/\Zsun$  &Stellar and interstellar metallicity ($Z = \Zism$) \\   
  \tauV & V-band attenuation optical depth\\
  $\mud$ & Fraction of attenuation arising in the diffuse ISM, fixed to 0.4\\
  $U_s$ & Effective gas ionization parameter\\
  \xid & Dust-to-metal mass ratio, fixed to 0.1\\
  $\nH/\mathrm{cm}^{-3}$ & Hydrogen gas density, fixed to 100\\
  (C/O)/(C/O)$_{\odot}$ & Carbon-to-Oxygen abundance ratio, fixed to solar, where (C/O)$_{\odot}=0.44$\\
  $m_{\mathrm{up}}/\Msun$ & Upper mass cutoff of the IMF, fixed to 100.\\
  \hline
\end{tabular}
\label{tab:physical_parameter_list}
\end{table*}

\begin{figure*}
  \centering
  \includegraphics[width=5in]{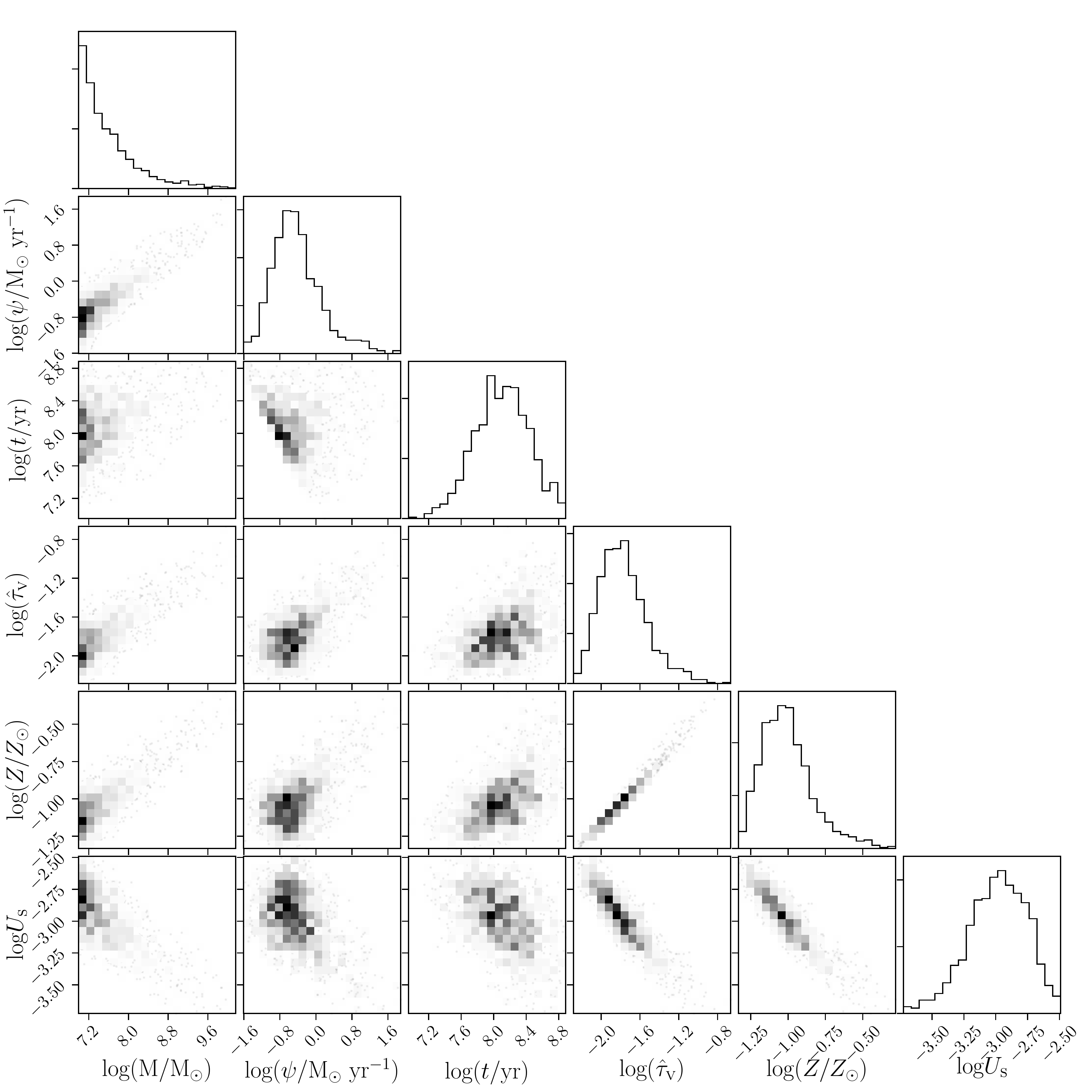}
  \caption{The distribution of physical parameters (physical meanings summarised in Table~\ref{tab:physical_parameter_list}) assigned when producing the mock photometry for the \constMF\ scenario.}
  \label{fig:scenario_iv_params}
\end{figure*}

\begin{figure}
  \centering
  \includegraphics[width=3.5in]{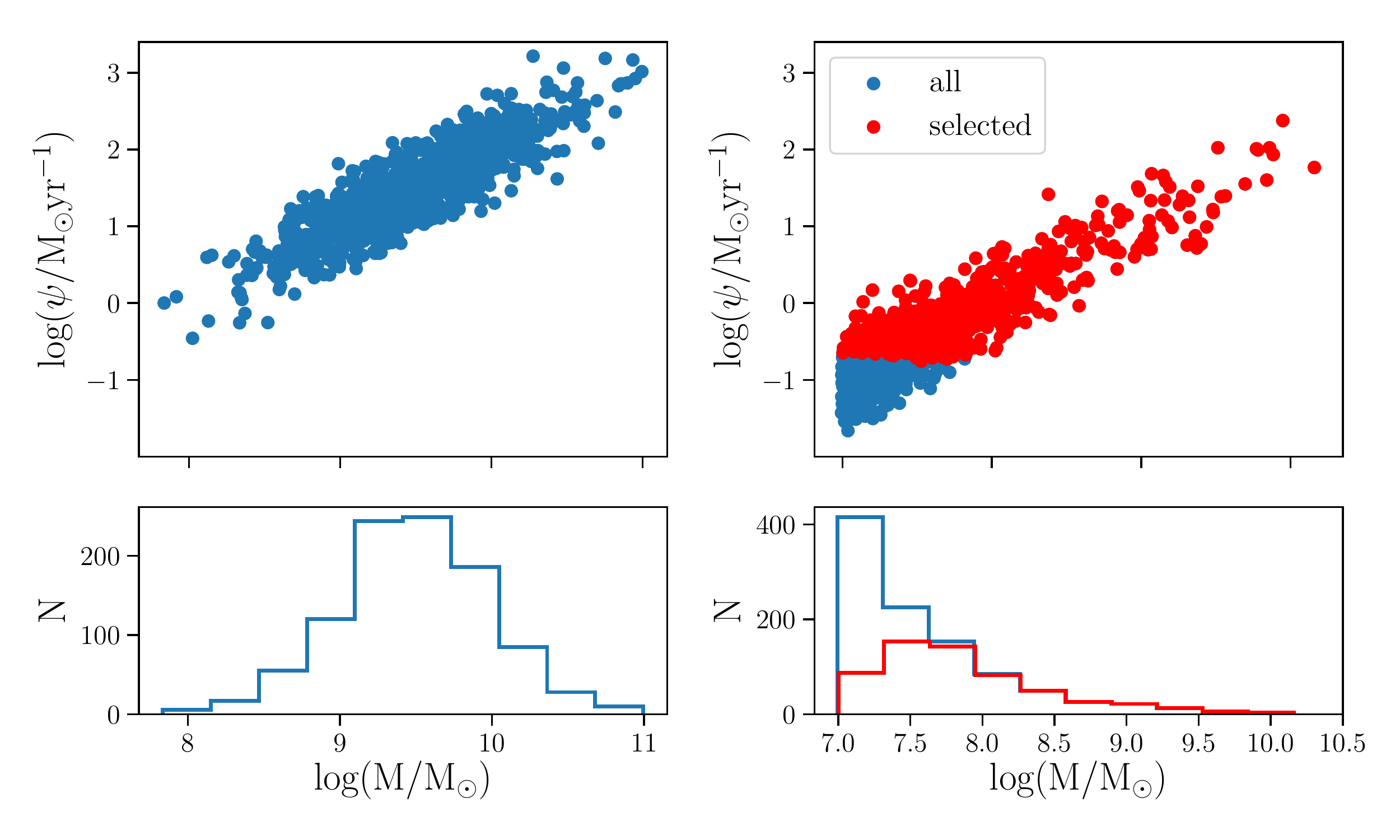}
  \caption{The distribution of \Mstar\ values [where \Mstar\ is defined as $\log(\MstarInLog/\Msun)$; see equation \ref{eq:mtot_and_sfr}] for the test scenarios. The \ideal, \constGauss\ and \delGauss\ scenarios are Gaussian distributed in \Mstar\ with $[\mu_{\mathrm{M}},\sigma_{\mathrm{M}}]=[9.5,0.5]$ (left panel) while the values of \Mstar\ in the \constMF\ scenario are drawn from the Duncan et al. (2014) measured mass function at $z\sim5$ (right panel).  All of the objects created for the first three scenarios are bright enough to be detected at the chosen filter depths, however, this is not true for objects in the fourth scenario.  The objects with a 5$\sigma$ detection in at least five filters are shown as red points (see text for details).}
  \label{fig:Testing_input_distribution}
\end{figure}

In order to produce mock photometry with \beagle\ for the \constGauss, \delGauss\ and \constMF\ scenarios, we are required to choose a set of photometric filters and a redshift.  Given the difficulties of obtaining stellar-mass estimates at high redshifts using current facilities, mainly because of the lower sensitivity of \Spitzer\ compared to \HST, we produce the mock photometry using a set of \JWST/NIRCam broad-band filters.  We fix the redshift to $z=5$, since this is the lowest redshift for which the Near-Infrared Camera (NIRCam) on board \JWST\ will cover the entire rest-frame wavelength range from UV ($\lambda\sim1100\,$\AA) to optical ($\lambda\sim8300\,$\AA).  At higher redshifts, objects will be fainter and thus constraints on \Mstar\ and \sfr\ will be more challenging, while at lower redshifts constraints on \Mstar\ and \sfr\ will depend on the variable depths of complementary \HST\ observations sampling the rest-frame ultra-violet of \JWST-observed sources. In practice, we consider a set of \JWST/NIRCam broad and medium-band filters to produce noiseless mock photometry, to which we add Gaussian noise to mimic the depths summarised in Table~\ref{tab:delayed_const_test_photom}. The depths of the broad-band filters correspond to the predicted 5$\sigma$ point source limits of the medium-depth imaging to be taken as part of the joint NIRCam/NIRSpec GTO survey \citep[e.g.][]{Williams2018}. The depths for the four medium bands we explore are set to the approximate depths of the two medium bands to be employed in this survey (F335M and F410M).  It is worth noting that galaxies at these redshifts are likely to be marginally resolved with \JWST, at least in the shortest wavelength bands, and as such these depths are somewhat optimistic for the medium portion of the survey. However, our aim is not to predict exactly what we can observe with this particular survey, but to set up a realistic combination of filters and associated depths. We use the set of broad-band filters as the base set and explore parameter constraints with the addition of the medium bands.  

Another way that the scenarios increase in complexity is by the distribution of \Mstar\ values.  The \Mstar\ values in the \ideal, \constGauss\ and \delGauss\ scenarios are Gaussian-distributed with mean ($\mu_\textrm{M}$) and standard-deviation ($\sigma_\textrm{M}$) of $[\mu_\textrm{M},\sigma_\textrm{M}]=[9.5,0.5]$, as shown in the left-hand panel of Fig.~\ref{fig:Testing_input_distribution} (hence the `G' in the labels).  In the \constGauss\ and \delGauss\ scenarios, where we produce mock photometry using \beagle, the input distribution of \Mstar\ chosen above allows all objects to be detected at the chosen photometric depths in all bands.  For the \constMF\ scenario, however, \Mstar\ values are drawn from the $z\sim5$ \cite{Duncan2014} measured stellar mass function (hence `MF', Fig.~\ref{fig:Testing_input_distribution}, right panel) and in this case we must also account for survey selection effects  by keeping only those galaxies with at least 5 bands detected at the $>5\sigma$ level. This simple approach neglects the photometric redshift uncertainty which is larger at low stellar masses and which can scatter objects into different redshift bins  (see, e.g. \citealt{Kemp2019} for a quantification of this effect on stellar mass function estimates).

In all scenarios, \sfr\ values are drawn from the corresponding \MstarSFR\ relation defined by the set of values $[\intercept,\slope,\scatter] = [-8,1,0.3]$.  
To produce the mock SEDs in the \constGauss\ and \constMF\ scenarios, we adopt a constant SFH, for which a combination of the model parameters \Mstar\ and galaxy age, \t, determine the \sfr\ of each galaxy.\footnote{To define the input \MstarSFR\ relation, the mass fraction returned to the ISM can be used to determine \MtotInLog\ for a given \MstarInLog, and hence the age, \t\ required to produce the object with given \sfrInLog\ via $\sfrInLog = \MtotInLog/\t$.}  We chose the input \MstarSFR\ relation such that the required values of \t\ would be less than the age of the Universe at $z\sim5$.   In the \delGauss\ scenario, we adopt a delayed exponential SFH, as the name would suggest, for which \sfr\ depends on the combination of \tausfr\ (the timescale of star formation for a delayed exponential SFH as defined in Table~\ref{tab:physical_parameter_list}) and \t.     Table~\ref{tab:test_scenario_beagle_parameters} summarizes the full set of physical parameters required to produce the mock SEDs, as well as the standard priors used in \beagle\ fitting (whose physical meaning is reported in Table~\ref{tab:physical_parameter_list}). We then fit each scenario using the same model assumptions with which it is built. When fitting the mock photometry, we include a 1\% minimum relative error, added in quadrature to the original photometric error.  This is a standard method used to reduce the impact of biases introduced into photometric measurements by systematic uncertainties. The four scenarios are summarised in Table~\ref{tab:four_test_scenarios}.

\begin{table}
\centering
  \caption{NIRCam filters and depths used to produce mock photometric catalogues for 
  the \constGauss, \delGauss\ and \constMF\ scenarios.}
  \begin{tabular}{c|c|c}
  \hline
  Filter & 5$\sigma$ depth & description\\
  \hline
  F090W & 29.4 & Broad\\
  F115W & 29.6 & Broad\\
  F150W & 29.7 & Broad\\
  F200W & 29.8 & Broad\\
  F277W & 29.4 & Broad\\
  F356W & 29.4 & Broad\\
  F444W & 29.1 & Broad\\
  F300M & 28.8 & Medium\\
  F335M & 28.8 & Medium\\
  F410M & 28.8 & Medium\\
  F430M & 28.8 & Medium\\
  \hline
\end{tabular}
\label{tab:delayed_const_test_photom}
\end{table}

\begin{table*}
\centering
\caption{The parameters used to produce mock photometry, and then fitted to it in the \constGauss, \delGauss\ and \constMF\ scenarios.   Mock photometry and photometric fits are produced with \beagle.  The central column describes the distribution of the parameter (right column) used to produce the mock photometry while the left column describes the prior set on that parameter when fitting.   }
\begin{tabular}{ C{0.5\columnwidth} C{0.7\columnwidth} C{0.7\columnwidth} } 
\toprule

 \multicolumn{1}{c}{Parameter}	  & \multicolumn{1}{c}{Mock distribution} & \multicolumn{1}{c}{Fitted prior} \\
 
\midrule

$\Mstar=\log(\MstarInLog/\Msun)$ & Drawn from Gaussian /mass function distribution & Uniform $\in [7,10.5]$\\

$\sfr=\log(\sfrInLog/\Msun \yr^{-1})$ & Drawn from [\intercept,\slope,\scatter]$=[-8,1,0.3]$ & Not fitted\\

$\log(\Z/\Zsun)$ & Dependent on \Mstar\ and \sfr\ according to equation~\eqref{eq:FMR} & Uniform $\in [-2.2,0.24]$\\

$\log(\tausfr/\yr)^a$ & Fixed to give \sfr & Uniform $\in [7,10.5]$\\

$z$ & Fixed 5.0 & Uniform $\in [4.5,5.5]$ \\

$\log(\t/\yr)$ & Fixed to give \sfr & Uniform $\in [6,8.96]$\\

\logUs & Dependent on Z according to equation~\eqref{eq:logU_Z} & Uniform $\in [-4,-1]$\\

\xid & Fixed 0.1 & Uniform $\in [0.1,0.5]$\\

\tauV & Dependent on \sfr\ and Z following method of \cite{Williams2018} &  Exponential exp(-\tauV), truncated $\in [0, 2]$ \\

\mud & Fixed 0.4 & Fixed 0.4\\

\bottomrule
\end{tabular}
\raggedright $^a$Parameter used only with delayed exponential SFHs.
\label{tab:test_scenario_beagle_parameters}
\end{table*}

\begin{table*}
\centering
\caption{Summary of four test scenarios.  The columns give, in turn, the test label, the form of $\sfr-\Mstar$ uncertainties, whether-or-not \beagle\ is used to produce and fit to mock photometry in the test, the SFH used to produce the mock photometry and subsequently fitted to it, the distribution in \Mstar\ and whether-or-not selection effects are accounted for in the test.  }
\begin{tabular}{ L{0.3\columnwidth} C{0.35\columnwidth} C{0.1\columnwidth} C{0.2\columnwidth}  C{0.5\columnwidth} C{0.15\columnwidth}} 
\toprule

 \multicolumn{1}{c}{Scenario}	  & \multicolumn{1}{c}{\sfr-\Mstar\ uncertainties} & \multicolumn{1}{c}{\beagle\ fit} & \multicolumn{1}{c}{SFH}  & Mass distribution & Selection effects\\
 
\midrule

\ideal & Homoskedastic, covariant Gaussians & & N/A  & Gaussian $\mu_\textrm{M},\sigma^2_\textrm{M} = 9.5,0.25$ & \\

\constGauss & Heteroskedastic & \checkmark & constant & Gaussian $\mu_\textrm{M},\sigma^2_\textrm{M} = 9.5,0.25 $& \\

\delGauss & Heteroskedastic & \checkmark & delayed exponential & Gaussian $\mu_\textrm{M},\sigma^2_\textrm{M} = 9.5,0.25$& \\

\constMF & Heteroskedastic & \checkmark & constant & Mass function with characteristic mass, faint-end slope and normalisation $\log($M$^*),\alpha_\textrm{MF},\phi^*/$Mpc$^{-3}=10.68,-1.74,1.24\times10^{-4}$& \checkmark\\

\bottomrule
\end{tabular}
\label{tab:four_test_scenarios}
\end{table*}

\section{Measuring the \MstarSFR\ relation}
\label{section:modelling}

Given a sample of galaxies with stellar mass and SFR estimates, the simplest way to obtain the slope and intercept of the main sequence would be to use ordinary linear regression.  In this case, the scatter can be estimated from the standard deviation of the offsets in \sfr\ from the relation.  However, ordinary linear regression assumes that the `true' values of the dependent and independent variables only deviate from the line because of random Gaussian fluctuations of the dependent variable, and that these fluctuations have a constant variance (i.e. they are homoskedastic).\footnote{See \citet{Hogg2010} for a description of the assumptions of ordinary linear regression, and possible Bayesian solutions to scenarios where they no longer apply.} If these random fluctuations are due to homoskedastic measurement errors then the underlying model requires that the `true' values are drawn from a very tight linear relationship. A further assumption of ordinary linear regression is that the distribution of the independent variable is uniform. The above assumptions do not hold in our case, since (i) the uncertainties in the measurements of \sfr\ \textit{and} \Mstar\ can be correlated, and non-Gaussian; (ii) the uncertainties tend to be heteroskedastic, as they are larger at low \Mstar\ where objects are fainter;  (iii) the distribution of \Mstar\ values from a flux-limited survey is not uniform, but rather a multiplication of the stellar mass function and the mass completeness of the survey; (iv) it would be a strange Universe indeed if \Mstar\ and \sfr\ lay along a very tight linear relationship. Given these considerations, SFR values are therefore better modelled as distributed about the linear relationship with some scatter, while also explicitly modelling the measurement uncertainties.

In the present work, we address the above limitations of ordinary linear regression by implementing a fully Bayesian Hierarchical method for constraining the intercept \intercept, slope \slope\ and scatter \scatter\ in the presence of heteroskedastic, co-varying errors and a non-uniform distribution of stellar mass values. Whether or not the intrinsic scatter is uniform with stellar mass at high redshift is still unknown, although San17 find tantalising evidence of an increase of the intrinsic scatter toward low stellar masses. Some simulations also suggest that the intrinsic scatter increases toward low stellar masses because of short timescale variations of star-formation (e.g. \citealt{Sparre2017,Matthee2019}).  Here, we  choose to model the relation with a scatter that is constant with stellar mass, but we will discuss in Section~\ref{section:mstar_sfr_constraints} the prospects for extending this model to account for a non-constant scatter. We compare the results obtained with our hierarchical method with those obtained using ordinary linear regression, and also study the impact of different methods used in the literature to estimate \sfr.
We summarise the assumptions held by each of these methods in Table \ref{tab:methods}.

\begin{table}
\label{tab:methods}
\caption{Summary of the assumptions held by the different methods employed here to measure the main parameters of the \MstarSFR\ relation.} 
\centering
\def\arraystretch{1.5}
\begin{tabular}{ C{0.4\columnwidth} C{0.07\columnwidth}  C{0.07\columnwidth}  C{0.07\columnwidth}  C{0.07\columnwidth} } 
\toprule
Method:	  & \PM & \PMSalmon & \PMSantini & \Kelly \\
\midrule
Dis-entangles intrinsic from observed scatter? & & & & \checkmark\\
Accounts for heteroskedastic errors? &  & & & \checkmark \\
Errors between \Mstar\ and \sfr\ co-variant? & \checkmark & & & \checkmark \\
Accounts for co-varying errors? & & & & \checkmark\\
$\prob(\Mstar)$ modelled? &  & & & \checkmark\\
\bottomrule
\end{tabular}
\end{table}

\subsection{The \PM\ method: linear regression to joint posterior medians}
\label{section:medians}

The simplest way to derive \intercept, \slope\ and \scatter\ is to perform ordinary linear regression (via ordinary least-squares) on point-wise estimates of \Mstar\ and \sfr. For this we use the median of the marginalised posterior probability from our \beagle\ fits,  which we will refer to as $\langle\Mstar\rangle_\textrm{med}$ and $\langle\sfr\rangle_\textrm{med}$ respectively.\footnote{We will use the notation $\langle x\rangle_\textrm{med}$ to denote the posterior median value from the \beagle\ fit of a given parameter, $x$.} As described above, linear regression assumes that the true values deviate from a straight line with random Gaussian fluctuations in the dependent variable only.  

Given a sample of galaxies with stellar mass and SFR estimates, we provide an estimate of \scatter\ from the standard deviation of residuals from the best-fit linear relation.  In essence this assumes that any deviation from the straight line is due to intrinsic scatter and as such should provide an estimate of the convolution of the intrinsic scatter and the scatter due to measurement uncertainties in \Mstar\ and \sfr.  As such, this method does not rigorously treat uncertainties in either \sfr\ or \Mstar, and does not allow for covarying errors or a non-uniform \Mstar\ distribution. We will refer to this method throughout the paper as the \PM\ (Posterior Median) method. 

\subsection{The \PMSalmon\ method: linear regression with Sal15 \sfr\ measurements}
\label{section:methodSal15}

Sal15 provided the first estimates of intrinsic scatter in the \MstarSFR\ relation to high redshift ($z\gtrsim4$).  They derive \Mstar\ adopting a constant SFH, and \sfr\ with an updated form of the \cite{Kennicutt1998} UV-luminosity to SFR conversion
(adjusted to a \citealt{Chabrier2003} IMF). 
The original \cite{Kennicutt1998} conversion assumes there has been at least 100 Myr of constant star formation in the galaxy.  By 100 Myr the relative contribution to the rest-frame UV of very young, hot and luminous stars to slightly older, cooler stars becomes fairly stable.  At younger ages, however, there are fewer intermediate-age stars and therefore a higher SFR is needed to produce the same UV luminosity.  The updated conversion by \cite{Reddy2012} takes account of the higher SFR-to-UV ratio found in young galaxies by including a dependence on age but is explicitly calculated for solar metallicity. 

To test the SFR estimation employed by Sal15, we take the observed flux closest to rest-frame 1500\,\AA\  for each simulated galaxy and correct for dust using $\langle\Auv\rangle_\textrm{med}$, the median of the marginalised attenuation at 1500\,\AA.  Sal15 uses both the \cite{Calzetti2000} and an SMC-like dust attenuation curve in their analysis, while here we are using the CF00 two-component dust model, which implies a dependence of the effective \Auv\ on a galaxy's SFH.   Once the dust-corrected 1500\,\AA\ flux is converted to luminosity (via $\langle z\rangle_\textrm{med}$), we use $\langle t\rangle_\textrm{med}$ of each galaxy to estimate \sfr\ using the \cite{Reddy2012} UV-to-SFR conversion. 

To determine the parameters \intercept, \slope\ and \scatter\ describing the \MstarSFR\ relation, Sal15 apply a weighted linear regression algorithm to \sfr\ values binned according to \Mstar.  The weights and reported scatter, \scatter, are given by the standard deviation of \sfr\ values in each bin. This definition of \scatter\ differs from that in equation~\eqref{eq:main_sequence}, as the variation of \sfr\ as a function of \Mstar\ within each \Mstar\ bin will provide an additional scatter component in their measurements.  Using \Auv\ and age derived from SED fits introduces correlated uncertainties between \Mstar\ and \sfr, and Sal15 provide estimates of the intrinsic scatter after accounting for these uncertainties with Monte Carlo simulations. Here, we do not attempt to reproduce the exact measurement procedure employed in Sal15, but estimate \intercept, \slope\ and \scatter\ using the approach described in Section~\ref{section:medians}.  The only difference with respect to the method of Section~\ref{section:medians} is hence in the way we estimate \sfr.  In this way, we can study to what extent the \sfr\ estimation affects derived properties.   We will refer to this method throughout the paper as the \PMSalmon\ method.

\subsection{The \PMSantini\ method: linear regression with San17 \sfr\ measurements}

San17 use data from the Hubble Frontier Fields \citep{Lotz2017} to probe the \MstarSFR\ relation to lower masses than in the Sal15 study by exploiting the magnification of galaxies by foreground clusters.  Their method to measure the main-sequence parameters also relies on a dust-corrected \sfr\ estimate. The attenuation-corrected UV-luminosity is estimated using the \cite{Meurer1999} prescription.  This correction essentially assumes that every galaxy has the same intrinsic UV slope of $-2.23$, and any deviation from this slope is caused by dust.  Using the \cite{Meurer1999} relation achieves minimal covariance between \Mstar\ and \sfr\ estimates, but likely at the expense of encompassing the true variation in UV slopes in the underlying population. Using the \cite{Calzetti2000} dust curve, San17 derive a dust-corrected UV-luminosity value, which they then convert to \sfr\ using the \cite{Kennicutt2012} conversion.  This conversion updates the \cite{Kennicutt1998} calibration given new stellar models and is based on a \cite{Chabrier2003} IMF, but unlike the \cite{Reddy2012} calibration adopted by Sal15, it does not include the effect of stellar population age on the UV-to-SFR conversion. 

As in the case of the \PMSalmon\ method, we estimate \intercept\, \slope\ and \scatter\ in the same way as for the \PM\ method, but using \sfr\ estimated with the San17 approach.  We will refer to this method as the \PMSantini\ method from now on. This method differs from the full method of San17, who provide estimates of the intrinsic scatter after accounting for the measurement uncertainties.  Their method also carefully accounts for the complicated selection function, including the effects introduced when more numerous, low-mass objects are scattered into the selected sample because of measurement uncertainties, than are less numerous, higher-mass objects are scattered out of it. We note that our approach of varying only the way in which to derive \sfr\ between the \PM, \PMSalmon\ and \PMSantini\ methods allows us to isolate the effects of \sfr\ estimation on the derived \MstarSFR\ relation.

\subsection{The \Kelly\ (Bayesian Hierarchical) method: the Kelly (2007) model}

As discussed at the beginning of this section, the assumptions of ordinary linear regression are not appropriate in the case of measuring the main sequence.  \citet[hereafter K07]{Kelly2007} presented a Bayesian Hierarchical model which self-consistently derives the posterior probability of the intercept, slope and intrinsic scatter between variables with correlated, heteroskedastic errors and a non-uniform distribution of covariate values.  We therefore adapt this model to work with output joint posteriors on \Mstar and \sfr\ derived from SED fitting with \beagle.  The K07 model is general and not specifically written for modelling the main sequence, but we will summarize key components of the model here using \Mstar\ as the independent variable, and \sfr\ as the dependent variable.  

Before describing the K07 model, it may be helpful to refer to Bayes theorem, in particular to the definitions of posterior probability, likelihood and prior probabilities. The Bayes equation can be written as:

\begin{equation}
\begin{aligned}
\conditional{\bmath{\Theta}}{\txn{\textbf{D}},H} = \frac{\conditional{\Db}{\modelThetaB,H}\conditional{\modelThetaB}{H}}{\int\conditional{\Db}{\modelThetaB,H}\conditional{\modelThetaB}{H}d\modelThetaB}
\end{aligned}
\end{equation}
\label{eq:Bayes}

\noindent
where $\conditional{\modelThetaB}{\txn{\textbf{D}},H}$ is the posterior probability distribution of the model parameters \modelThetaB\ given a set of data, \Db\ and a model (or hypothesis) H.  In this equation:  $\conditional{\Db}{\modelThetaB,H}$ is the likelihood function, which quantifies the statistical agreement between model and data for a fixed set of parameters; $\conditional{\modelThetaB}{H}$ expresses the prior probability, which encapsulates our knowledge on the model parameters \textit{before} analysing the observations; the denominator is the evidence (or marginal likelihood), a normalization factor often written as $\conditional{\Db}{H}$. 

A Bayesian Hierarchical model is one that is structured over different `levels'.  The highest level of the K07 model describes the distribution of \Mstar\ values, which do not have to be uniform.  This can be written as:
\begin{equation}
    \Mstar\sim\,\conditional{\Mstar}{\boldsymbol{\eta}}\label{eq:Mstar_GMM}
\end{equation}
which denotes that stellar masses have some distribution given by $\conditional{\Mstar}{\boldsymbol{\eta}}$, where $\boldsymbol{\eta}$ is a set of variables specifying the shape of that distribution. In the K07 sampler, this distribution is a weighted linear combination of a set of Gaussians, or a Gaussian mixture model (GMM).

The second level describes the distribution of \sfr\ values given \Mstar, which is modelled as a linear relation with some Gaussian scatter:
\begin{equation}
    \sfr|\Mstar\sim\,\intercept+\slope\Mstar + \mathcal{N}(0,\scatter^2)\label{eq:sfr_given_Mstar}\\
\end{equation}
(note that this describes the same relationship as in equation~\ref{eq:main_sequence} but written in a different way here for clarity). 

The lowest level of the model describes the measurements of \Mstar\ and \sfr\ of the individual objects.  The K07 model  assumes that one has point-wise estimates of the two variables of interest, and some Gaussian errors on these estimates, which may be correlated and are described by a covariance matrix $\boldsymbol{\Sigma}$.  Taking a measurement in this way can be thought of as drawing the measured values from a bi-variate Gaussian distribution (with some covariance $\boldsymbol{\Sigma}$) centred on the `true' values of \Mstar\ and \sfr. We will label these point-wise estimates $x$ and $y$:
\begin{equation}
x,y|\Mstar,\sfr\sim\,\mathcal{N}_2([\Mstar,\sfr],\boldsymbol{\Sigma})\,.
\label{eq:measurements_given_everything}
\end{equation}

Modelling the \Mstar\ and \sfr\ estimates in this way is not wholly appropriate for our needs as we derive posterior probability distributions of \sfr\ and \Mstar\ using \beagle.  We will describe how we incorporate the \beagle\ estimates at the end of this section but proceed to describe the model assuming we can take direct measurements of \Mstar\ and \sfr\ (following the model presented in K07).

With this model, we wish to calculate the posterior probabilities of the main-sequence parameters (\intercept, \slope, \scatter), as well as the parameters describing the distribution in \Mstar\ ($\boldsymbol{\eta}$ from equation~\ref{eq:Mstar_GMM}), given the measurements and the `true' \Mstar\ and \sfr\ values.  We can write this posterior probability using the chain rule   as:
\begin{equation}\label{eq:full_data_likelihood}
\begin{aligned}
\prob(\intercept,&\slope,\scatter,\boldsymbol{\eta}|x,y,\Mstar,\sfr)\\
&\propto\conditional{x,y,\Mstar,\sfr}{\intercept,\slope,\scatter,\boldsymbol{\eta}}\prob(\intercept,\slope,\scatter,\boldsymbol{\eta})\\
&\propto\conditional{x,y}{\Mstar,\sfr}\conditional{\sfr}{\Mstar,\intercept,\slope,\scatter}\conditional{\Mstar}{\boldsymbol{\eta}}\prob(\intercept,\slope,\scatter,\boldsymbol{\eta})
\end{aligned}
\end{equation}

\noindent
where the first conditional probability on the right-hand side is described in equation~\eqref{eq:measurements_given_everything}, the second conditional probability describes the linear relation with Gaussian scatter given in equation~\eqref{eq:sfr_given_Mstar}, the third describes the distribution of \Mstar\ values (equation~\ref{eq:Mstar_GMM}) and $\prob(\intercept,\slope,\scatter,\boldsymbol{\eta})$ describes the prior probabilities assumed for the parameters of interest. Note that the proportionality comes from dropping the evidence when applying Bayes theorem (equation~\ref{eq:Bayes}).
However, we must note that under normal circumstances we do not know the `true' \Mstar\ and \sfr\ values for individual objects, which means that we cannot define these conditional probabilities.  K07 deals with this by treating the true values as missing data that must be marginalised over, thus obtaining the observed data likelihood function:

\begin{equation}\label{eq:margnialise}
\begin{aligned}
\conditional{x,y}{\intercept, \slope, \scatter} = \\
\int\conditional{x,y}{\sfr,\Mstar}\conditional{\sfr}{\Mstar,\intercept,\slope,\scatter}\conditional{\Mstar}{\boldsymbol{\eta}}\d\Mstar\d\sfr
\end{aligned}
\end{equation}
\noindent
This likelihood function can be used to compute the maximum-likelihood.  However, computing this integral is non-trivial and K07 proposes a Gibbs sampler.  A Gibbs sampler is a Markov Chain Monte Carlo (MCMC) algorithm that generates posterior samples by performing random draws from the full conditional probability distribution of each free parameter in turn. In our case, each free parameter from equation~\eqref{eq:full_data_likelihood} will be updated in turn within each MCMC iteration. In practice, treating the `true' values of \Mstar\ and \sfr\ as missing data means that these are treated as free parameters which will also be updated in each iteration of the sampler. We refer the reader to K07 for further details, in particular section 6.2.1 which lists the steps performed in each iteration.   We have employed the Python implementation of the K07 Gibbs sampler, as implemented by Josh Meyers.\footnote{https://github.com/jmeyers314/linmix.}

As stated above, rather than having direct measurements of \Mstar\ and \sfr\ (which we referred to as $x$ and $y$), we have \beagle-derived joint \Mstar-\sfr\ posterior distributions that we do not want to assume to be described by a single (bi-variate) Gaussian. We therefore extend the K07 Gibbs sampler to accept Gaussian mixture models of the \beagle-derived joint \Mstar-\sfr\ posterior probabilities.\footnote{The forked version can be found on GitHub https://github.com/eclake/linmix.}  Essentially, these models are given by the sum of $K$ bi-variate Gaussians, each with weights given by $\pi_k$, mean $\boldsymbol{{\mu}}_k = [\barMstar,\barsfr]$, and covariance matrix $\boldsymbol{\zeta}_k$:
\begin{equation}
\label{eq:Mstar_sfr_GMM}
    \conditional{\Mstar,\sfr}{\mathbf{F},\mathbf{E}} \approx \conditional{\Mstar,\sfr}{\boldsymbol{\gamma}} = \sum_{k=1}^{K}\pi_k\mathcal{N}_2(\boldsymbol{\mu}_k,\boldsymbol{\zeta}_k)
\end{equation}
In this equation, $\mathbf{F}=[f_1,f_2,...,f_N]$ and $\mathbf{E}=[e_1,e_2,...,e_N]$ are the sets of $N$ broad-band fluxes and flux errors, respectively, and $\boldsymbol{\gamma} = [\boldsymbol{\pi}(=[\pi_0,\pi_1,...\pi_k]),\boldsymbol{\mu}(=[\boldsymbol{\mu}_0,\boldsymbol{\mu}_1,...\boldsymbol{\mu}_k]),\boldsymbol{\zeta}(=[\boldsymbol{\zeta}_0,\boldsymbol{\zeta}_1,...\boldsymbol{\zeta}_k])]$ is the set of free parameters describing the GMM.  These joint posterior probability distributions \textit{are not} the same thing as direct measurements of \Mstar\ and \sfr\ as written in equation~\eqref{eq:measurements_given_everything}.  Our data are flux estimates and our posterior probabilities give the probability of the `true' \Mstar\ and \sfr\ values given the data, rather than the other way around.  Yet, within the Gibbs sampler, when we update the `true' \Mstar\ and \sfr\ values, we sample directly from the full conditional probabilities.  For \Mstar\ this will look like:
\begin{equation}
\begin{aligned}
    \conditional{\Mstar}{\mathbf{F},\mathbf{E},\intercept,\slope,\scatter,\boldsymbol{\eta},\sfr} = \\ \conditional{\Mstar}{\mathbf{F},\mathbf{E},\sfr}\conditional{\Mstar}{\sfr,\intercept,\slope,\scatter}\conditional{\Mstar}{\boldsymbol{\eta}}\,,
\end{aligned}
\end{equation}
where we can derive $\conditional{\Mstar}{\mathbf{F},\mathbf{E},\sfr}$ from the joint \Mstar-\sfr\ posterior probability distribution defined in equation~\eqref{eq:Mstar_sfr_GMM}.

 \begin{figure}
  \centering
  \includegraphics[width=3.5in]{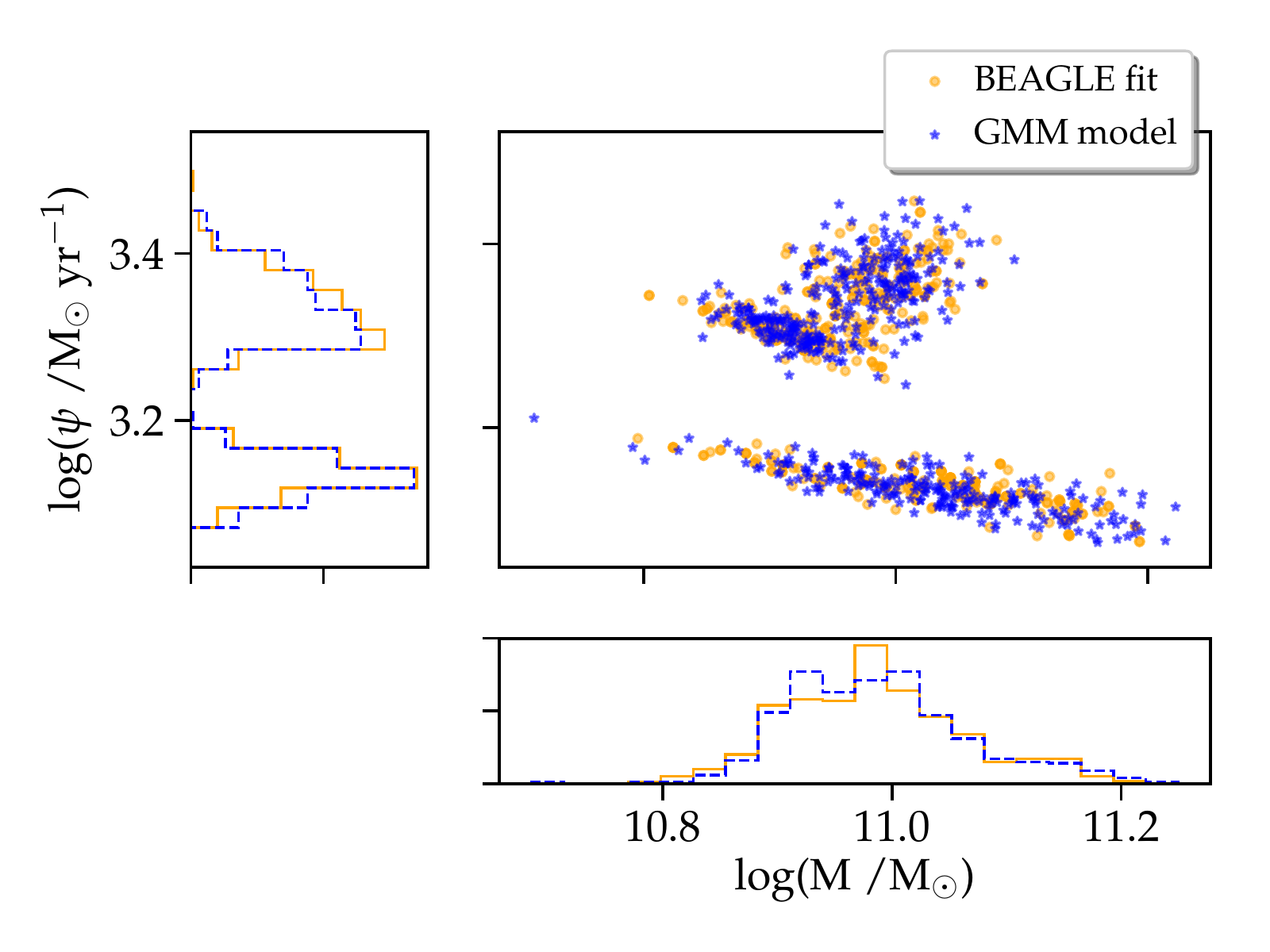}
  \caption{Samples from the joint $\sfr-\Mstar$ posterior from a \beagle\ fit to a galaxy produced with constant SFH (yellow), plotted with the corresponding samples from the GMM fit produced using the scikit GaussianMixture package (blue).  The bottom and left panels show the marginalised distributions on \Mstar\ and \sfr\ respectively.}
  \label{fig:GMM_example}
\end{figure}

 \begin{figure*}
  \centering
  \includegraphics[width=7in]{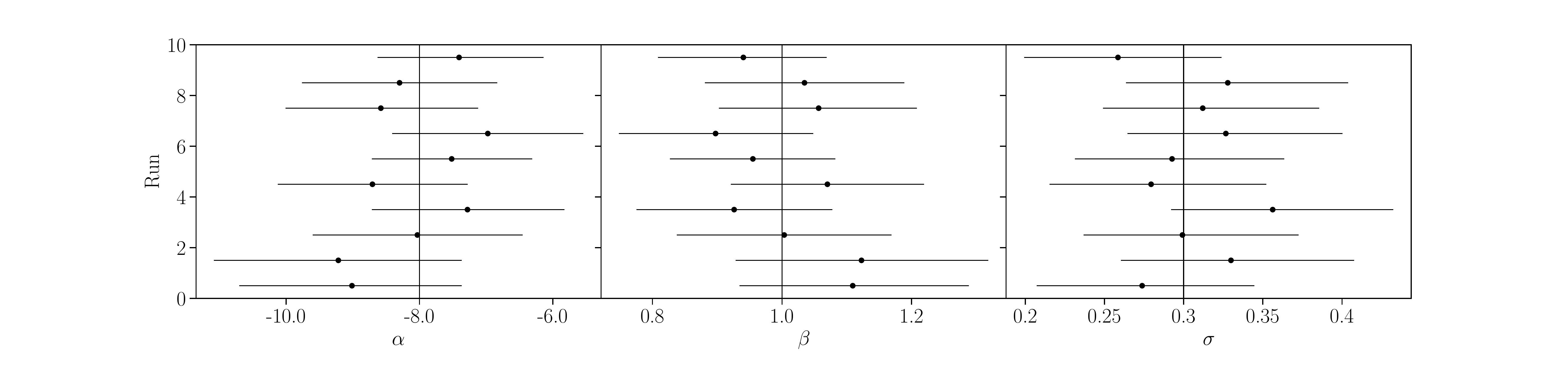}
  \caption{The results for the \ideal\ scenario.  The median and 95\% confidence limits from the derived posterior distribution using the \Kelly\ method for each independent sample of 100 objects (plotted on the y-axis) and for each parameter, \intercept, \slope\ and \scatter\ (the three panels from right to left).  The thick vertical black lines show the input values for the \MstarSFR\ distribution. }
  \label{fig:Testing_toy_model_summary}
\end{figure*}
 
 We fit GMM models to the \beagle-derived joint $\sfr-\Mstar$ posterior distributions using sklearn.mixture.GaussianMixture from the Python package scikit-learn \cite{scikit-learn}.   After some experimentation, we opted to fit 3 Gaussians to each posterior, finding that this prevented over-fitting to individual features, but was flexible enough to represent the range of mophologies in the output posterior distributions.  An example GMM fit is shown in Fig.~\ref{fig:GMM_example}, where we display samples from the joint $\sfr-\Mstar$ posterior probability distribution from the \beagle\ fit to a single object from the \constGauss\ scenario, which was fitted to with a constant SFH.  Over-plotted are the samples from the GMM fit to this $\sfr-\Mstar$ posterior.

As constructed, the Gibbs sampler self-consistently propagates the uncertainties on \sfr\ and \Mstar\ estimates of individual objects onto the uncertainties on main sequence parameters \intercept, \slope\ and \scatter.  By using \beagle-derived \sfr\ estimates, the full uncertainties in dust attenuation, intrinsic UV slope and UV-to-SFR conversion are self-consistently accounted for.  Additionally, the model explicitly accounts for a non-uniform distribution of \Mstar\ values.

\section{Results}
\label{section:results}

\begin{figure*}
  \centering
  \subfigure{\includegraphics[width=4.0in, trim={4cm 1.5cm 5cm 8cm},clip]{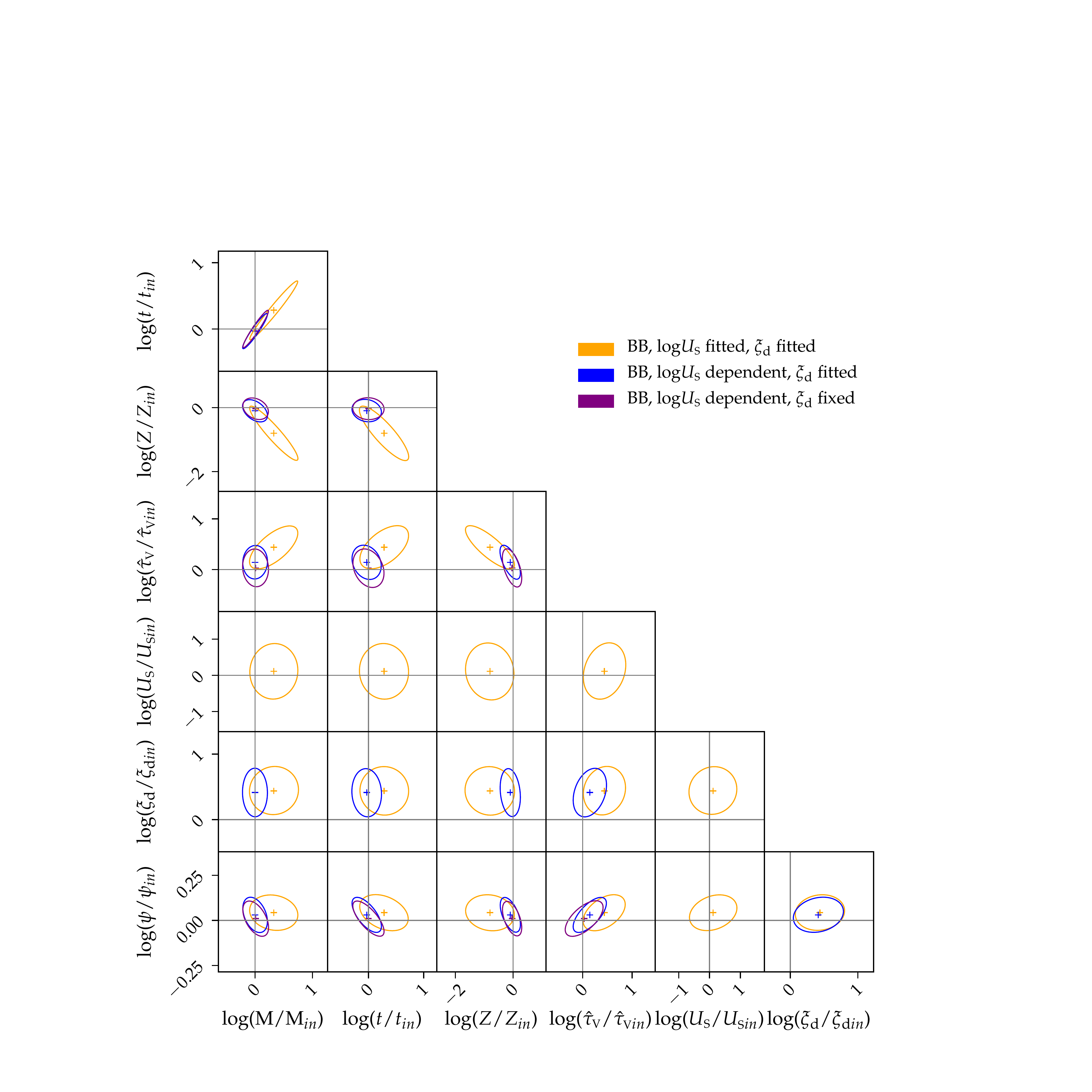}}
  \subfigure{\includegraphics[width=2.0in]{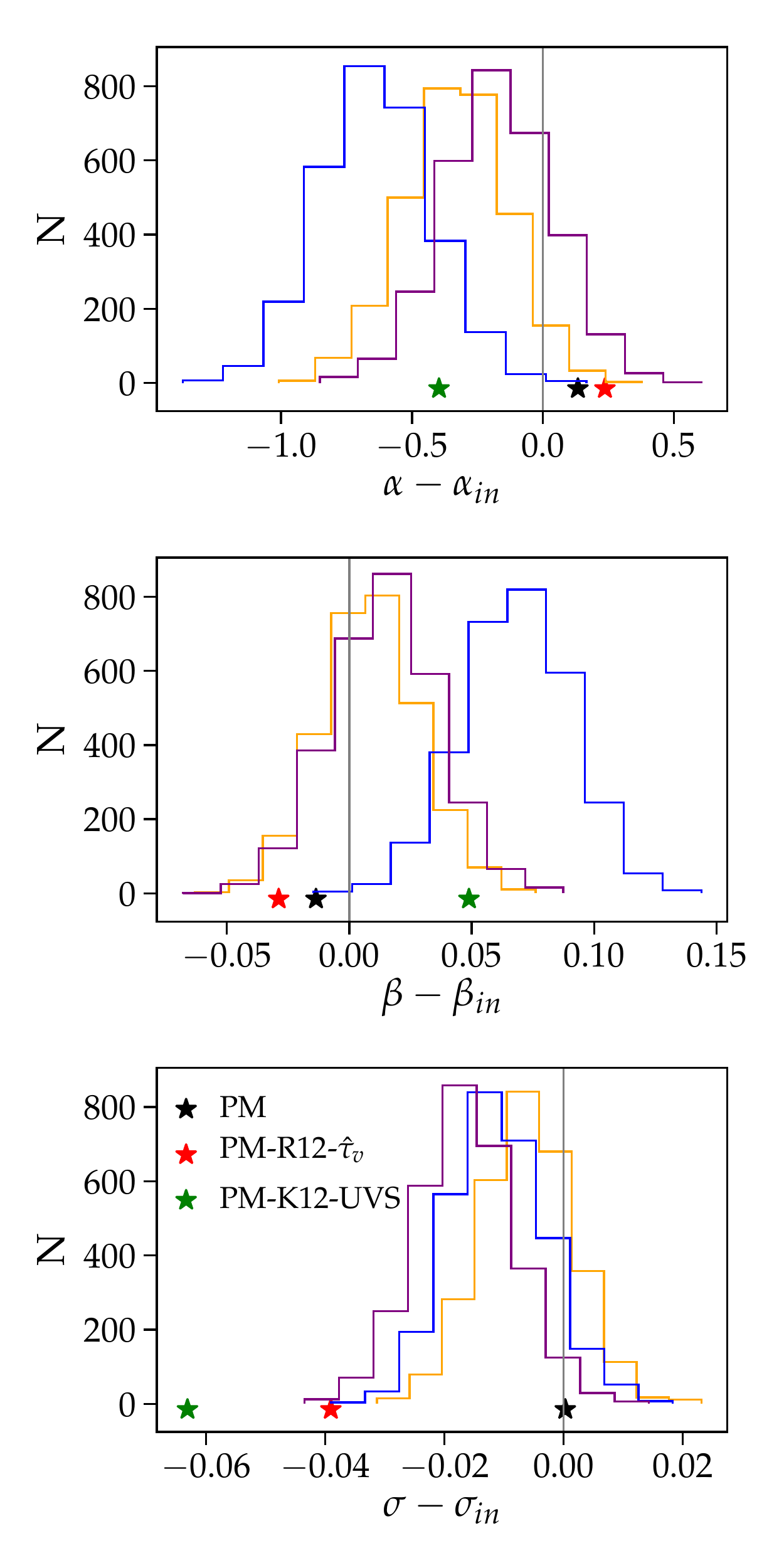}}
  \caption{ Results of \beagle\ fitting to the \constGauss\ scenario, where mock photometry is produced and fitted to using constant SFHs.  The results displayed employ the broad-band filters listed in Table~\ref{tab:delayed_const_test_photom}.  The results for three different \beagle\ parameter configurations are plotted in different colours, as defined in the legend (see text for details).   The left-hand panel displays the triangle plot of the average parameter constraints relative to the input values for the whole sample [where $\log(\MstarInLog/\MstarInLog_{in}) = \Mstar - \Mstar_{in}$ and $\log(\sfrInLog/\sfrInLog_{in})=\sfr - \sfr_{in}$].  The crosses show the average bias over all objects for each parameter pair, and the ovals show the 1$\sigma$ contour of the bi-variate Gaussian describing the average parameter constraints (see text for details on how this plot is made).  The right-hand panel displays the results of fitting the \MstarSFR\ relation to the samples fitted with the same \beagle\ parameter configurations. The histograms show the posterior probability distributions, derived with the \Kelly\ method, relative to the input values for the main-sequence parameters \intercept\ (top), \slope\ (middle) and \scatter\ (bottom).  The stars show the results of fitting with the \PM\ (black), \PMSalmon\ (red) and \PMSantini\ (green) methods, respectively, using the \Mstar\ constraints from the scenario with \logUs\ dependent on \Z\ and \xid\ fitted (purple).}
  \label{fig:constG_BB}
\end{figure*}

\begin{figure}
  \centering
  \includegraphics[width=3.5in]{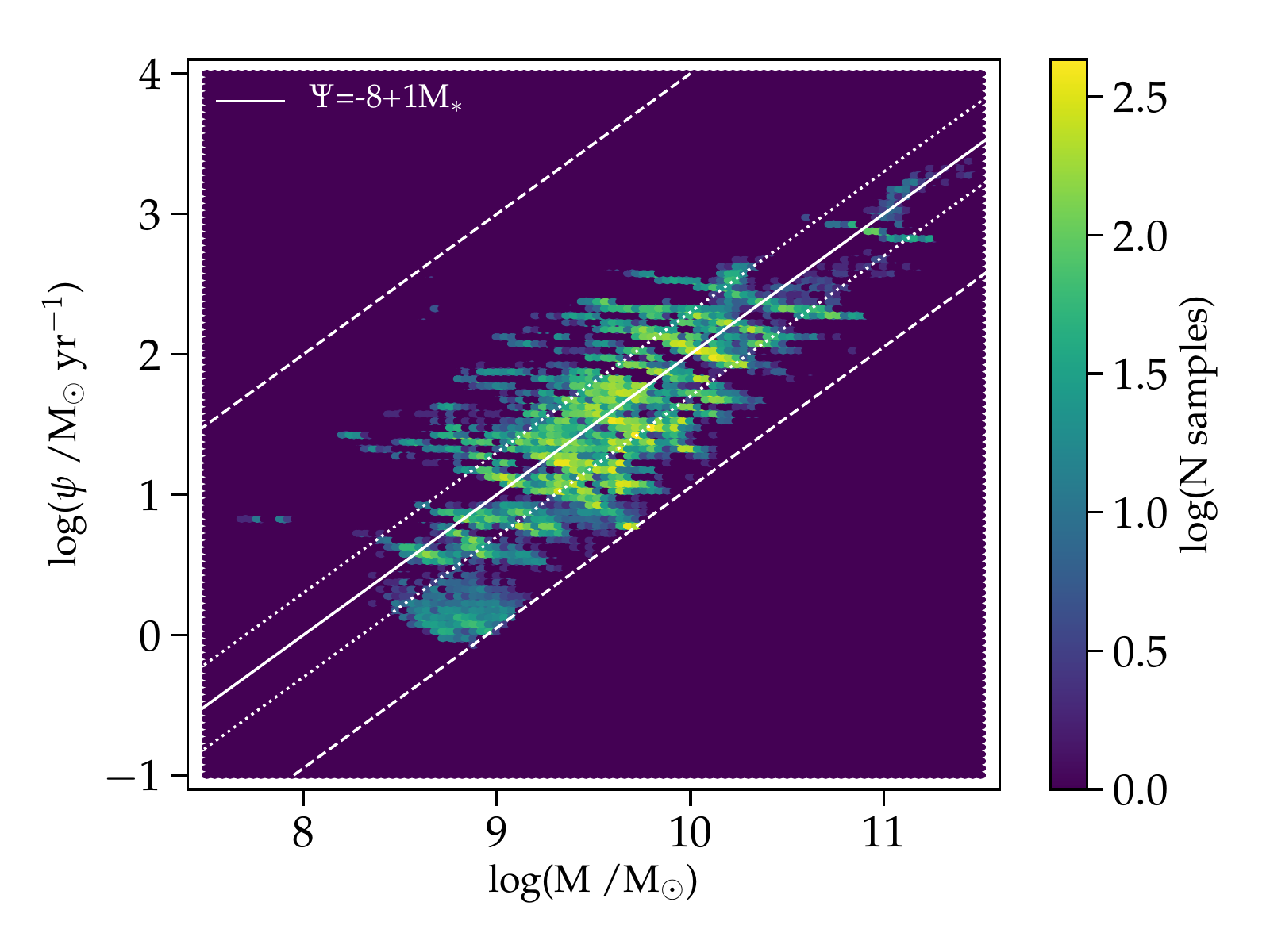}
  \caption{A heatplot of samples from the $\sfr-\Mstar$ posterior distributions from \beagle\ fits to the photometry of 100 mock galaxies in the \constGauss\ scenario, assuming  \logUs\ dependent on Z and \xid\ fixed to 0.1.  The colour coding indicates the density of samples.  The solid white line shows the input \MstarSFR\ relation while the dotted lines show $\pm1\sigma$ intrinsic scatter about the relation. The dashed lines show the hard limits in the prior on \sfr\ that are imposed by the age of the Universe at the redshift of observation (lower line) and the minimum age allowed in the fitting (upper line).}
  \label{fig:heatplot}
\end{figure}

\begin{figure}
  \centering
  \includegraphics[width=3.5in]{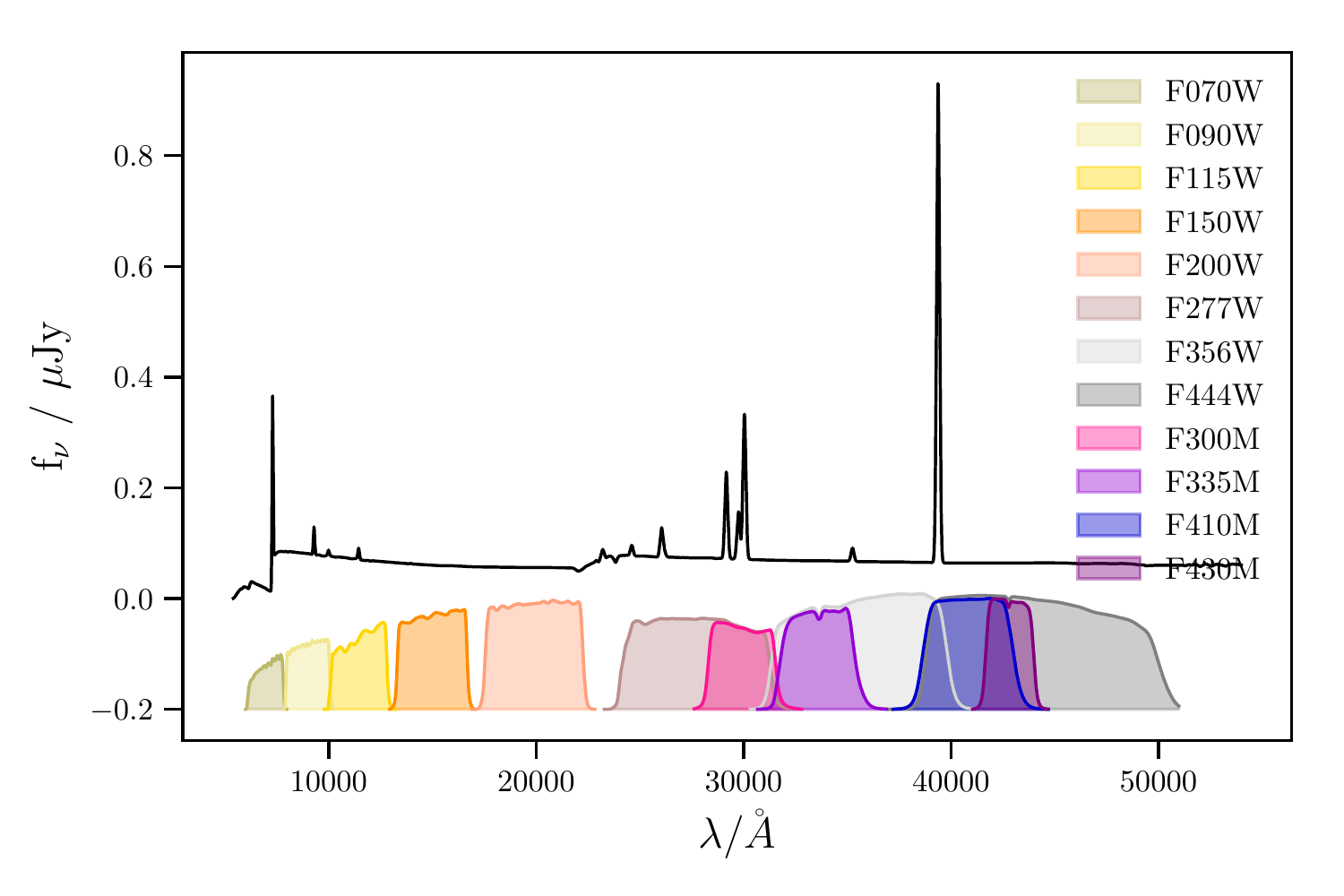}
  \caption{An example spectrum of a mock galaxy at redshift $z=5$ produced with a constant SFH as part of the \constGauss\ scenario.  Below the spectrum we plot the full set of NIRCam broad-band filter profiles as well as four of the red medium-band filters (see legend).  The profiles are fitted with arbitrary normalisation and offset from the spectrum for clarity.  We see that at $z\sim5$, the F300M filter covers the H$\beta$ and \OIII\ emission lines, and the F410M filter covers H$\alpha$.}
  \label{fig:SED_plot}
\end{figure}

\begin{figure*}
  \centering
  \includegraphics[width=3.5in,trim={4cm 1cm 5cm 8cm},clip]{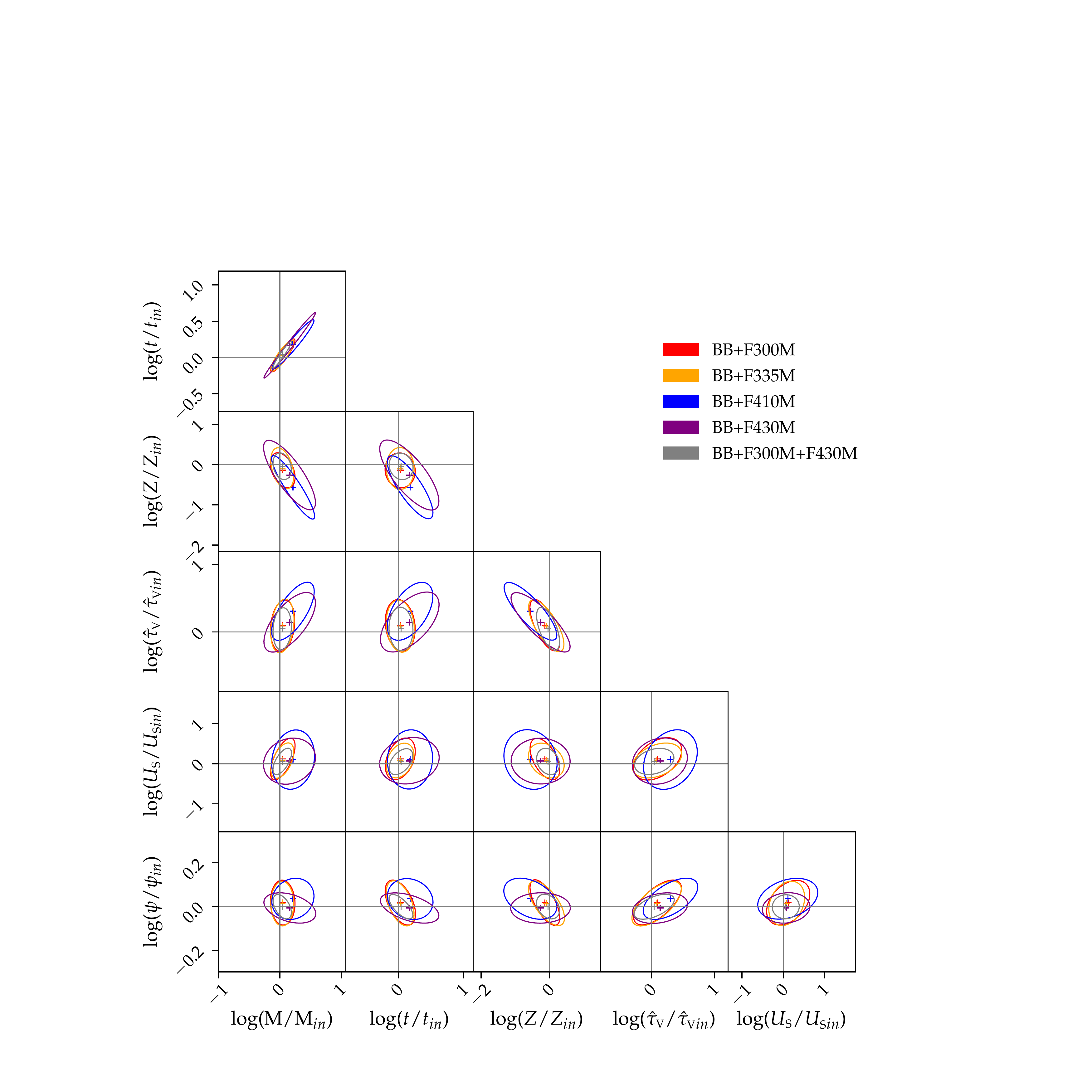}
  \subfigure{\includegraphics[width=1.8in]{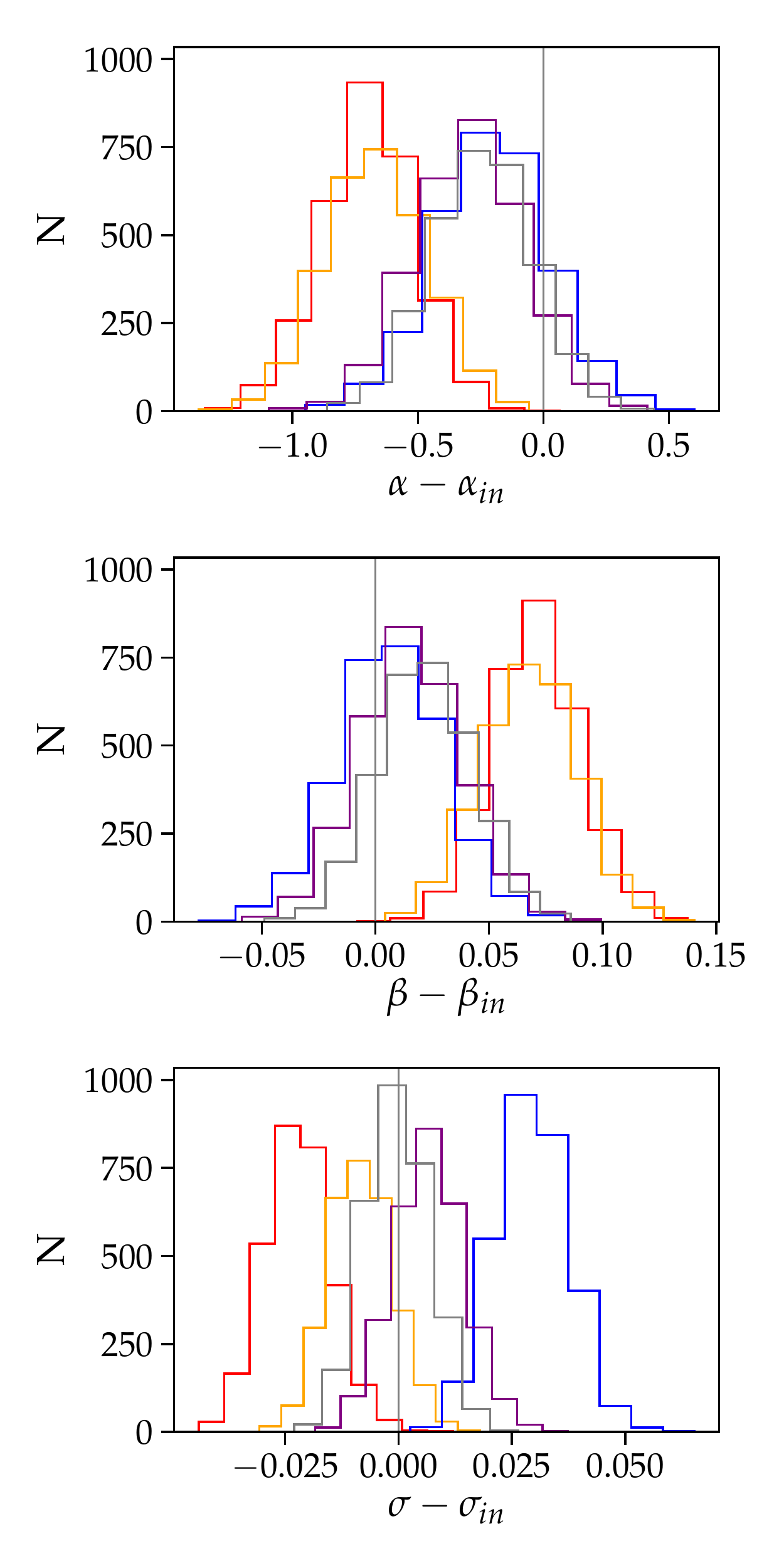}}
  \caption{As for Fig.~\ref{fig:constG_BB},but when including different medium-band fluxes in the \beagle\ fitting to the \constGauss\ scenario, as specified in the legend.  The fits are produced with \xid\ fixed to 0.1.}
  \label{fig:constG_med}
\end{figure*}

\begin{figure*}
  \centering
  \subfigure[Broad-band filters plus F300M]{\includegraphics[width=3.4in,trim={4cm 1cm 5cm 8cm},clip]{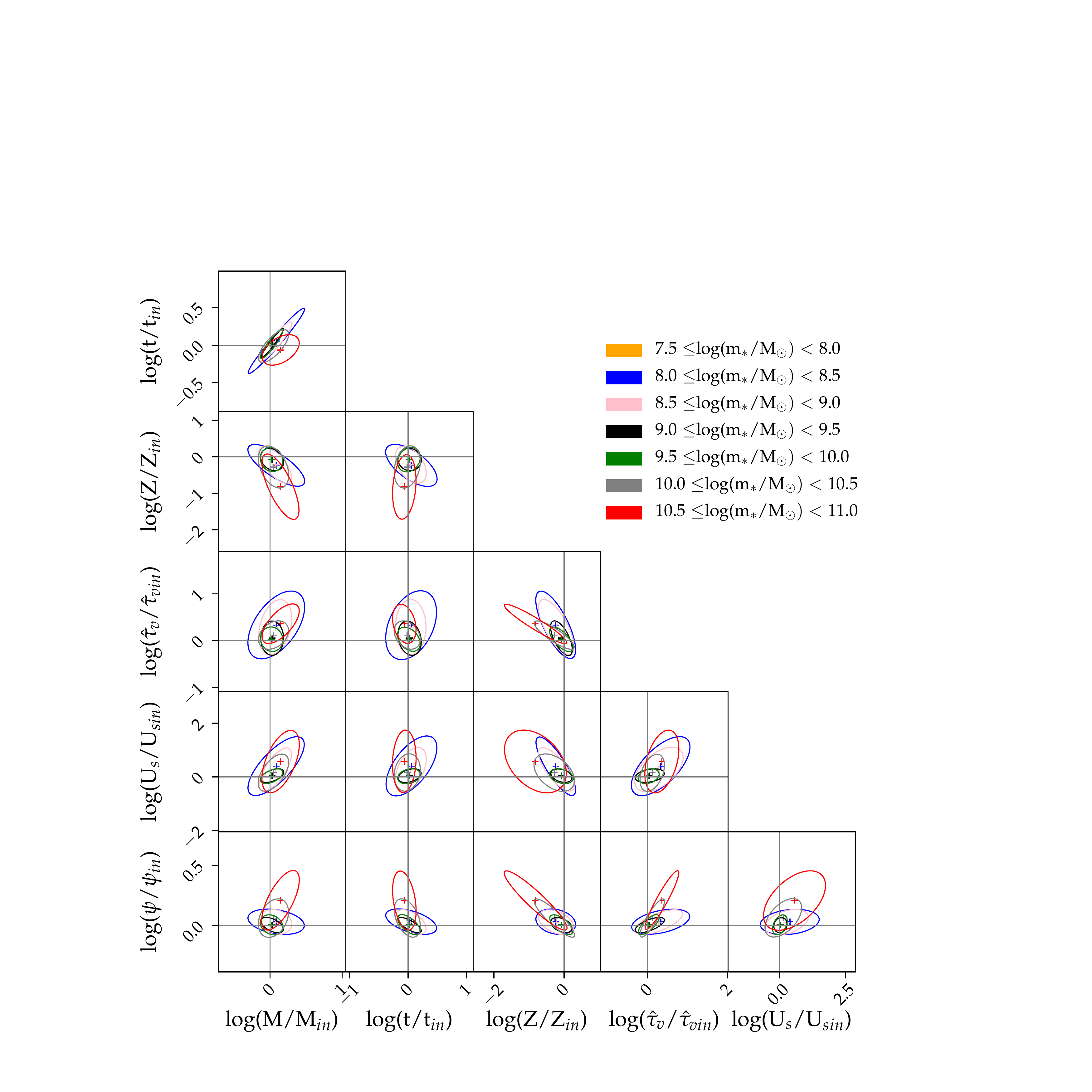}}
  \subfigure[Broad-band filters plus F430M]{\includegraphics[width=3.4in,trim={4cm 1cm 5cm 8cm},clip]{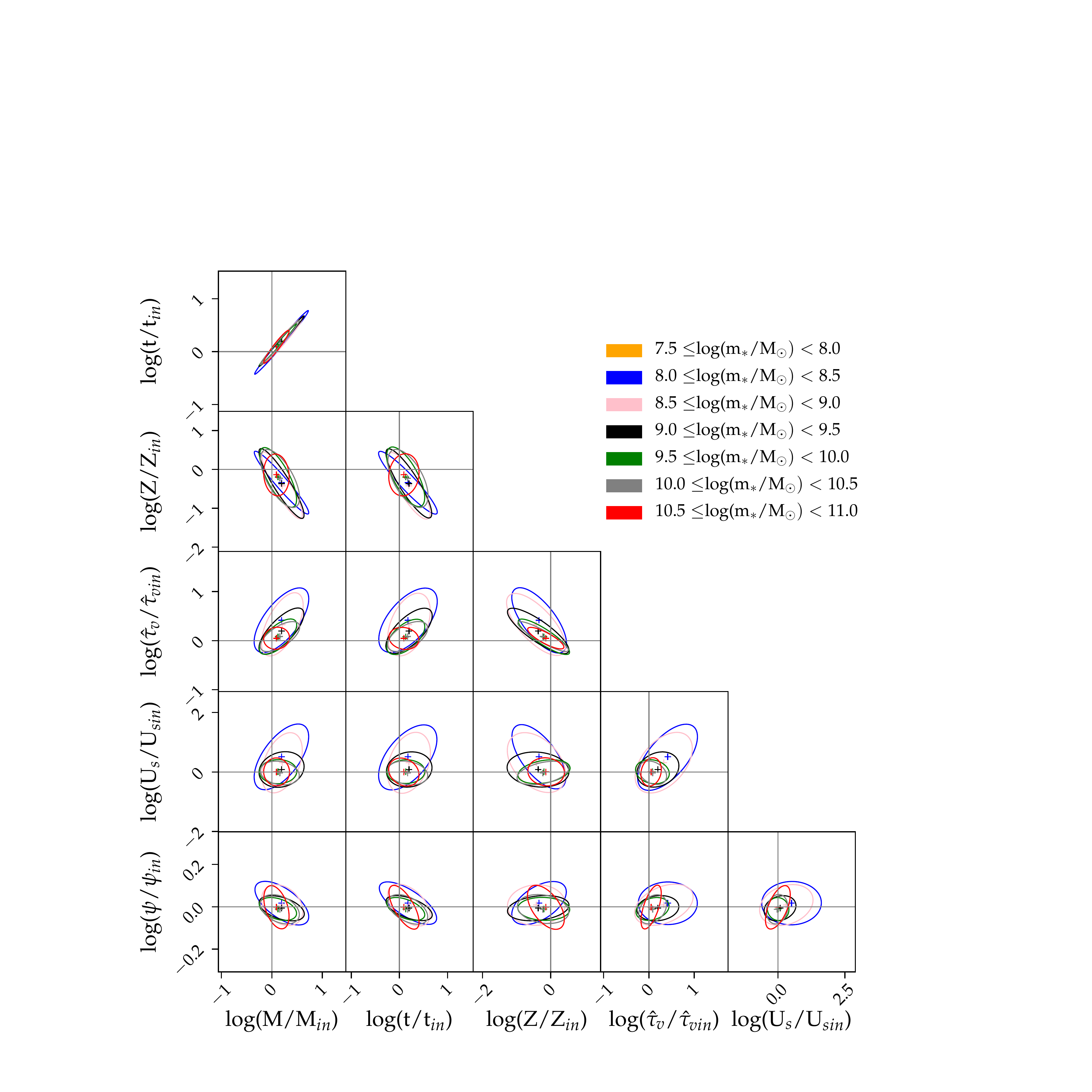}}
  \caption{Average parameter constraints derived from fits to the \constGauss\ scenario using mock photometry in broad-band filters plus the (a) F300M and (b) F430M medium-band filters.  The average constraints are shown in bins of measured \Mstar, as indicated in the legend.}
  \label{fig:constG_med_mass}
\end{figure*}

\begin{figure*}
  \centering
  \subfigure{\includegraphics[width=3.5in, trim={4cm 1cm 5cm 8cm},clip]{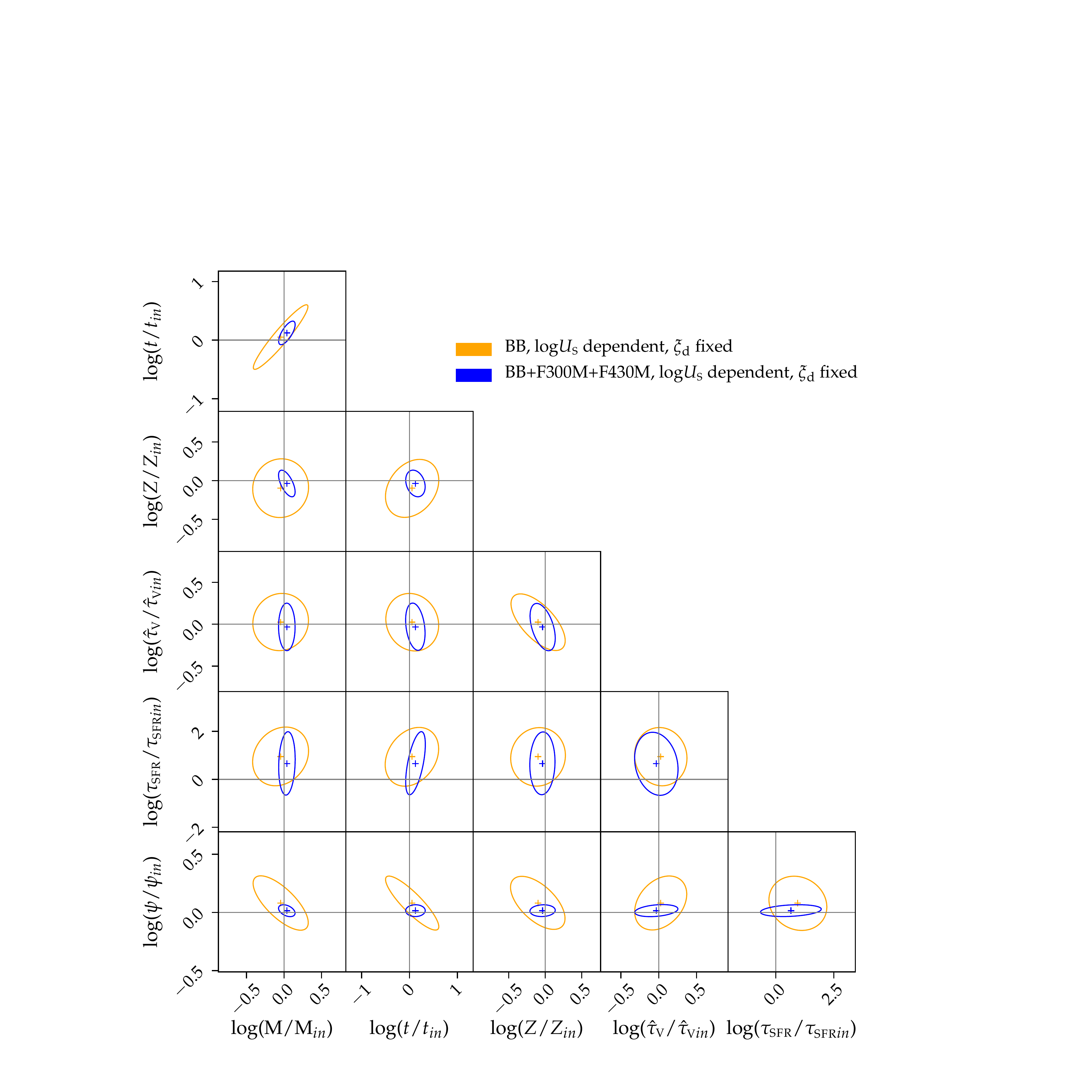}}
  \subfigure{\includegraphics[width=1.8in]{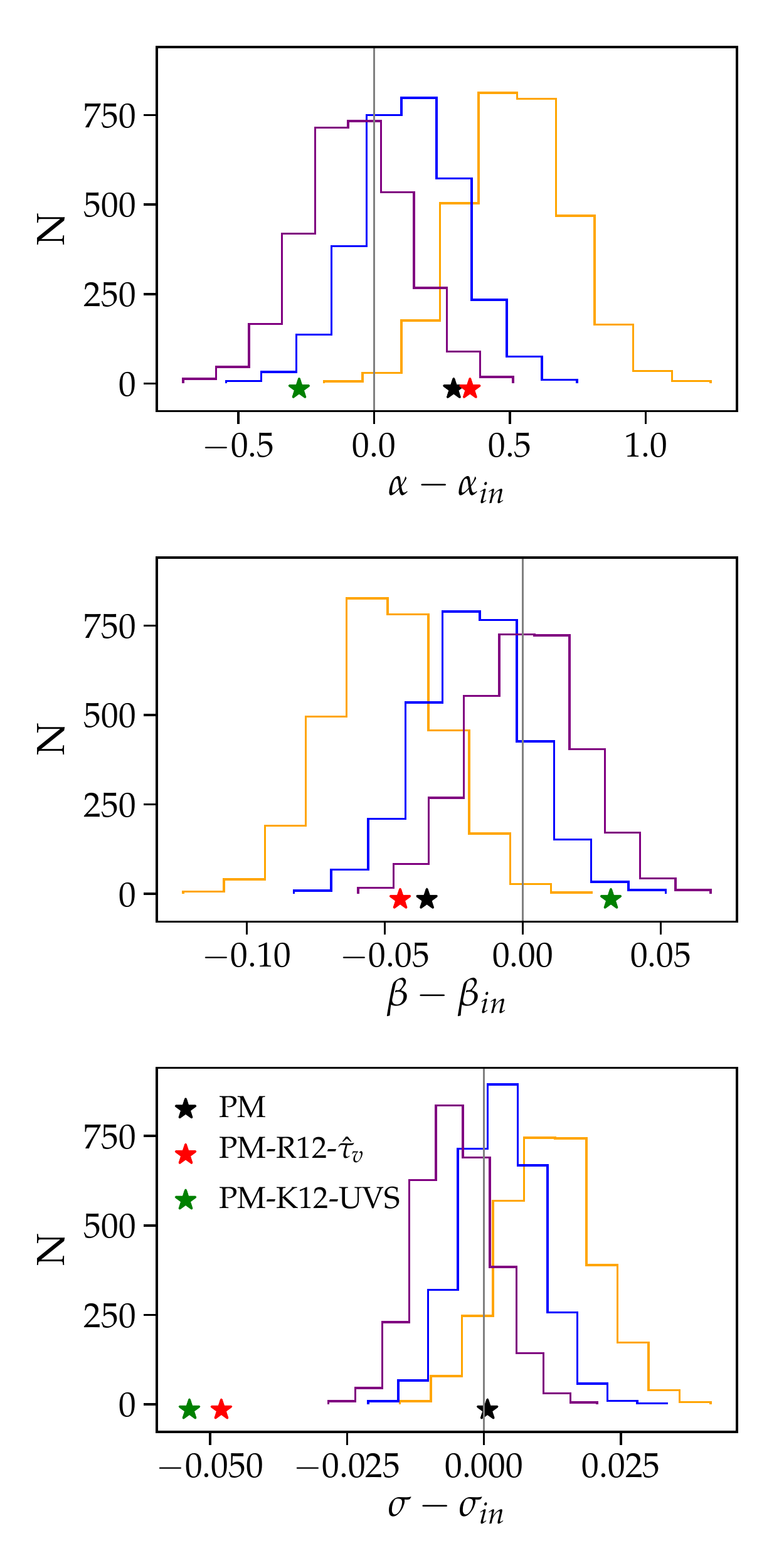}}
  \caption{As for Fig.~\ref{fig:constG_BB} but displaying the results of \beagle\ fitting to the \delGauss\ scenario, where mock photometry is produced and fitted to using delayed exponential SFHs.  The two \beagle\ fitting configurations consist of broad-band only fluxes (yellow) and broad-band fluxes plus F300M and F430M (blue).  Both configurations set \logUs\ dependent on Z and \xid\ fixed to 0.1.}
  \label{fig:delG}
\end{figure*}

\begin{figure}
  \centering
  \includegraphics[width=3.5in,trim={3cm 2cm 5cm 5cm},clip]{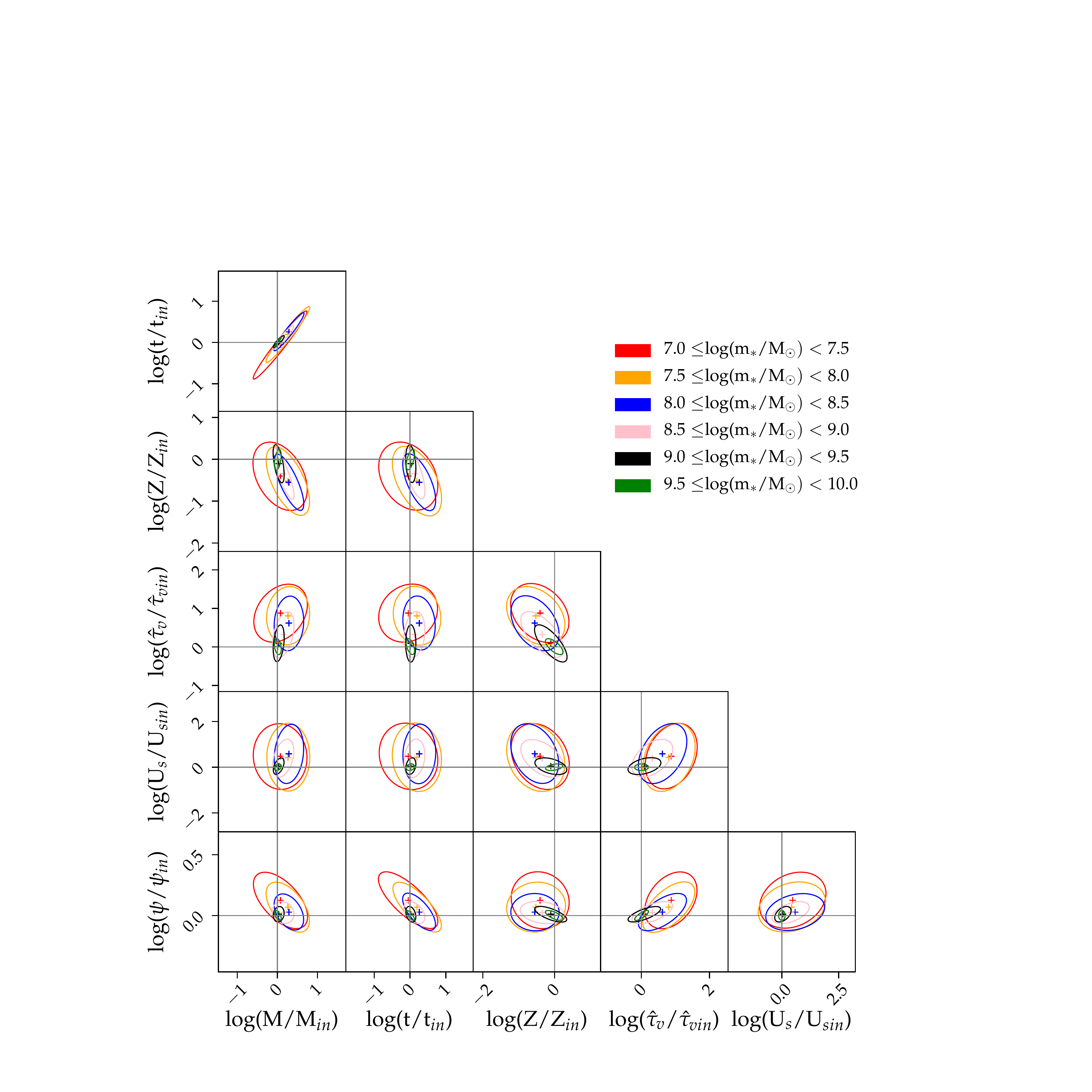}
  \caption{The triangle plot showing average parameter constraints to objects from the \constMF\ scenario fitted to with \beagle. The average constraints are shown as a function of measured \Mstar\ as indicated in the legend.}
  \label{fig:const_MF_triangle}
\end{figure}

\begin{figure}
  \centering
  \includegraphics[width=3.5in,trim={4.5cm 5cm 4.5cm 6cm},clip]{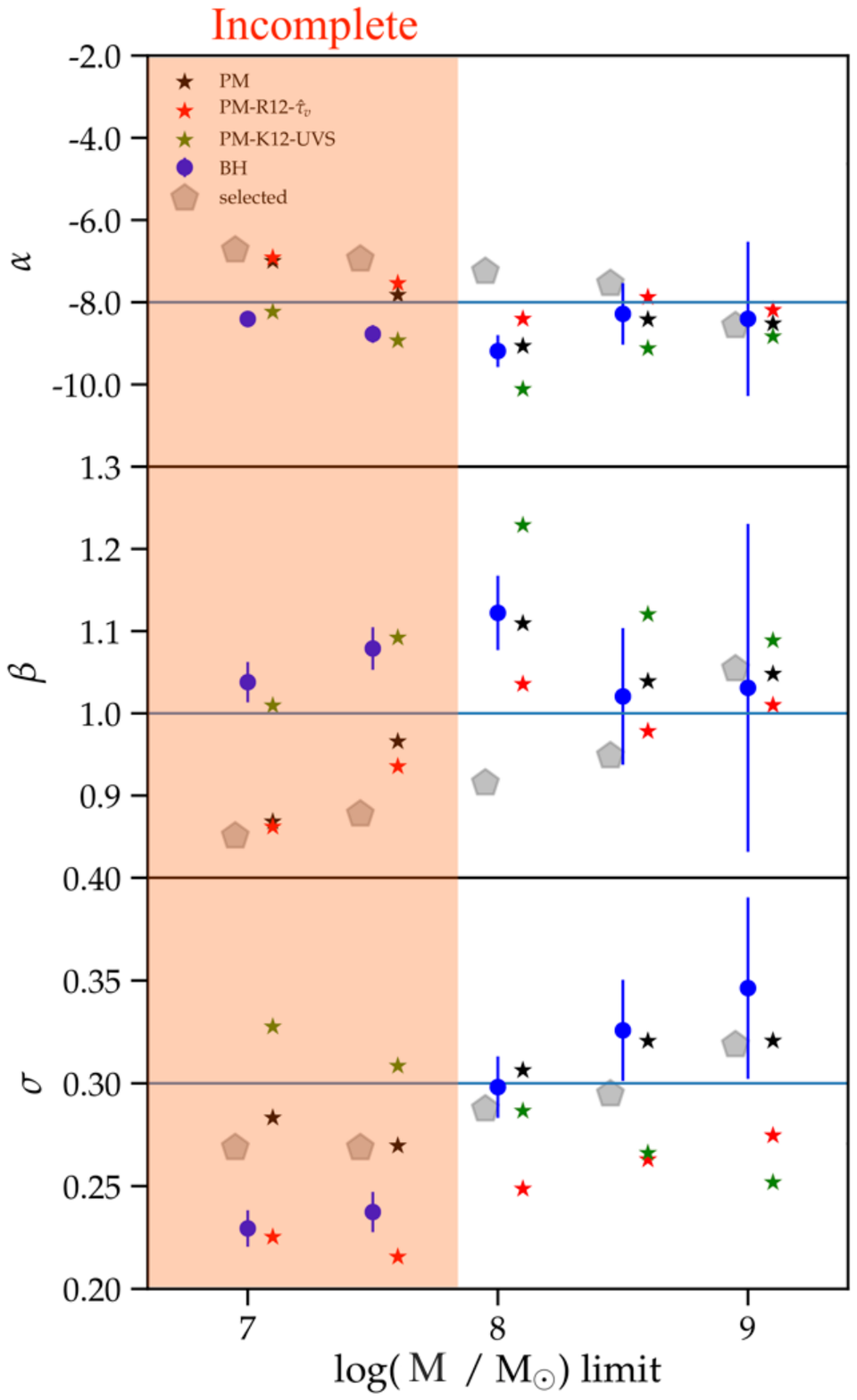}
  \caption{The results of fitting to the \constMF\ scenario, where mock photometry is produced and fitted to using constant SFHs, and mass values are drawn from a mass function (see Table \ref{tab:four_test_scenarios}).   The median and 68-per-cent credible interval of the posterior distribution for each parameter are plotted for the \Kelly\ method as filled circles with error bars, whereas the measurements performed with individual object $\sfr-\Mstar$ posterior medians and alternative \sfr\ estimates are plotted as stars.  The results are shown as a function of lower-limit in \Mstar, where this limit is applied based on measured median \Mstar\ value.  The blue horizontal lines show the true input values, while the grey pentagons show the  relation measured when including only galaxies that meet our selection criteria above the given \Mstar\ limit using the \textit{true} input values.  The orange shaded region delimits the measurements affected by incompleteness (see Fig.~\ref{fig:Testing_input_distribution}).}
  \label{fig:const_MF}
\end{figure}

The four test scenarios described in Section~\ref{section:scenarios} allow us to test the four methods outlined in Section~\ref{section:modelling}.  The \ideal\ scenario was constructed to demonstrate that the \Kelly\ method works as expected in the case of homoskedastic measurement uncertainties that are described by single, covariant Gaussians before considering the added complexity of using \Mstar\ and \sfr\ estimates derived with \beagle.  We show the results in Fig.~\ref{fig:Testing_toy_model_summary}, where the posterior median and 95~\% credible interval of the parameters \intercept, \slope\ and \scatter\ are plotted for the ten independent realizations of the \ideal\ scenario.  The results demonstrate that the \Kelly\ method allows us to recover the input parameters (intercept, slope and scatter of the \MstarSFR\ relation) in an unbiased way. 

The \constGauss, \delGauss\ and \constMF\ scenarios employ \beagle\ to fit to noisy mock photometry to obtain \Mstar\ and \sfr\ estimates.
Starting with the \constGauss\ scenario, we investigate which parameters can be constrained employing the broad-band filters listed in Table~\ref{tab:delayed_const_test_photom}.  The results are plotted in  Fig.~\ref{fig:constG_BB} with the left-hand panel showing a triangle plot of the average constraints obtained for the full set of fitted parameters (\t, \Mstar, Z, \tauV, \logUs\ and \xid) as well as the derived parameter, \sfr.  To produce this plot we first take the logarithm of the ratio between the derived parameter and its input value so that the the joint posteriors can be combined for different objects. For each object, we then fit a bi-variate Gaussian to the joint posterior of each parameter pair.  For each parameter-pair, we plot the mean of single-object Gaussian centres as crosses, and define a sample-wide, bi-variate Gaussian for which each entry in the covariance matrix is the mean of the corresponding entries in the individual-object covariance matrices.  This is an approximation of the average constraints on each parameter from the whole sample, and it is important to point out that this method is not always a good representation of the results.  For example, \xid\ is seldom constrained in these fits, and in most cases, individual-object posterior distributions would be better modelled as uniform between the limits of the prior.  Additionally, the joint posterior is not always well described by a single bi-variate Gaussian, as shown in Fig.~\ref{fig:GMM_example}. However, bearing these caveats in mind, the representation is still useful for displaying average biases and degeneracies between measured parameters.  

The yellow ovals in the triangle plot of Fig.~\ref{fig:constG_BB} show the 1$\sigma$ contour of the sample average bi-variate Gaussians for \beagle\ fits to the photometry of mock galaxies with each free parameter (\t, \Mstar, Z, \tauV, \logUs\ and \xid) allowed to vary within its prior range in Table ~\ref{tab:test_scenario_beagle_parameters}. These fits show \Mstar, \sfr, \t, \tauV\ and \xid\ biased high, and Z biased low while \logUs\ has very large uncertainties but is not obviously biased. In addition, uncertainties on \Mstar\ are positively correlated with \t\ and \tauV, but negatively correlated with Z, and uncertainties on Z are also negatively correlated with \tauV. After experimenting, we found that poor constraints on \logUs\ are driving the observed biases.  Setting \logUs\ to depend on Z according to $\logUs = -3.638+0.055\Z+0.68\Z^2$ (similar to equation~\ref{eq:logU_Z}, but without the added scatter) vastly improves the constraints as shown by the blue ovals.  Fixing \xid\ makes little difference to the average parameter estimates (purple ovals).  The ionisation parameter, \logUs, has a strong impact on the nebular continuum and line emission, of which the contribution to the SED cannot be constrained by broad-band filters alone, as demonstrated by these results.  When using \beagle, the existence of biases in \Mstar\ and \sfr\  arising from this issue can be diagnosed from poor \logUs\ constraints, and can be mitigated by setting a suitable prior on \logUs.

Fig.~\ref{fig:heatplot} shows a `heatplot' of samples randomly drawn from the joint posterior probability of $(\Mstar, \sfr)$ for \beagle\ fits to the photometry of 100 mock galaxies in the \constGauss\ scenario, assuming \logUs\ dependent on Z and \xid\ fixed to 0.1. From this plot we see that the $\sfr-\Mstar$ uncertainties include an imprint of the priors associated with other physical parameters that were sampled over in the \beagle\ fits. In particular, limits in the prior on galaxy age \t\  can impose sharp cutoffs in the allowed \sfr \ (shown as white dashed lines in Fig.~\ref{fig:heatplot}).

The results of fitting to the \MstarSFR\ relation with the \Kelly\ method are displayed as histograms in the right-hand panel of Fig.~\ref{fig:constG_BB}.  The posterior probability distributions of \intercept, \slope\ and \scatter\ are shown relative to the true input values.  When all free parameters are varied in the fits (yellow histograms), \intercept\ is biased low whereas \slope\ and \scatter\ are unbiased.  The biased \Mstar\ and \sfr\ estimates act to shift the relation to the right parallel to the true input relation.  It is not obvious that \slope\ would remain unbiased if the slope of the cutoff imposed by the maximum age limit were not similar to the slope of the input relation (see Fig.~\ref{fig:heatplot}).    \beagle\ fits with \logUs\ dependent on Z while still fitting for \xid\ lead to a measured \MstarSFR\  relation with lower intercept and a steeper slope (blue histograms), while fixing \xid\ to the correct value gives unbiased estimates of slope and intercept.  The interplay between \xid\ and other parameters is complex, and the biases introduced in the retrieved main-sequence parameters when \xid\ is unconstrained cannot be easily explained by looking at the sample-averaged constraints alone.  This highlights the necessity of attempting to recover the input \MstarSFR\ relation itself, as well as assigning suitable priors, or fixed values, to parameters that cannot be constrained by the data set.  The fits with fixed \xid\ and \logUs\ dependent on Z, in contrast, lead to \scatter\ being only marginally underestimated.  Fig.~\ref{fig:heatplot} shows that the correlated uncertainties in \Mstar\ and \sfr\ are significant in extent compared to the intrinsic scatter in the underlying mock (as indicated by the dotted lines). K07 show that the method can start to under-estimate the intrinsic scatter in this regime (their fig.~6). 

Fig.~\ref{fig:constG_BB} also shows the results obtained with the \PM, \PMSalmon\ and \PMSantini\ methods as coloured stars (black, red and green respectively). For each of these measurements, we use the \Mstar\ (and \sfr\ for the \PM\ method) from \beagle\ fits with fixed \xid\ and \logUs\ dependent on Z.  The \PM\ method gives good estimates of \intercept, \slope\ and \scatter.  The \PMSalmon\ method returns reasonable estimates of \intercept\ and \slope\ (with small offsets likely due to the difference in UV-to-SFR calibration between R12 and the stellar+nebular models used here, see Section~\ref{section:salmon}),  but somewhat underestimates \scatter.  The \PMSantini\ method, however, measures steeper slopes with lower intercept and significantly under-estimates the intrinsic scatter.  The biased slope and intercept estimates are likely due to the mass-dependent metallicities of the underlying mock (equation~\ref{eq:FMR}).  Objects at low \Mstar\ have lower metallicities and hence steeper UV slopes than objects at high \Mstar. The correction  for dust attenuation using the \cite{Meurer1999} prescription will therefore increase with \Mstar, leading to higher corrected \sfr\  at high \Mstar, and hence a steeper measured slope. Also, the UV-to-SFR calibrations employed in the \PMSalmon\ and \PMSantini\ methods do not account for variability due to metallicity, age in the case of \PMSantini, or the contribution of nebular continuum emission to the rest-frame UV, all of which will act to reduce the measured intrinsic scatter compared to the underlying relation.

We investigate whether \logUs\ can be constrained with the addition of medium band filters in the fitting.  Fig.~\ref{fig:SED_plot} shows an example spectrum of a mock galaxy at redshift $z=5$ from the \constGauss\ scenario, with the NIRCam broad-band filters  and four medium-band filters (F300M, F335M, F410M, F430; see Table~\ref{tab:delayed_const_test_photom}) shown at the bottom of the plot.  The F300M filter contains flux contributions from H$\beta$ and the \OIII\ emission lines, while the F335M filter constrains the continuum just red-ward of these lines.  F410M contains flux from H$\alpha$ and F430M samples the continuum just red-ward of H$\alpha$.  We add one medium-band constraint at a time to the fits with \beagle, while allowing \logUs\ to vary but keeping \xid\ fixed to 0.1. The average \beagle\ parameter constraints, and corresponding main-sequence parameter constraints, are shown in Fig.~\ref{fig:constG_med}.  The least biased estimates of \Mstar, \t, Z and \tauV\ are achieved when the F335M or F300M filters are used, likely due to a combination of better constraints on the \logUs-dependent \OIII\ flux, as well as improved constraints on the shape of the Balmer break (by breaking degeneracies between emission-line or stellar contribution to the broad-band fluxes). 

The right panel of Fig.~\ref{fig:constG_med} displays the constraints on the main-sequence parameters when including one medium-band filter at a time, as well as with both the F300M and F430M filters.  The configurations with a single medium-band constraint that give the least biased \beagle\ parameter estimates on average (i.e., adding F300M or F335M fluxes to the fitting; left panel) lead to estimates of the main sequence that are steeper and with lower intercept than the true input values. Conversely, the configurations that give more biased \beagle\ parameter estimates on average (broad-band plus F410M or F430M fluxes), provide less biased estimates of \intercept\ and \slope.  To understand the origin of this finding, we examine in Fig.~\ref{fig:constG_med_mass} how the average \beagle\ parameter estimates depend on \Mstar\ for the broad-band plus F300M and F430M configurations (Figs~\ref{fig:constG_med_mass}a and b, respectively). When the F300M filter is included in the fitting, \logUs, Z and \tauV\ become increasingly biased and poorly constrained at high mass (grey and red contours, Fig.~\ref{fig:constG_med_mass}a).  This is because of the presence of mass-dependent dust in the model SEDs (see Fig.~\ref{fig:scenario_iv_params}): more dust is allowed at high \Mstar, which, in turn, allows a greater range of \logUs\ in the fitting (as described in Section~\ref{section:modelling}). At $z=5$, the F300M and F335M filters provide constraints on H$\beta$+\OIII, and so constraints on \logUs\ (indirectly from \OIII) and \tauV\ (indirectly from H$\beta$) are degenerate.  This leads to estimates on \Mstar\ and \sfr\ that are biased high at high \Mstar, unlike in the case when F430M (or F410M) provide constraints on H$\alpha$ at high \Mstar\ (Fig.~\ref{fig:constG_med_mass}b). Fig.~\ref{fig:constG_med_mass}b shows that \Mstar\ is biased high at low \Mstar, but this does not translate into such biased estimates of the main sequence.  This is likely because the prior on age limits the level of bias that can be reached. These results demonstrate that to obtain unbiased \beagle\ parameter estimates \textit{and} main-sequence parameter constraints, two medium band filters are needed, one providing constraints on H$\beta$ plus \OIII, and one providing constraints on H$\alpha$.  The results including F300M and F430M constraints are displayed in Fig~\ref{fig:constG_med}, and we trialed different configurations (F300M plus F410M, F335M plus F410M and F335M plus F430M) finding similar results.

The results for the \delGauss\ scenario are shown in Fig.~\ref{fig:delG}.  Here we tried fitting with \beagle\ to the broad-band fluxes setting \logUs\ dependent on Z and $\xid=0.1$ (the optimal configuration when fitting to broad-band fluxes in the \constGauss\ scenario).  However, with this  configuration, \sfr\ and \tausfr\ are biased high.  The constraints on \tausfr\ are very poor, with the 68-per-cent credible interval spanning $\sim2$ dex.  The corresponding \intercept, \slope\ and \scatter\ estimates derived from fitting to the \MstarSFR\ relation with the \Kelly\ method are biased toward high intercept, shallow slopes and over-estimates of the intrinsic scatter.  Adding the F300M and F430M filters to the fitting significantly improved constraints on \t, \Mstar\ and \sfr, leading to unbiased constraints on \intercept\ and \slope\ and only marginally over-estimated \scatter. Although not shown on the figure, we verified that \logUs\ remains unconstrained when medium bands are included in the fits, meaning that a prior on \logUs\ (as well as the additional medium-band information) is required to provide the unbiased \intercept\ and \slope\ estimates.  It is worth noting that we fit the \MstarSFR\ relation to instantaneous \sfr\ estimates, rather than SFRs averaged over a given timescale (e.g. 10\,Myr or 100\,Myr).  Averaging \sfr\ estimates in this way would act to lower the upper limit in the \MstarSFR\ prior space (this limit is approximately the same for the delayed-exponential SFH as for the constant SFH and is shown as the upper dashed-white line in Fig.~\ref{fig:heatplot}), which can in turn lead to under-estimates of the scatter.

In the \constMF\ scenario we consider a more physically-motivated distribution of \Mstar\ values, i.e. with a larger proportion of objects with low stellar masses.  The lower mass objects have significantly poorer \Mstar\ and \sfr\ constraints than higher-mass objects.  This can be best appreciated from Fig.~\ref{fig:const_MF_triangle}, which shows the average parameter constraints for mock galaxies in different mass ranges.  This plot shows that the constraints on \Mstar\ and \sfr\ are much poorer at low than at high \Mstar.  In fact, at low stellar masses, \Mstar, \t, \logUs, \tauV\ and \sfr\ are biased high, while Z is biased low. The \constMF\ scenario also includes selection effects, since some objects are too faint to be detected at the chosen survey depths. Fig.~\ref{fig:Testing_input_distribution} shows the input \MstarSFR\ distribution, as well as the objects entering our selection.  The distribution of selected objects is characterised by SFRs that are higher than the average population at low \Mstar.  

In practical situations, stellar mass cuts are imposed on samples of galaxies to ensure that either the samples are complete in stellar mass, or only objects with good-enough constraints on stellar mass and \sfr\ are used to derive population-wide relations.  In Fig.~\ref{fig:const_MF}, we hence adopt a similar approach and show the constraints on \intercept, \slope\ and \scatter\ obtained for the \constMF\ scenario when imposing different minimum stellar mass cuts.  The \beagle\ fits used in this analysis include the F300M band, allow \logUs\ to vary but fix \xid\ to 0.1 (the optimal fitting strategy identified for the \constGauss\ scenario, although the results do not change significantly when fitting without the medium-band filters and setting \logUs\ to depend on Z). We also show the result of ordinary linear regression to the `true' \Mstar\ and \sfr\ values of the galaxies selected at each mass cut (grey pentagons). This allows us to identify how the selection effects themselves are biasing the measurements of the \MstarSFR\ relation.  In particular, the intercept is progressively over-estimated and the slope progressively under-estimated from the true \Mstar\ and \sfr\ values with decreasing \Mstar\ cut.  This indicates that the objects entering the selection based on the posterior median \Mstar\ value have true values below the nominal \Mstar\ cut, but the covariant form of the uncertainties mean that they preferentially have higher \sfr\ than the average. Fig.~\ref{fig:const_MF} shows that the constraints on \intercept\ and \slope\ obtained with the \Kelly\ method for stellar mass cutoffs $\gtrsim10^{8.5}\Msun$ agree to within the uncertainties with the underlying input relation, once selection effects are accounted for. For lower mass cutoffs, however, the \Kelly\ method overestimates \slope\ and underestimates \intercept, and at the lowest \Mstar\ cuts also underestimates \scatter. This indicates that including objects with poorer parameter constraints biases the measured relation to steeper slopes and lower intercepts in our idealised scenario, and this is shown to be occurring above the nominal completeness limit.

The measurements of the \MstarSFR\ relation from the \PM, \PMSalmon\ and \PMSantini\ methods for high \Mstar\ cuts follow similar trends to that seen in the \Kelly\ method. Specifically, the \PMSantini\ method measures steeper slopes with lower intercept and scatter than the input relation, while \PMSalmon\ provides reasonable estimates of slope and intercept, but still under-estimates the scatter.  However, when low \Mstar\ cuts are imposed, the \PM\ and \PMSalmon\ methods measure shallower slopes with higher intercepts than measured by the \Kelly\ method. This indicates that the posterior medians roughly follow the underlying input values, but that there is significant posterior probability below the input relation at low \Mstar\ that acts to bias the \intercept\ and \slope\ estimates derived with the \Kelly\ method.  The trend of the \scatter\ estimates obtained with the \PMSantini\ method is also of note, with increasing measured \scatter\ when lower \Mstar\ limits are applied.  This is because the intrinsic UV slope estimates used to correct the rest-frame UV for dust become increasingly uncertain.  The UV slope uncertainty is accounted for in the San17 work.

\section{What can we measure with JWST}

We have seen from the results of fitting to the \constGauss, \delGauss\ and \constMF\ scenarios in Section~\ref{section:results} that the priors imposed in the SED fitting, as well as poor parameter constraints, can significantly bias measurements of the \MstarSFR\ relation.  In this section, we explore how we can avoid these biases while also allowing for the possibility of measuring any mass dependence of the intrinsic scatter, in particular if the mass dependence is due to stochastic star formation.  We then investigate how far down the mass function we might probe at $z\sim5$ with \JWST\ with simple exposure time calculations.

\subsection{Stellar-mass and SFR estimates and their impact on the search for mass dependence of the intrinsic scatter}
\label{section:mstar_sfr_constraints}

We found that employing the UV-based \sfr\ estimates (the \PMSalmon\ and \PMSantini\ methods) requires simplifying assumptions that cause \scatter\ to be biased.  One alternative would be to use stellar-mass estimates from SED fitting along with completely independent \sfr\ estimates, e.g. from H$\alpha$ corrected for dust using the H$\alpha$/H$\beta$ ratio, as used in the work of, e.g., \cite{Shivaei2015b}, however even this approach may prove problematic.

Considering first the stellar-mass estimates, even if the photometry has high-enough S/N to avoid biases, the effective priors on SFR imposed by a constant or rising SFH can \textit{still} affect the mass estimates.  Although we have not directly investigated rising SFHs, it is relatively simple to calculate the lower limit imposed by the age of the Universe in the \MstarSFR\ plane for a given parametrization (lower white-dashed line in Fig.~\ref{fig:heatplot} for constant SFHs).  For example, if an object with a somewhat bursty SFH sits (in between bursts) below the lower limit imposed by the age of the Universe at the time of observation and is fitted to with a constant or rising SFH, the derived stellar mass will be biased low and/or the SFR biased high. The extent of either bias will depend on the relative S/N between rest-frame optical and UV bands.  
These effects would mask any increase in scatter to low mass present in the population. 

\begin{figure*}
  \centering
  \subfigure[constant SFH]{\includegraphics[width=3.4in,trim={4.5cm 2cm 4cm 5cm},clip]{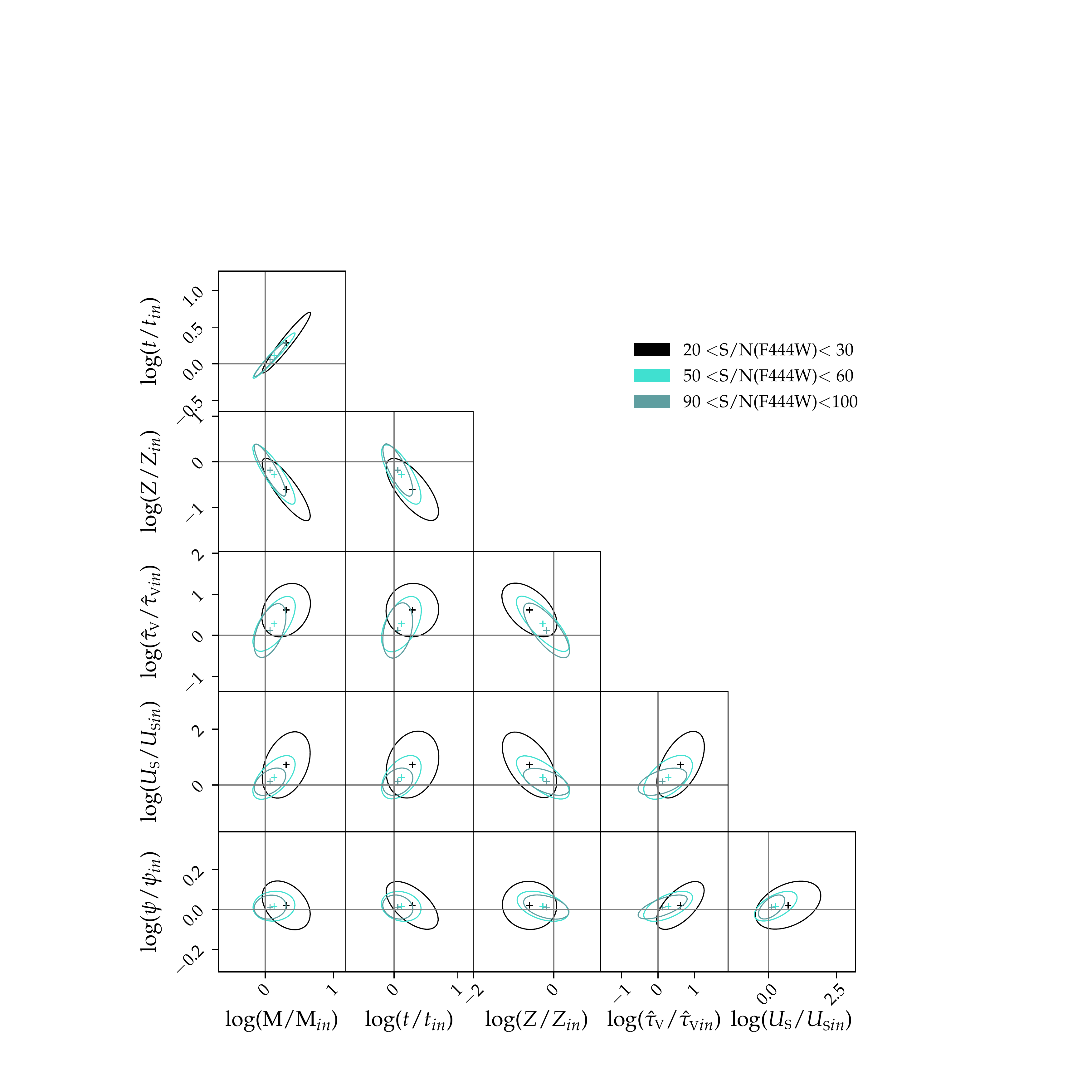}}
  \subfigure[\sfrFree\ SFH]{\includegraphics[width=3.4in,trim={4.5cm 2cm 4cm 5cm},clip]{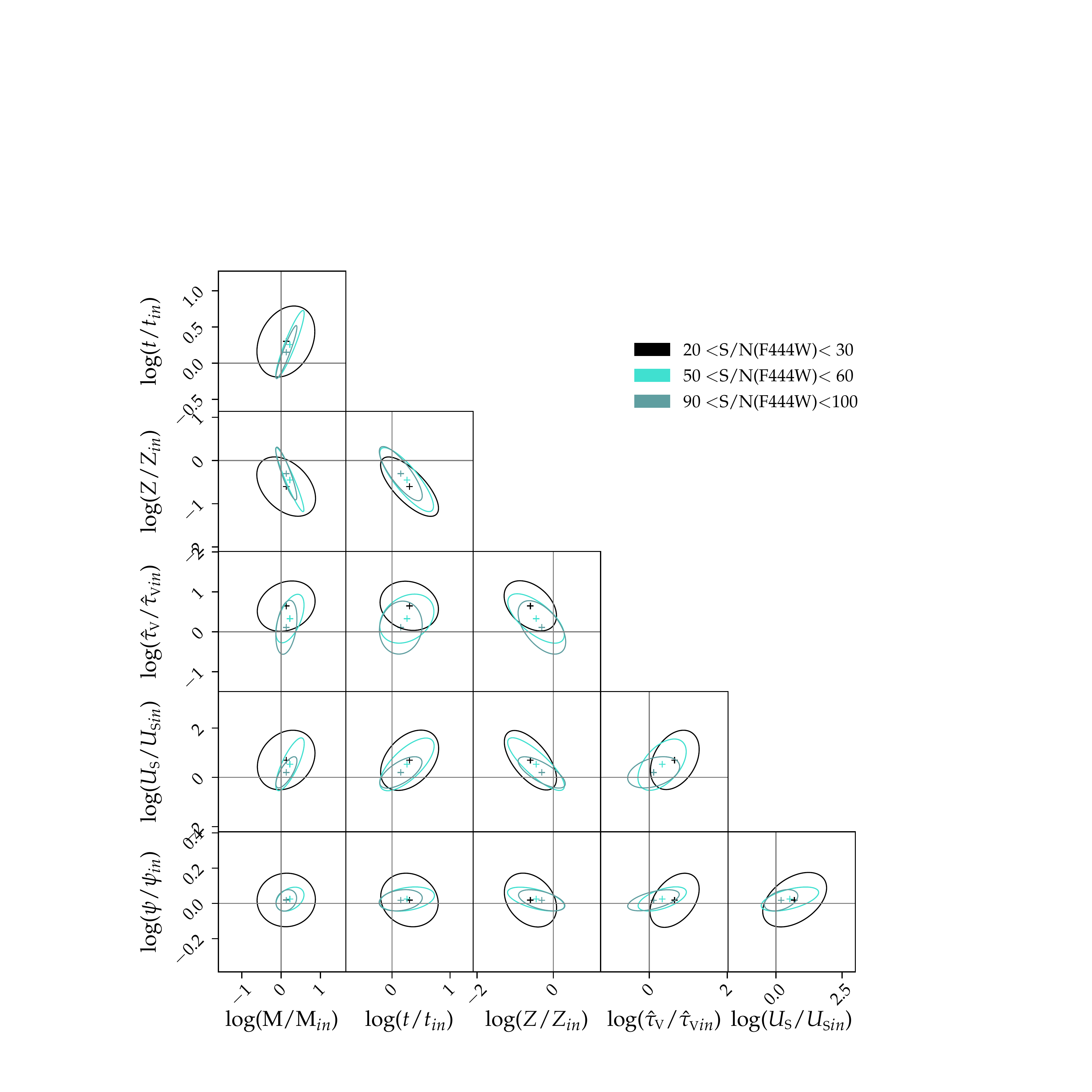}}
  \caption{Average parameter constraints derived from fits to the \constMF\ scenario using mock photometry in broad-band filters plus two medium bands, F335M and F410M: (a) with a constant SFH; (b) with a constant SFH with freely varying SFR in the last 10\,Myr (\sfrFree\ SFH). The average parameter constraints are displayed for three bins of F444W flux S/N.}
  \label{fig:sfrFree_vs_const}
\end{figure*}

A delayed exponential SFH \textit{does} allow for the scenario where SFR was higher in the past than at present, therefore allowing the constraints in \sfr\ to reach low values at given \Mstar.  However, it is not generally used to characterise SFR variations on short timescales and has been used because it describes the general bell-like curve of SFHs of high-mass galaxies on longer timescales \citep{Pacifici2013, Iyer2019}.  A rising SFH might be the general trend for average SFHs at low mass, at both low and high redshifts \citep{Salmon2015}.  Indeed, San17 constrain their fits with a delayed exponential history to be rising prior to $z\sim4$. The choice of a smoothly varying SFH is generally justified by the assumption that constraints on the SFR from photometric fits are driven by the rest-frame UV, which varies on timescales of $\sim30$--100\,Myr (e.g., Sal15). However, our results show (as do many works prior to this, e.g. \citealt{Curtis-Lake2013,Smit2014,Marmol-Queralto2016}) that at high redshifts, photometry \textit{is} sensitive to star formation on shorter timescales, primarily because emission lines contribute a considerable fraction of the broad-band flux.  A delayed exponential, constant or rising SFH does not account for SFR variations on short timescales. 

A significant body of work is currently exploring how constraints on galaxy parameters may depend on SFH parametrization, from possible biases imposed by analytic SFHs \citep{Carnall2019a}, to new constraints from more flexible histories \citep{Leja2019a,Leja2019}, and innovative parametrizations using Gaussian Processes \citep{Iyer2017,Iyer2019}.  We offer a simple parametrization that mitigates some of the issues we find for measuring the \MstarSFR\ relation in particular (rather than for deriving SFHs of individual galaxies), by decoupling current SFR from the previous SFH.  For example, one can construct a SFH in which the last 10\,Myr have constant SFR, while the previous history be described by a burst, delayed exponential  or constant history \citep[e.g.][]{Curtis-Lake2013,Chevallard2018}. 

In Fig.~\ref{fig:sfrFree_vs_const}, we show the average parameter constraints for three different bins in F444W S/N for \beagle\ fits to the \constMF\ scenario using a constant SFH (Fig.~\ref{fig:sfrFree_vs_const}a) and a constant SFH with freely varying SFR in the last 10\,Myr (we shall call this a \sfrFree\ SFH; Fig.~\ref{fig:sfrFree_vs_const}b). We find that \sfr\ is un-constrained for broad-band fluxes using a \sfrFree\ SFH, but including medium bands in the fitting provides constraints on \Mstar\ and \sfr\ comparable to the fits using constant SFH.  The fits shown in Fig.~\ref{fig:sfrFree_vs_const} have \xid\ fixed to 0.1, allow \logUs\ to vary freely and include the F335M and F410M filters.   In fact, the \sfrFree\ parametrization avoids biases in \Mstar\ from the fits using the constant SFH at low F444W S/N, which are introduced by degeneracies between \t, Z and \tauV. The \sfrFree\ is still quite restrictive, however, and one may investigate the change in parameter constraints with different prescriptions for the past SFH segment (e.g. with a delayed-exponential history to allow for variation between past star formation at intermediate [$30\lesssim{t}\lesssim100$] and older [$\gtrsim100$] ages).

We note that the \Mstar\ and \sfr\ constraints obtained with the \sfrFree\ SFH at S/N(F444W)$\sim20$ are still relatively poor compared to the expected level of intrinsic scatter we hope to measure in the \MstarSFR\ relation (we verify that the same is true if binning based on S/N in the F090W filter).  It may be advantageous to obtain independent constraints on the SFR from spectroscopy with NIRSpec.  

Considering the requirement for an un-biased estimate of the SFR, correcting H$\alpha$ for dust attenuation using the H$\alpha$/H$\beta$ flux ratio (or Balmer decrement) is a standard procedure. To obtain such measurements, we must account for H$\beta$ and H$\alpha$ stellar absorption.  Under the assumption of smoothly-varying SFHs, the stellar absorption varies little, and the line emission may be corrected for stellar absorption using an average correction factor.  However, 
H-Balmer absorption increases from type O to B to A stars, and so increases significantly after the first population of O-type stars die.
If we relax the assumption of a smoothly-varying SFH, H$\beta$ and H$\alpha$ stellar absorption becomes highly uncertain. In Fig.~\ref{fig:Ha_Hb}, we demonstrate the level of variability in the measured
 Balmer decrement accounting from H-Balmer absorption in the stellar atmospheres.  The plot shows the intrinsic and measured H$\alpha$/H$\beta$ ratio as a function of $V$-band attenuation optical depth \tauV, for a set of mock spectra created with the \sfrFree\ SFH on a grid of input parameters (given in the figure caption), all with a current SFR\,$=1\Msun/$yr. Here we are assuming that spectra at these redshifts obtained with NIRSpec have insufficient continuum S/N or resolution to allow simultaneous fitting to the Balmer absorption and the emission line (a reasonable assumption for low-mass galaxies, where stochastic star formation may dominate).  The measured H$\alpha$/H$\beta$ ratio therefore includes the effects of underlying stellar absorption. H$\alpha$ cannot, therefore, be corrected for dust attenuation using H$\beta$ plus some average stellar absorption correction.  The measured H$\alpha$/H$\beta$ flux ratio will depend not only on dust attenuation, which it is being used to correct for, but \textit{also} on the ratio of recent to longer-term SFR, as well as the duration of the most recent burst.  In principle, fitting to the broad-band fluxes \textit{together with} the measured H$\alpha$ and H$\beta$ line fluxes with a two-component SFH would self-consistently account for the uncertainty in H$\beta$ and H$\alpha$ stellar absorption. 

\begin{figure}
  \centering
  \includegraphics[width=3.5in]{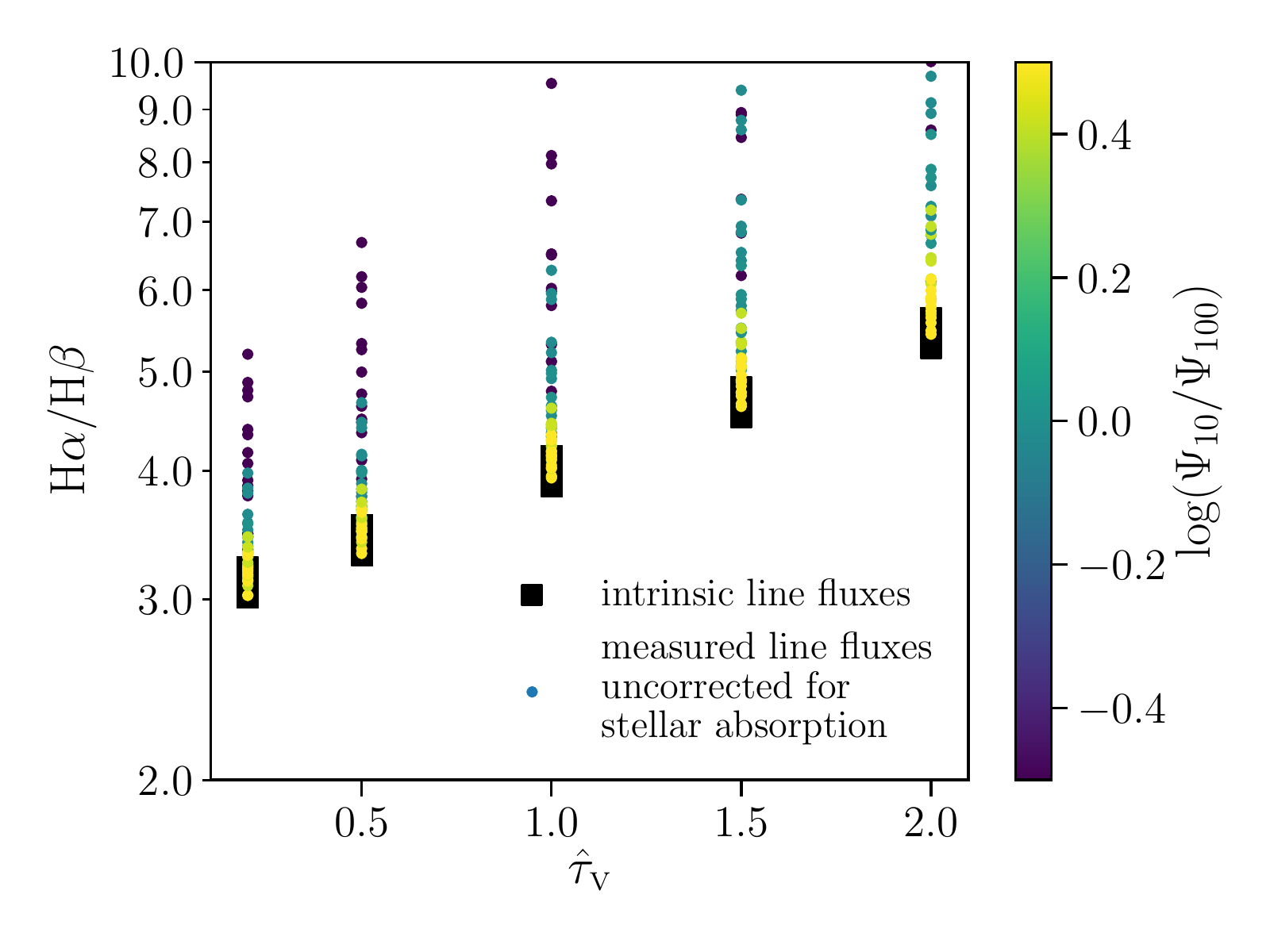}
  \caption{Balmer decrement plotted against \tauV\ for a grid of mock spectra produced with the \sfrFree\ SFH. Each spectrum is produced with the same current SFR (of 1$\Msun$/yr). The grid consists of log(Z/\Zsun)=$[-2,-1.5,-1,-0.5,0]$, log(\t/yr)$=[7.5,8,8.5,9,9.5]$, \tauV=$[0.2,0.5,1,1.5,2]$ and log(\MtotInLog/\Msun)=$[7.5,8,8.5,9,9.5,10]$, while \logUs\ is set to depend on Z and \xid\ fixed to 0.1.  The black squares show the intrinsic Balmer decrement, while the coloured circles show the range of Balmer decrements `measured' from H$\alpha$ and H$\beta$ line fluxes without correcting for underlying stellar absorption (see text for details). These are colour-coded by the logarithm of the ratio between the SFR averaged over the last 10\,Myr ($\Psi_{10}$), to the SFR averaged over the last 100\,Myr ($\Psi_{100}$).}
  \label{fig:Ha_Hb}
\end{figure}

\subsection{JWST exposure-time estimates}
\label{section:JWST}

As discussed in Section~\ref{section:results}, the S/N of the photometry fitted to has a direct impact on the tightness of derived constraints on \Mstar\ and \sfr, and hence, on whether or not the estimates will be biased. For example, the difference between output and input \Mstar\ and \sfr\  in the SED fits to the mock galaxies in the \constMF\ scenario described in Fig.~\ref{fig:sfrFree_vs_const} above depends on the S/N of the NIRCam F090W and F444W photometry, the estimates starting to become significantly biased when the S/N falls below $\sim50$.  However, a \sfrFree\ SFH provides unbiased \Mstar\ and \sfr\ estimates for S/N$\gtrsim20$, and we will take this as the minimum S/N for measuring the \MstarSFR relation.

\begin{table}
\centering
  \caption{\pandeia\ setup for exposure time calculations in Fig. \ref{fig:ET_plots}}
  \begin{tabular}{c|c|c}
  \hline
  & NIRCam & NIRSpec\\
  \hline \hline
  N Group & 6 & 19\\
  N Integration & 1 & 1\\
  Instrument setup & sw\_imaging F090W & H$\beta$: G235M/F170LP \\
                                 & lw\_imaging F444W & H$\alpha$: G395M/F290LP \\
  Source  & Point Source & Point Source\\
                &                        & line fwhm=50 km/s\\
  Extraction & aperture & 1x3 slitlet, Full MSA\\
  \hline
\end{tabular}
\label{tab:pandeia_setup}
\end{table}

\begin{figure*}
  \centering
  \subfigure{\includegraphics[width=3in]{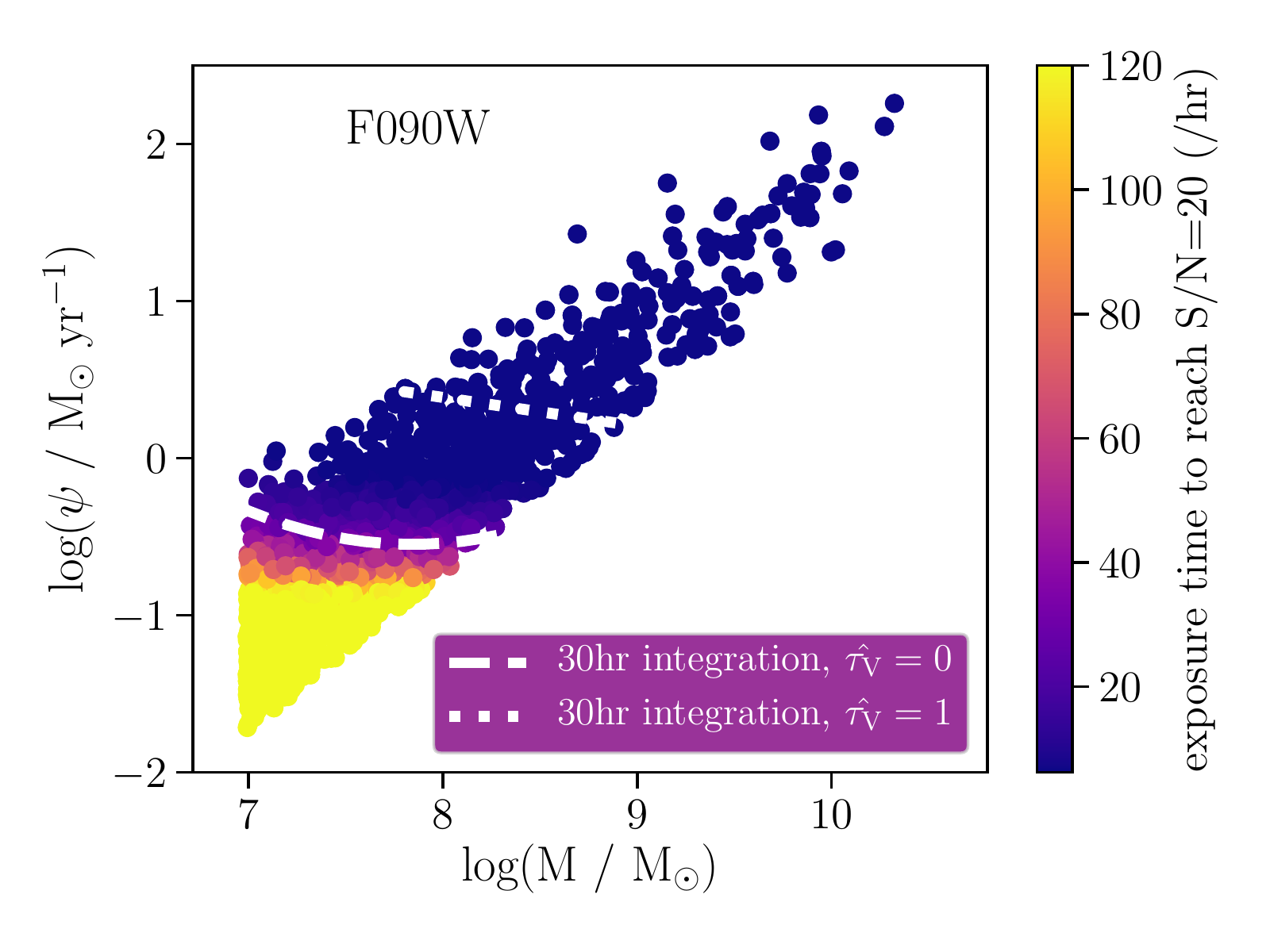}}
  \subfigure{\includegraphics[width=3in]{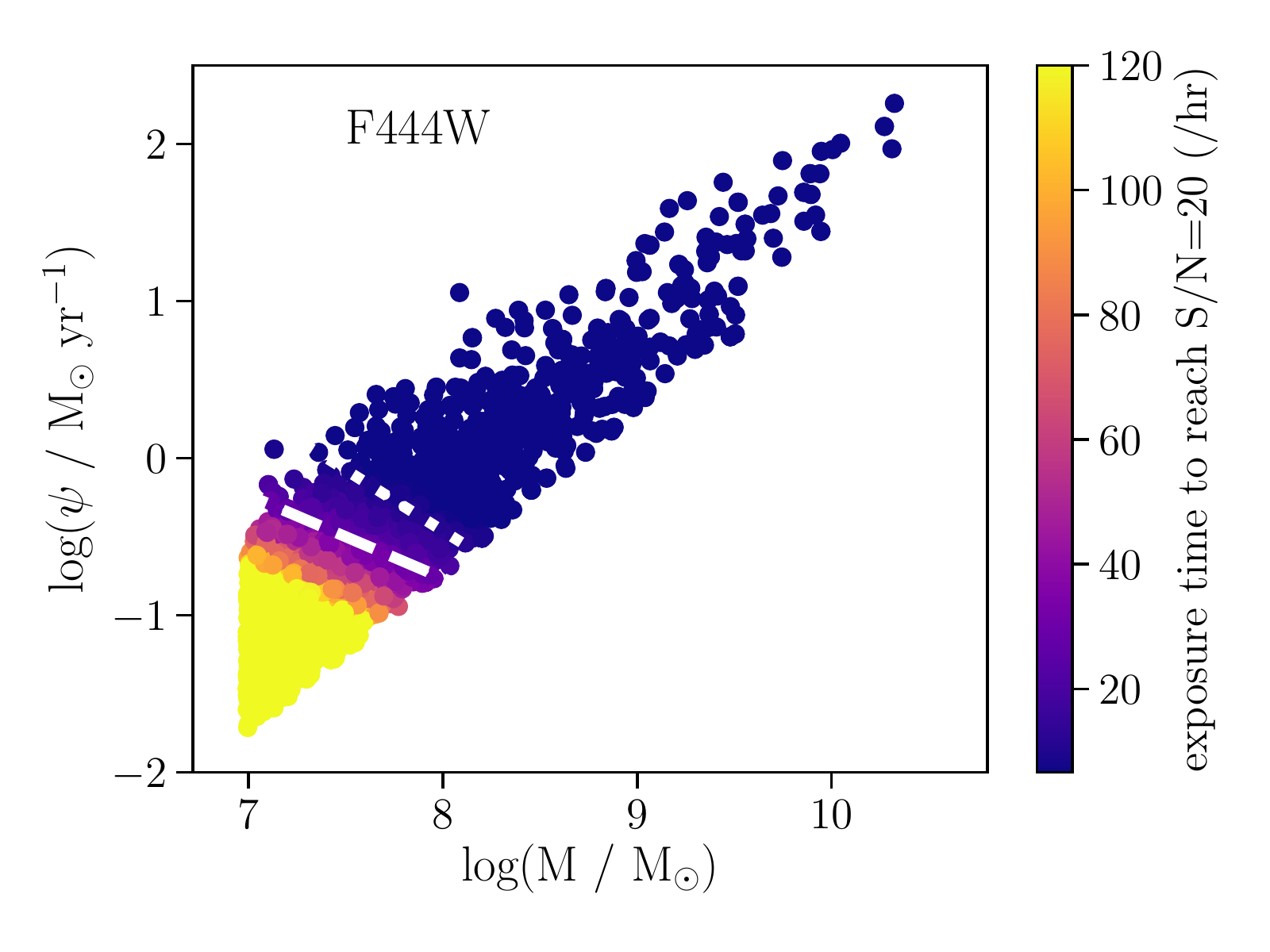}}
  \subfigure{\includegraphics[width=3in]{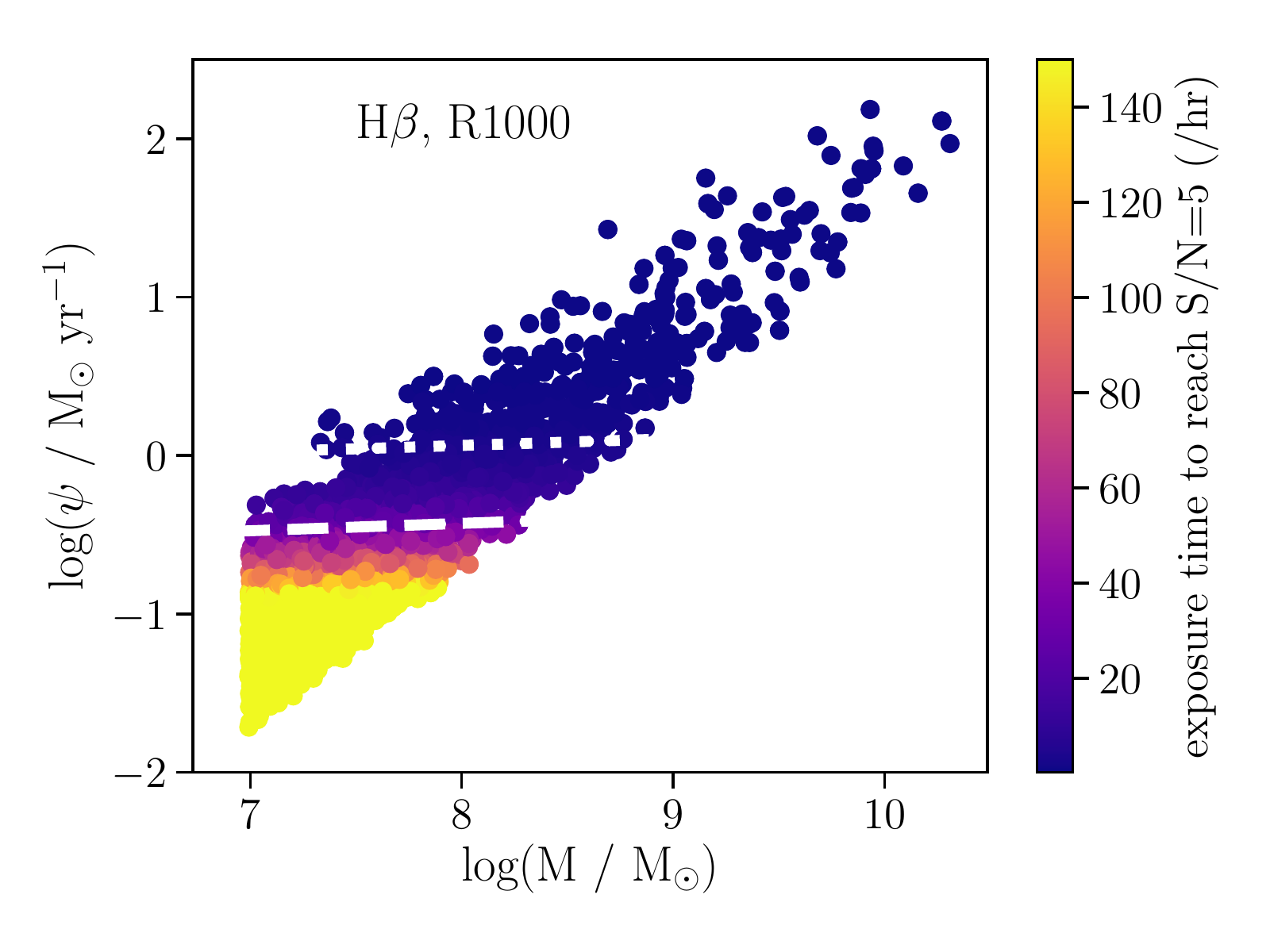}}
  \subfigure{\includegraphics[width=3in]{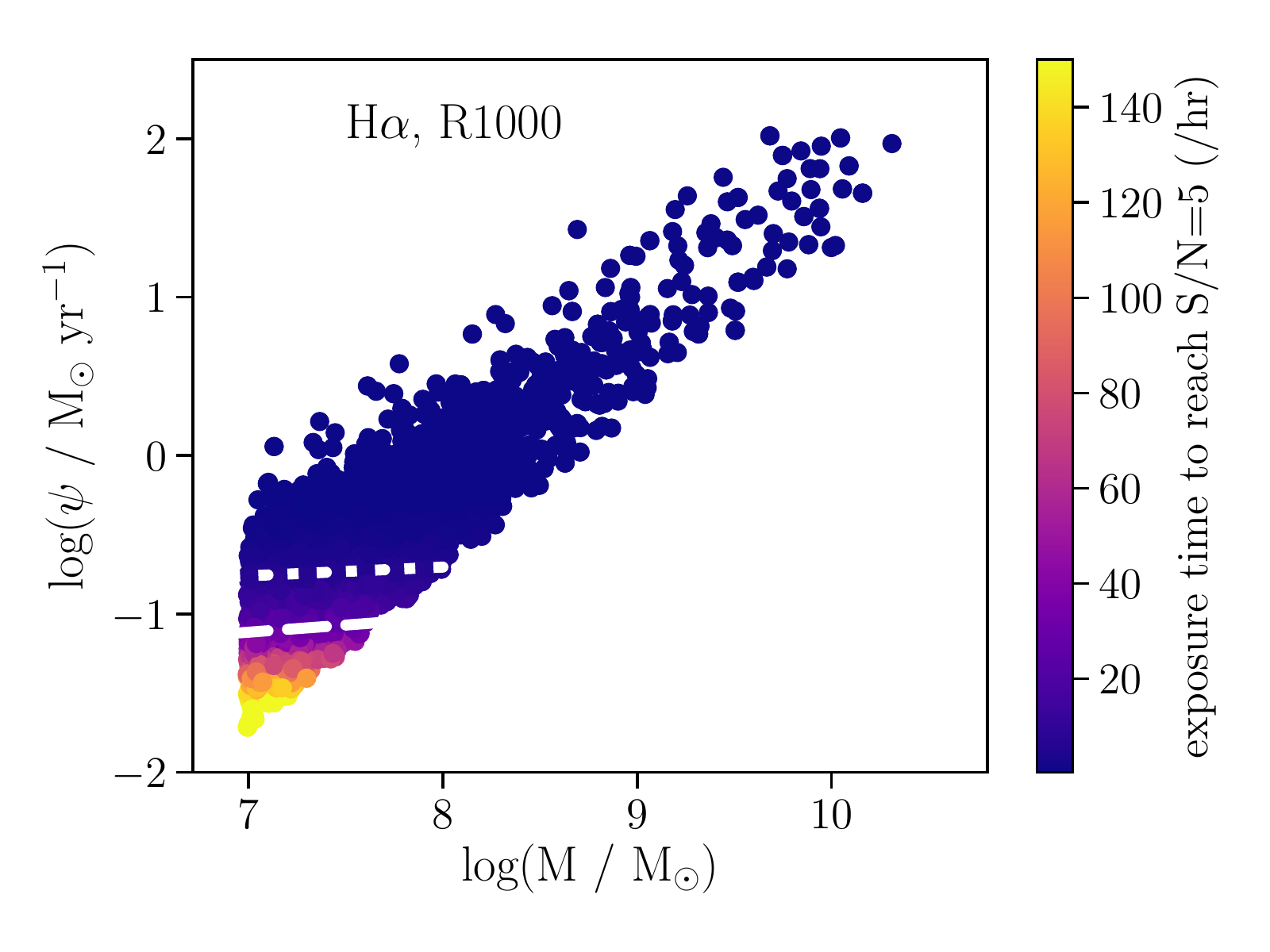}}
  \caption{ The \MstarSFR\ sequence with input intercept and slope chosen to match the San17 measurements at $z\sim5$ and a constant intrinsic scatter of 0.3\,dex. \beagle\ was used to generate a mock catalogue of 1000 galaxies drawn from the \protect\cite{Duncan2014} $z\sim5$ mass function, using constant SFHs and no dust.  The points in each panel are colour-coded by the exposure time required to meet a given S/N in different key observables.  The top two panels are colour-coded by the exposure time required to reach S/N$\sim$20 in the \JWST/NIRCam F090W and F444W filters, respectively.  The bottom two panels are colour-coded by the exposure time required to reach S/N$\sim$5 in integrated H$\beta$ and H$\alpha$ measurements at R$\sim1000$ with \JWST/NIRSpec. In each diagram, the white dashed line marks the lower range of the \MstarSFR\ relation probed in 30\,hr of exposure time, while the white dotted line shows how this range would change in the presence of dust attenuation with $\tauV=1$.}
\label{fig:ET_plots}
\end{figure*}

We use the Python version of \pandeia\ \citep{Pontoppidan2016}, the \JWST\ exposure time calculation (ETC) tool, to translate these required S/N values to required exposure times in the two NIRCam filters.  The tool provides S/N for a given exposure setup, so we invert the calculation to provide the exposure time at given S/N by calculating the S/N on grids of flux and exposure time. One is required to set various exposure settings within the tool, which we summarise in Table~\ref{tab:pandeia_setup}.\footnote{The readout patterns for NIRCam and NIRSpec are described in the respective jdocs pages,  https://jwst-docs.stsci.edu/display/JTI/NIRCam+Detector+Readout+Patterns and https://jwst-docs.stsci.edu/display/JTI/NIRSpec+Detector+ Recommended+Strategies.} The top two panels of Fig.~\ref{fig:ET_plots} display the \MstarSFR\ sequence colour-coded by the exposure time to reach a S/N of 20 in the F090W and F444W bands.  This main sequence was constructed from a mock catalogue of 1000 galaxies with \Mstar\ drawn from the \cite{Duncan2014} $z\sim5$ stellar mass function, and \sfr\ drawn from the \MstarSFR\ relation measured by San17 ($\intercept=-7.748$, $\slope=0.94$), with a constant intrinsic scatter of $\scatter=0.3$. The mock galaxies were constructed with constant SFHs without any dust.  We see that, in the absence of dust, NIRCam should reach mass-complete samples with the required S/N down to $\Mstar\sim8$ in 30\,hr per filter.  It is likely that a similar exposure time is required in the full complement of NIRCam filters, although we have not investigated the goodness-of-fit with variable depths in each filter.  We indicate how this exposure time is affected by the presence of dust with the dotted white line in Fig.~\ref{fig:ET_plots}, which shows which objects would require 30\,hr of integration to reach S/N$\sim20$ when dust is added to the SEDs using the CF00 dust prescription with $\tauV=1$ and $\mud=0.4$.

The bottom two panels in Fig~\ref{fig:ET_plots} show the exposure times required to reach S/N$\sim5$  in H$\alpha$ and H$\beta$ with NIRSpec at medium spectral resolution ($\mathrm{R}\sim1000$. The exposure times were estimated assuming the sources are point sources and centred on the central shutter of a $1\times3$ slitlet of open shutters in the micro-shutter array (MSA). These plots indicate that the constraints required to probe the intrinsic scatter of the \MstarSFR\ relation down to $\Mstar\sim8.5$ could be obtained with NIRSpec within 30\,hr of integration.  The R100 mode on NIRSpec observes the full 0.6--5\,$\mu$m range in a single exposure, but at R1000, the full spectral range is covered with three different filters. Thus, reaching the required S/N on both H$\alpha$ and H$\beta$ requires $\sim30$\,hr of integration in two different NIRSpec filters. We note that, although this is a highly idealised scenario, we are most interested in how far we can push measurements of intrinsic scatter down the mass function. At low stellar masses, we are more likely to meet the conditions of low amounts of dust and small galaxies that are only partially resolved with \JWST.

\section{Discussion}
\label{section:Discussion}

In this paper we have presented a Bayesian Hierarchical approach to modelling the main sequence that self-consistently propagates the uncertainties on \Mstar\ and \sfr\ estimates onto the uncertainties about the intercept, \intercept, slope, \slope\ and scatter, \scatter\ of the \MstarSFR\ relation.  We considered a set of four scenarios with increasing complexity to test key aspects of this approach and compared the results to three other methods based on ordinary linear regression.  In addition, these scenarios allowed us to test standard SED-fitting procedures and their impact on main sequence parameter retrieval. 

Our results show that poor physical parameter constraints can lead to biased estimates of \Mstar\ and \sfr.  Thus, the fraction of objects entering the sample with poor \Mstar\ and \sfr\ constraints is one of the most important factors in main-sequence parameter estimation. Simple estimates of \sfr\ based on UV-to-SFR calibrations can be used to avoid these issues, but may be biased themselves if the calibrations are not suitable for the populations under study, causing the simplifying assumptions to bias the measurement of intrinsic scatter.  Of similar importance to considerations of bias due to template degeneracies is the treatment of sample selection. The simple scenarios considered here show that explicitly modelling the distribution of \Mstar\ values is of secondary importance (as illustrated for example by the similarity of the results obtained with the \constGauss\ and \constMF\ scenarios), although this is likely due to low numbers of objects with good parameter constraints at any of the mass cuts probed.  

In this section, we further discuss our results in the context of previous measurements of the main sequence at high redshifts from photometric data as well as best practices for modelling the main sequence in order to derive estimates of the intrinsic scatter. 

\subsection{Measurements of intrinsic scatter at high redshift ($z\gtrsim3.5$) from the literature}

\subsubsection{Salmon et al. (2015)}
\label{section:salmon}

We find that \scatter\ is somewhat under-estimated when using the \PMSalmon\ method employed by Sal15 to estimate \sfr.  As mentioned in Section~\ref{section:methodSal15}, Sal15 additionally quantify the scatter introduced by uncertainties in \Mstar\ and \sfr\ measurements, and subtract this in quadrature from the measured scatter.  This approach would therefore further under-estimate the intrinsic scatter. 

The under-estimation of \scatter\ using the \PMSalmon\ SFRs in the \constGauss\ scenario is likely due to the difference in age-dependence of the \cite{Reddy2012} UV-to-SFR calibration, compared to the UV-to-SFR age-dependence of the \cite{Gutkin2016} stellar plus nebular models we use. However, we must note that the \cite{Reddy2012} calibration does not account for the stellar-metallicity dependence of the calibration, which would likely further under-estimate \scatter\ for a population of galaxies with a range of metallicities. In Fig.~\ref{fig:SFR_to_UV}, we plot the age-dependent calibration of \cite{Reddy2012}, converted to a \cite{Chabrier2003} IMF, compared to the calibration derived for the \cite{Gutkin2016} models for solar and 0.01 times solar metallicities, with and without nebular emission.  We set \logUs\  to follow the metallicity dependence of equation~\eqref{eq:logU_Z} without scatter, and $\xid=0.1$.  As discussed in Section~\ref{section:scenarios}, we self-consistently include the effects of dust in the \HII\ regions.  Very high values of \logUs\ give fainter luminosity at given SFR, because higher dust levels in the HII regions attenuate the stellar spectra.  Raising \xid\ also increases the attenuation of young stars from the surrounding \HII\ region.  The \logUs--Z relation avoids unphysical regions of the parameter space with high dust attenuation in the \HII\ regions, in particular a maximum $\tauHII=0.025$ is reached at $\log(Z/\Zsun)=-2$, $\logUs=-3.64$. A calibration based solely on the stellar continuum plus nebular continuum emission will not include this dust term.  The metallicity dependence of the rest-frame UV emission, and the contribution to it by the nebular continuum emission, should be accounted for in calibrations of SFR to UV luminosity, in particular when comparing samples over a wide redshift range.

\begin{figure}
  \centering
  \includegraphics[width=3.5in]{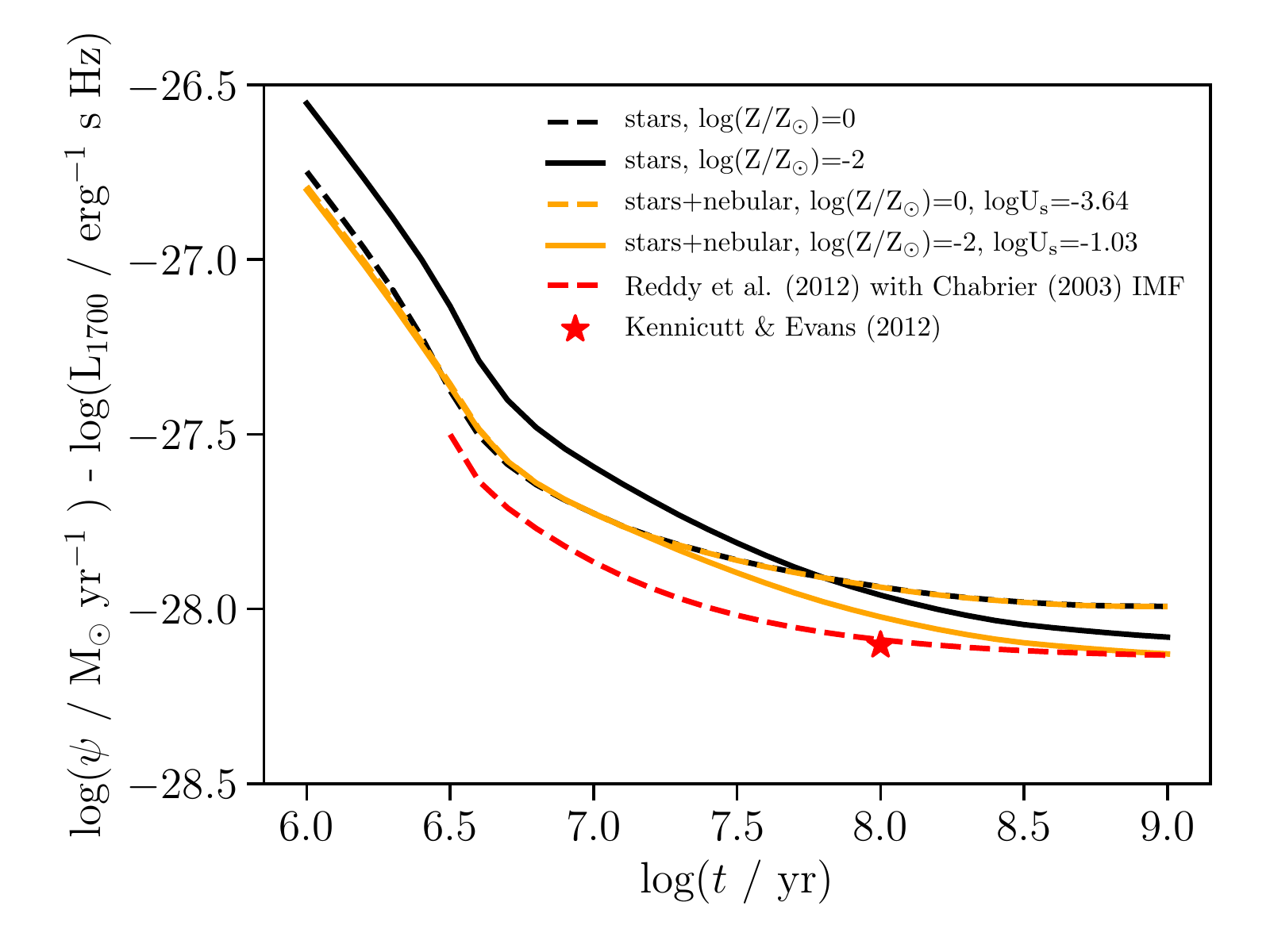}
  \caption{Logarithm of the ratio between SFR and luminosity at 1700\,\AA\ plotted against age for a constant SFH.  We show the conversion for the stellar models used in this work at solar metallicity (dashed black line), as well as the Reddy et al. (2012) conversion, converted to a Chabrier (2003) IMF (dashed red line).  The red star shows the Kennicutt \& Evans (2012) conversion.  We also display the influence on the conversion of a change in metallicity [dashed black line: $\log(\Z/\Zsun)=-2$] and of accounting for the contribution by nebular emission to the rest-frame UV luminosity [the orange (black) lines corresponding to the stellar+nebular (pure stellar) case]. For the models including nebular emission, \logUs\ values are assigned following the dependence on metallicity given by equation~\eqref{eq:logU_Z} (without scatter) to avoid unphysical \logUs\ values at a given metallicity.}
  \label{fig:SFR_to_UV}
\end{figure}

\subsubsection{Santini et al. (2017)}

San17 employ a sophisticated forward modelling approach to mitigate the effects imposed by their complicated selection function, which results from selecting magnified galaxies within the  Hubble Frontier Fields.  When using the method of San17 to estimate \sfr\ (the \PMSantini\ method), we find that \scatter\ is significantly under-estimated when the UV slope (required to correct the UV luminosity for dust) is well constrained (see Fig.~\ref{fig:constG_BB}), while it is over-estimated  when flux errors dominate UV-slope measurements (Fig.~\ref{fig:const_MF}).  San17 take these increased uncertainties in \sfr\ estimates toward low-S/N UV measurements into account in their analysis. However, we find that any increase in intrinsic scatter to low stellar masses is unlikely to be reliably measured with this method, as the underlying assumption that all objects have the same intrinsic UV slope will break down in practice.  The distribution of UV slopes is taken into account to some extent in their analysis by adding an uncertainty in quadrature to other uncertainties, but this approach is appropriate only if the distribution of values arises from measurement errors rather than intrinsic properties of the population.

\subsubsection{Kurczynski et al. (2016)}

 \cite{Kurczynski2016a} measure the intrinsic scatter at redshifts $0.5<z<3$ using the same model as we employ for the main sequence (equation~\ref{eq:main_sequence}).  They estimate the output joint uncertainties on \Mstar\ and \sfr\ as single, multivariate Gaussians (whereas we find that single Gaussians are not always a good representation of the uncertainties on \Mstar\ and \sfr; see Fig.~\ref{fig:GMM_example}), and take these into account when fitting the model parameters \intercept, \slope\ and \scatter.   Our tests indicate that \Mstar\ and \sfr\ are very poorly constrained from broad-band fluxes alone at $z\sim5$, and this can be diagnosed by investigating the dependence of the results on the prior on \tausfr. However, we note that the lower redshifts of the measurements presented in \cite{Kurczynski2016a} mean that emission-line contribution to broad-band fluxes is significantly lower than in our scenarios, potentially allowing for better constraints to be placed on \Mstar\ and \sfr\ from broad bands alone.

\subsubsection{Feldmann (2019)}
 
\cite{Feldmann2019} introduces \leopy, a package to estimate likelihoods for data while modelling the correlation between variables in a flexible way. \leopy\ accounts for complex measurement uncertainties, censored and missing data, as well as arbitrary marginal distributions of the variables.  This is in many respects an extension of K07, and can be used with a likelihood maximisation algorithm or as the likelihood estimate within a sampler to determine the posterior probabilities of the model parameters.  Although non-Gaussian uncertainties are accepted, accounting for multiple peaks in the uncertainty structure is not accounted for, whereas our extension to the K07 sampler can deal with this scenario.  For the current implementation, the accounting for multiple peaks in the joint \sfr--\Mstar\ posterior is of relatively minor importance. However, this point will be more important in instances where multi-modal posteriors are expected, e.g., when including the full photometric redshift uncertainty.  In addition, \leopy\ does not handle truncated data.  We have illustrated the importance of accounting for selection effects (i.e. data truncation), and discuss this further in Section~\ref{section:selection_effects}.

\leopy\ is particularly suited to applications where measurements involve data sets covering very different observational regimes.  \cite{Feldmann2019} presents models of the main sequence at $0.01<z<0.05$ using the extended GALEX Arecibo SDSS Survey \citep[xGASS, ][]{Catinella2018} catalogue. 
\cite{Feldmann2019} uses a zero-inflated gamma distribution to model the distribution of \sfr\ at given \Mstar, following the method proposed in \cite{Feldmann2017a}.  The gamma distribution accounts for the asymmetric \sfr\ distribution observed at low redshifts and avoids the problem of arbitrarily defining which galaxies inhabit the main sequence and which galaxies lie off it, either quenched or in the process of being so.  Our simple model of the main sequence does not account for an asymmetric distribution in \sfr. We discuss this point further in Section~\ref{section:modelling_complexity}.

\subsubsection{Boogaard et al. (2018)}

\cite{Boogaard2018} measure the low mass end ($7<\Mstar<10.5$) of the \MstarSFR\ relation at $0.11<z<0.91$ using dust-corrected SFRs based on H$\alpha$, H$\beta$ and H$\gamma$ lines detected with MUSE.  SFRs are corrected for dust using H$\alpha$/H$\beta$ for galaxies at $z<0.41$ and H$\beta$/H$\gamma$ at higher redshifts. \cite{Boogaard2018} model the \MstarSFR\ relation including redshift evolution and fit the intrinsic scatter perpendicular to the hyperplane defined by \Mstar, \sfr\ and log$(1+z)$ using the method of \cite{Robotham2015}.  This method self-consistently accounts for heteroskedastic measurement uncertainties.  The scatter perpendicular to the hyperplane is converted to an estimate of the intrinsic scatter in the \sfr\ direction. However, this re-projection neglects the non-uniform distribution of \Mstar\ values (see Section~\ref{section:limitations}). In our scenarios, a comparison between the \PM\ method (from ordinary linear regression that neglects the non-uniform distribution of \Mstar\ values) and the \Kelly\ method suggest that accounting for the distribution of \Mstar\ values is of secondary importance to correcting for selection effects.  However, we cannot comment directly on the relative importance of modelling the \Mstar\ distribution for the \cite{Boogaard2018} sample.

\cite{Boogaard2018} account for selection effects by simulating mock distributions of galaxies and applying a selection function based on emission-line flux to the mocks.  From the measured vs. true \MstarSFR\ relations, they define an affine transformation based on these mocks to apply to the measured relation parameters.  The mock galaxies and original SFR estimates do not account for variation in H$\alpha$-to-SFR (or H$\beta$-to-SFR) conversion with other parameters, such as metallicity or ionisation parameter, which may introduce systematic biases due to, e.g., the mass-metallicity relation.  Using \beagle\ to estimate SFRs from Balmer lines would allow one to marginalise over these uncertainties.

\subsection{Selection effects}
\label{section:selection_effects}

We have quantified two different types of selection effects on the recovery of \MstarSFR\ parameters in the results of the \constMF\ scenario (shown in Fig.~\ref{fig:const_MF}).   Mass incompleteness in the selection biases the measured slope, \slope, to shallower values and the intercept, \intercept, to higher ones, when these measurements are based on the `true' underlying values.  This is because the galaxies preferentially detected below the mass-completeness limit have high SFR (Fig.~\ref{fig:Testing_input_distribution}).  At the same time, adopting a mass limit based on consideration of the mass completeness limit using only estimated stellar masses can potentially introduce additional biases.  This is because parameter constraints are usually poorest at the low-mass end, and these poor constraints can lead to biased estimates of \Mstar\ and \sfr. The combination of correlated uncertainties in \Mstar\ and \sfr, which are a natural consequence of using an analytical SFH that does not explicitly decouple the current SFR from the accumulation of stellar mass, and biased estimates implies that galaxies can be preferentially scattered into the selection from a particular region about the \MstarSFR\ relation.  In our \constMF\ scenario, \sfr\ and \Mstar\ can be biased high at low stellar mass (Fig.~\ref{fig:const_MF_triangle}), meaning that including galaxies with poor parameter constraints bias estimates of the main sequence to steeper slopes and lower intercepts. 

One can account for these biases by explicitly modelling the effects of the selection criteria, measurement uncertainties and biases.  The K07 full model includes the possibility to model selection effects (see K07, section 5.1). However, the implemented model is only appropriate for the case when the two measurements are independent. This is not useful for our purpose, as the selection function depends on both \Mstar\ and \sfr, and the uncertainties are correlated between these two parameters and subject to biases.  Extending the model to allow for correlated errors in \Mstar\ and \sfr\ is discussed in K07, section 5.1.2. Although implementation of this model is beyond the scope of the present paper, we will investigate this type of modelling, as well as accounting for the effects of biases, in future work.  San17 \textit{do} explicitly account for the full complexity of the selection function, but their method for deriving \sfr\ biases the measurement of \scatter. To obtain unbiased constraints on the intrinsic scatter, we either need to explicitly model the selection effects, Eddington bias and biases in the \Mstar\ and \sfr\ estimates themselves, or we need to minimise this effect by choosing a mass cutoff that selects galaxies with tight constraints on \Mstar\ and \sfr.

\subsection{Modelling complexity in SED fitting}
\label{section:modelling_complexity}

Increasing modelling complexity in SED fitting introduces degeneracies that can hamper physical parameter determination.  It can be advantageous to avoid these degeneracies by, for example, decoupling the estimation of \sfr\ from the attenuation and age estimates derived from SED fitting.  Doing so requires simplifying assumptions, as in the method used by San17 to estimate SFR. However, our results show that when searching to constrain the level of \textit{variety} in a population, rather than population mean relations, this approach is flawed.

We note that including more uncertain physics introduces a risk of over-fitting to the data, and also diluting the information the data contain in the sea of uncertainties.  
We should include here at least enough variation to allowinvestigation ofthe intrinsic scatter in the main sequence, and the possible dependence on stellar mass, namely:
\begin{enumerate}
\item Variation in the ratio of past to present SFR;
\item Variation in stellar and nebular emission due to metallicity, \logUs\ (and \xid, if using constraints from lines other than hydrogen);
\item Variation in the dust attenuation curve.
\end{enumerate}
\noindent
We describe how our implementation includes variation in the dust attenuation curve via the implementation of the CF00 dust model in Appendix~\ref{app:dust}.

Different spectral synthesis models allow one to explore different underlying physical models such as: inclusion or not of binary stars, which have been incorporated into the BPASS models \citep[Binary Population and Spectral Synthesis;][]{Eldridge2017,Stanway2018}; stellar rotation, which can be explored using \starburst\ \citep{Leitherer1999}, using the Geneva stellar evolutionary tracks for single rotating stars \citep{Ekstrom2012,Georgy2013}; treatment of hot stellar atmospheres, in particular those of hot O- and B-type stars, and Wolf-Rayet stars for which the \tlusty\ model grid of \citet[for O- and B-type stars]{Hubeny1995}, and \powr\ library of \citet[for Wolf-Rayet stars]{Hamann2004} are incorporated into the stellar plus nebular models of \cite{Gutkin2016}.  The differences between evolutionary stellar spectral synthesis models likely cause systematic differences in our inferences.  However, the underlying physical models are not generally parameterized in such a way that they can be sampled over continuously, which would allow the associated uncertainties to be subsumed in our measurement uncertainties. The best way to encompass those uncertainties currently would be to compare main-sequence determination obtained with different model assumptions and see to what extent they provide the same qualitative evolution and intrinsic scatter estimates.

\subsection{Limitations of this analysis }
\label{section:limitations}

The model adopted in this paper to describe the main sequence (equation \ref{eq:main_sequence}) is quite simple, primarily because current data at high redshift have had limited constraining power for more complicated models.  In this context, the shape of the main sequence at low redshift, as measured from SDSS or from simulations of galaxy formation and evolution, can provide insight into how our model may need to be improved once more complete high-redshift samples and better \sfr\ constraints are available.

For example, at low redshift, there is evidence for a flattening of the relation \citep{Whitaker2014}, and potentially a rise in the intrinsic scatter toward high stellar masses \citep{Guo2013a}, where feedback from the growth of a central super-massive black hole can quench star formation \citep{Matthee2019}. Our current model does not account for this type of mass dependence in the intrinsic scatter, and we have only discussed the possibility of measuring an increase in scatter at the low-mass end. 

To a certain extent, the inferred form of the main sequence at high mass is sensitive to how one selects star-forming vs. non star-forming galaxies, as well as how the main sequence is defined. For example,  \cite{Renzini2015} offer an objective definition of the main sequence as the ridge in the mass-SFR-number density plot.  By this definition the relation is measured from the mode in the SFR distribution at given stellar mass, which is less sensitive to the definition of what constitutes a star-forming galaxy than the median or mean.   \cite{Renzini2015} see no flattening of the main sequence to high stellar masses when it is measured in this way.  In essence, our model is consistent with this definition of the main sequence.  High-mass sources with lower number density that sit below the main relation will not strongly impact the derived \MstarSFR\ properties, unless the mass completeness limits themselves are very high and so only these objects are included.  As such, the model we employ is not suitable for tracing the flattening of the main sequence or the mass dependence of the scatter at high stellar mass due to AGN feedback.  An extension to the model would have to allow for asymmetric scatter at the high-mass end.

It is also worth noting that we have investigated here a model with intrinsic scatter only in the direction of \sfr, which is standard for studies of the main-sequence at high redshift.  At low redshift, however, it is often standard to measure the scatter in specific SFR, or even perpendicular to the main-sequence.  For a population with uniform distribution in the independent variable and Gaussian scatter about a linear relation in the dependent variable, the corresponding scatter perpendicular to the relation will also be Gaussian.  As such, modelling the scatter in the y-direction or perpendicular to the relation are essentially equivalent to each other. However, this is no longer the case when including the number density distribution imposed by a stellar mass function.  Specifically, if the intrinsic scatter is Gaussian perpendicular to the relation, the scatter will be skewed to large SFR values in the y-direction.   
One way to mitigate the biases introduced by an incorrect assumption of the form of the intrinsic scatter would be to allow the scatter to be non-symmetric, e.g. by introducing skew to the Gaussian scatter.

We have not considered so far the photometric-redshift quality in the selection criteria and ignored its influence on physical parameter uncertainties (we only allow redshift to vary over the range $4.5<z<5.5$ in the SED fitting).  As shown by \cite{Kemp2019}, photometric-redshift uncertainties impact stellar-mass estimations significantly for faint objects or objects with significant dust attenuation. 

All the tests presented in this paper were performed measuring \Mstar\ and \sfr\ with the `correct' SFH (i.e. adopting the same SFH in the fitting as used to produce the spectra).  We also fit with the same stellar and nebular emission models that were used to produce the spectra.  We will break these criteria in future work.

Finally, we note that the nebular-emission models of \cite{Gutkin2016} assume ionization-bounded nebulae opaque to hydrogen-ionizing photons ($f_\mathrm{esc}=0$ where $f_\mathrm{esc}$ is the escape fraction of ionizing photons).  \cite{Plat2019} investigate the effects of density-bounded nebulae allowing for some escape of ionizing photons. Incorporating these models into our analysis is beyond the scope of the present paper.

\section{Conclusions}

We have presented a Bayesian Hierarchical approach, based on the model of \cite{Kelly2007}, to self-consistently propagate the full measurement uncertainties on stellar mass and SFR derived from SED fitting to the derived population-wide main-sequence parameters (intercept, \intercept, slope, \slope\ and intrinsic scatter, \scatter).  The Bayesian Hierarchical model takes as input the joint posterior probabilities on \Mstar\ and \sfr\ from \beagle, or any other Bayesian SED-fitting code, and in turn provides self-consistent posterior probability distributions on \intercept, \slope\ and \scatter.  In this work we model the main sequence as a linear relation with uniform, Gaussian intrinsic scatter (see equation~\ref{eq:main_sequence}) and test the Bayesian Hierarchical approach on a set of idealised scenarios, while comparing to standard methods employed in the literature. It is possible to extend the model to allow a dependence of the intrinsic scatter on stellar mass, and we will explore this in future work.

With idealised scenarios for which we know the input main sequence, we show that the Bayesian Hierarchical method provides robust determinations of the main sequence parameters and their uncertainties when fitting to data using a constant SFH and when photometric constraints are good.  However, broad-band fluxes alone cannot easily constrain the contribution from nebular emission lines at high redshifts.  This means that, as the equivalent width of emission lines increase with redshift, providing a higher fractional contribution to broad-band fluxes, SFR estimates derived from photometric fitting are not primarily derived from the UV slope.  This in turn results in SFR estimates derived from photometric data being sensitive to star formation on short timescales, which is not generally the case at lower redshifts.

When fitting to photometric data with flexible stellar+nebular models at high redshifts, it is important to determine how well the parameters driving nebular-continuum and emission-line strength (i.e. \logUs\ and \xid\ in the case of the \citealt{Gutkin2016} models) are constrained. If they are unconstrained, the derived stellar masses and star formation rates can be biased, the extent of the bias depending on the underlying priors.  In the idealistic scenarios considered here, the bias is toward over-estimating stellar mass and SFR.  When these parameters are unconstrained, a suitable prior can be set to mitigate such biases, although the results will then become dependent on the chosen prior.  NIRCam medium-band filters provide the necessary constraints on nebular-emission contribution to broad-band fluxes to mitigate these biases without the need for a prior on \logUs\ when fitting with constant SFHs. We find that with broad-band photometric constraints alone, delayed exponential SFHs have poor \Mstar\ and \sfr\ constraints that lead to biased estimates of \intercept, \slope\ and \scatter\ in our idealised scenarios, even when constraining priors on \logUs\ and \xid.  NIRCam medium-band filters will mitigate these biases.

Decoupling the stellar-mass and SFR estimates by estimating the SFR from a dust-corrected rest-frame UV luminosity may reduce the impact of biases due to un-constrained parameters, but at the price of neglecting variation among the population.  For example, not accounting for the variations in UV-to-SFR calibrations driven by metallicity and stellar age leads to under-estimates of the intrinsic scatter.  When investigating variation within the population, the modelling must include enough complexity to replicate the likely variation we expect to see.  Specifically, correcting MUV for dust using the \cite{Meurer1999} prescription will wrongly attribute variations in UV slope arising from age, present-to-past star formation activity or metallicity, to dust attenuation.  

The biases in \Mstar\ and \sfr\ estimates for individual galaxies, introduced when photometric constraints are poor, also impact which galaxies would be selected when imposing a given cut on measured \Mstar. Thus, sample selection based on expected mass completeness alone is not sufficient to prevent bias of the intrinsic scatter measurement.  

In addition to the above considerations, standard SED-fitting assumptions of constant or rising SFHs are not appropriate for investigating whether the intrinsic scatter increases to low stellar masses.  In particular, the underlying prior on age restricts \Mstar\ and \sfr\ estimates to an increasingly constricted region of the \MstarSFR\ plane with increasing redshift.  For this reason, it is not advised to use these SFHs to provide the stellar-mass estimates if aiming to investigate any potential mass dependence of the intrinsic scatter, even if they were to be used with independent SFR estimates.  A delayed exponential SFH is flexible enough to fill in the \MstarSFR\ parameter space below the maximum age-of-the-universe constraint inherent with constant SFHs. However, given that photometric data \textit{are} sensitive to SFR variations on short timescales at high redshifts, we propose the use of  two-component star formation histories (e.g. constant or even delayed exponential SFH with a 10\,Myr burst of recent star formation) to be more appropriate to investigate increasing scatter at low stellar mass, if it is due to stochastic star formation.

We find that for the filter set and corresponding depths probed with our idealised scenarios (including medium-band constraints), a S/N$\sim20$ in NIRCam F090W and F444W filters is sufficient to avoid biases with a \sfrFree\ SFH.  Based on simple point-source exposure time estimates calculated with \pandeia\ \citep{Pontoppidan2016}, we find that NIRCam should reach S/N$\sim20$ down to $\log(\MstarInLog/\yr)\sim8$ at $z\sim5$ in F090W and F444W within 30 hours of integration per filter.  When using H$\alpha$ and H$\beta$ to provide SFR estimates, care is needed with the uncertain stellar Balmer absorption in the case of stochastic star formation.  R1000 NIRSpec spectroscopy will reach S/N$\sim5$ down to $\log(\MstarInLog/\yr)\sim8$ in H$\alpha$ within 30 hours, and down to $\log(\MstarInLog/\yr)\sim8.5$ in H$\beta$ (assuming point sources and no dust attenuation).

\section*{Acknowledgments}
The authors would like to thank the anonymous referee whose skepticism of the original draft helped us significantly improve our treatment of dust in \HII\ regions within \beagle.  The paper is significantly improved thanks to their rigor and attention during the referee process.  ECL gratefully acknowledges the abundant and fruitful discussions with David Stenning and Tom Charnock as well as the invaluable support of Olga Degtyareva et al. ECL, JC, and SC acknowledge support from the European Research Council (ERC) via an Advanced Grant under grant agreement no. 321323-NEOGAL.  ECL and LS additionally acknowledges support from the  ERC Advanced Grant 695671 ‘QUENCH’ as well as the Science and Technology Facilities Council (STFC). This research made use of Astropy, a community-developed core PYTHON package for astronomy \citep{Robitaille2013,TheAstropyCollaboration2018}, NumPy and SciPy \citep{Oliphant2007}, Matplotlib \citep{Hunter2007}, IPython \citep{Perez2007} and NASA’s Astrophysics Data System Bibliographic Services. 

\section*{Data Availability}
The simulated data on which the results in this article will be shared on reasonable request to the corresponding author.

\bibliographystyle{mn2e}

\bibliography{library}

\appendix

\section{Variation in the dust attenuation curve from the CF00 dust model}
\label{app:dust}

Attenuation by dust is a key uncertainty in galaxy spectral fitting. \cite{Reddy2012} find that using the \cite{Calzetti2000} attenuation curve when fitting to photometry of high-redshift galaxies can lead to un-physically low ages.  More physically plausible ages are obtained by using a steeper  attenuation curve similar to the SMC extinction\footnote{We follow the standard nomenclature and refer by attenuation to the combined effects of absorption and scattering in and out of the line of sight to a galaxy caused by both local and global geometric effects, while the term extinction is reserved for photon absorption along and scattering out of a single line of sight \citep[e.g.][]{Charlot2000}.}  curve \citep{Pei1992}.

Here we propose to use the CF00 dust prescription to provide a way of marginalising over the uncertainty in dust attenuation curves in a physically-consistent way.  As described in Section~\ref{section:scenarios}, the CF00 two-component model accounts for the extra attenuation of light from young stars that still reside in their birth clouds compared to that from older stars floating in the diffuse ISM.  The ISM has an attenuation curve which is modelled with a power-law slope of 0.7, while the power-law slope of modelled attenuation curve of the birth clouds is 1.3. A result of this prescription is an effective, galaxy-wide attenuation curve that is dependent on the distribution of stellar ages in the galaxy.  This is best appreciated in Fig.\ref{fig:CF00}, showing that varying current-to-past star-formation rate provides effective, galaxy-wide attenuation curves that approximate both the \cite{Calzetti2000} and SMC dust curves.  This plot was produced by sampling 1000 random objects with \sfrFree\ SFH, and \Z, \logUs\ sampled from the full available parameter space of the models.

\begin{figure}
  \centering
  \includegraphics[width=3.5in]{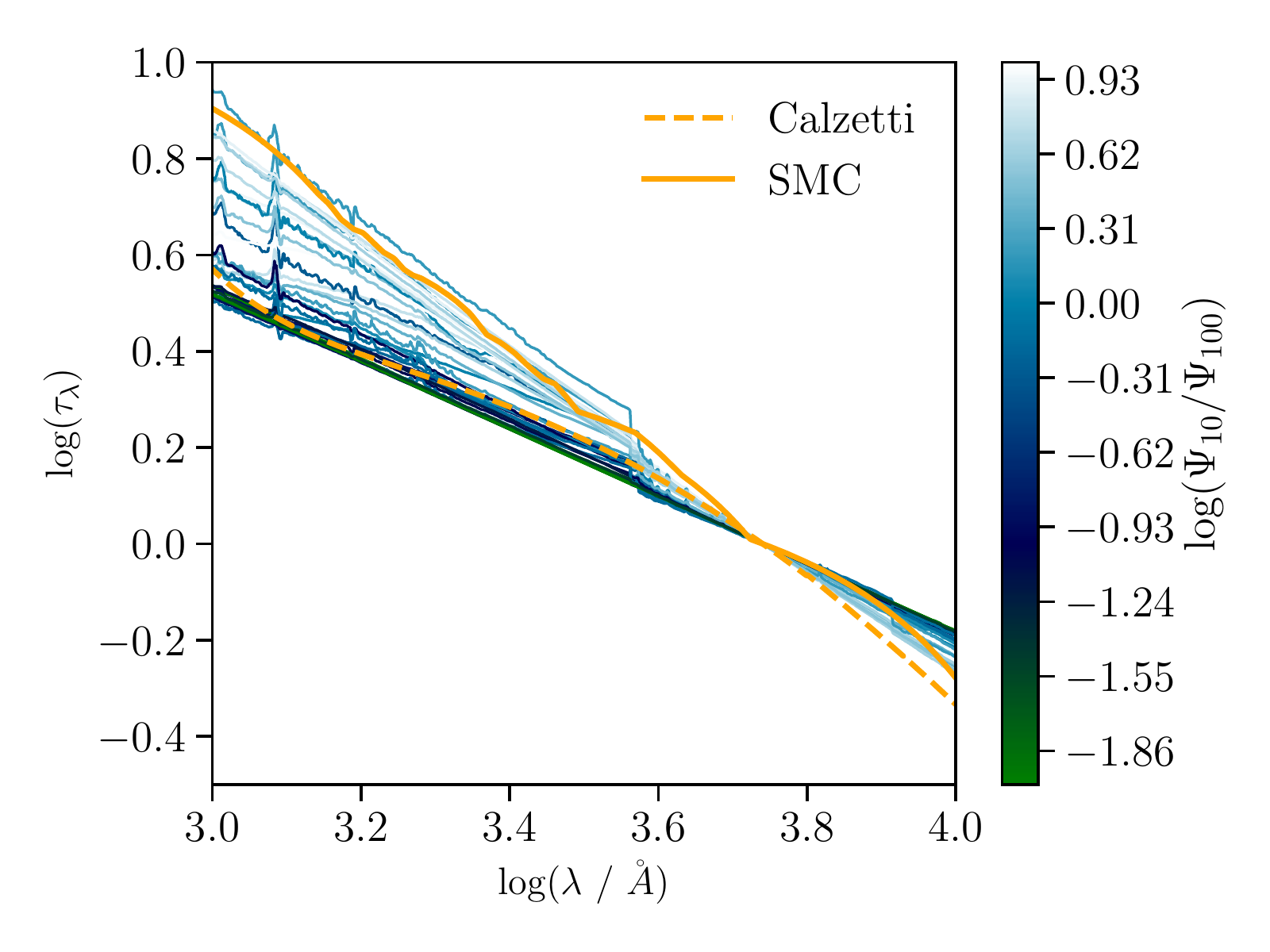}
  \caption{Effective galaxy-wide attenuation curves of 1000 galaxy templates with two-component SFHs (constant SFH with recent 10\,Myr burst of star formation). The templates were drawn from the parameter space used to produce the \constMF\ scenario, with the recent 10\,Myr of star formation allowed to vary in the range $-4<\log(\sfrInLog/\Msun\,\mathrm{yr}^{-1})<4$.  Each line is colour-coded by $\sfr_{10}/\sfr_{100}$, where $\sfr_{10}$ is \sfr\ averaged over the last 10\,Myr, and $\sfr_{100}$ is \sfr\ averaged over the last 100\,Myr.  Also plotted in the thick, solid, orange line is the SMC attenuation curve as measured by \protect\cite{Pei1992}, while the dashed, orange line shows the \protect\cite{Calzetti2000} attenuation curve.}
  \label{fig:CF00}
\end{figure}

\label{lastpage}

\end{document}